\documentclass[preprint2]{aastex6}  

\usepackage{natbib}
\usepackage{amsmath}
\usepackage{graphicx}
\usepackage{subfig}

\newcommand{\chandra}{\emph{Chandra}}

\begin{document}


\title{VLA Radio Observations of the HST Frontier Fields Cluster Abell 2744: The discovery of new radio relics}
\shorttitle{VLA observations of Abell~2744}
\shortauthors{Pearce et al.}

\author{C.~J.~J.~Pearce\altaffilmark{1,2}, R.~J.~van~Weeren\altaffilmark{1}$^{\star}$,  F.~Andrade-Santos\altaffilmark{1}, C.~Jones\altaffilmark{1}, W.~R.~Forman\altaffilmark{1}, M.~Br\"uggen\altaffilmark{3}, E.~Bulbul\altaffilmark{4}, T.~E.~Clarke\altaffilmark{5}, R.~P.~Kraft\altaffilmark{1}, E.~Medezinski\altaffilmark{6}, T.~Mroczkowski\altaffilmark{7},  M.~Nonino\altaffilmark{8}, P.~E.~J.~Nulsen\altaffilmark{1,9}, 
S.~W.~Randall\altaffilmark{1}, K.~Umetsu\altaffilmark{10} 
}


\email{connorjjpearce@gmail.com}

\altaffiltext{1}{Harvard-Smithsonian Center for Astrophysics, 60 Garden Street, Cambridge, MA 02138, USA}
\altaffiltext{2}{School of Physics and Astronomy, University of Southampton, Highfield, Southampton SO17 1BJ, UK}
\altaffiltext{3}{Hamburger Sternwarte, Universit\"at Hamburg, Gojenbergsweg 112, D-21029 Hamburg, Germany}
\altaffiltext{4}{Kavli Institute for Astrophysics and Space Research, Massachusetts Institute of Technology, \\77 Massachusetts Avenue, Cambridge, MA 02139}
\altaffiltext{5}{U.S. Naval Research Laboratory, Remote Sensing Division, 4555 Overlook Ave SW, Washington, D.C. 20375, USA}
\altaffiltext{6}{Department of Astrophysical Sciences, Princeton University, Princeton, NJ 08544, USA}
\altaffiltext{7}{European Organization for Astronomical Research in the Southern hemisphere, Karl-Schwarzschild-Str. 2, D-85748 Garching b.
M\"unchen, Germany}
\altaffiltext{8}{INAF-Trieste Astronomical Observatory, via Bazzoni 2, 34124 Trieste, Italy} 
\altaffiltext{9}{ICRAR, University of Western Australia, 35 Stirling Hwy, Crawley WA 6009, Australia}
\altaffiltext{10}{Institute of Astronomy and Astrophysics, Academia Sinica, PO Box 23-141, Taipei 10617, Taiwan\\ \\ $\star$ Clay Fellow}

\begin{abstract}
Cluster mergers leave distinct signatures in the ICM in the form of shocks and diffuse cluster radio sources that provide evidence for the acceleration of relativistic particles. However, the physics of particle acceleration in the ICM is still not fully understood. Here we present new 1--4 GHz Jansky Very Large Array (VLA) and archival \chandra\ observations of the HST Frontier Fields Cluster \object{Abell~2744}. 
In our new VLA images, we detect the previously known $\sim2.1$~Mpc radio halo and $\sim1.5$~Mpc radio relic.  We carry out a radio spectral analysis from which we determine the relic's injection spectral index to be $\alpha_{\rm{inj}} =  -1.12 \pm 0.19$. This corresponds to a shock Mach number of $\mathcal{M}$ = 2.05$^{+0.31}_{-0.19}$ under the assumption of diffusive shock acceleration. We also find evidence for spectral steepening in the post-shock region. We do not find evidence for a significant correlation between the radio halo's spectral index and ICM temperature. In addition, we observe three new polarized diffuse sources and determine two of these to be newly discovered giant radio relics. These two relics are located in the southeastern and northwestern outskirts of the cluster.  The corresponding integrated spectral indices measure $-$1.81 $\pm$ 0.26 and $-$0.63 $\pm$ 0.21 for the SE and NW relics, respectively. From an X-ray surface brightness profile we also detect a {possible} density jump of $R=1.39^{+0.34}_{-0.22}$ co-located with the newly discovered SE relic. This density jump would correspond to a shock front Mach number of $\mathcal{M}=1.26^{+0.25}_{-0.15}$.

\vspace{5mm}
\end{abstract}




\section{Introduction}
In the standard $\Lambda$CDM cosmological model, galaxy clusters form via a hierarchical sequence of merging events of smaller structures, growing from small groups of galaxies to major clusters. Cluster mergers, driven by the gravitational interaction of the dominant dark matter, can release up to $10^{63}-10^{64}$ ergs of energy on timescales of ${\sim}$~Gyr (one cluster crossing time). Shocks are driven into the ICM during such mergers which dissipate the energy into heating the gas, along with the subsequent turbulent ICM motions that follow.

Part of the energy that is dissipated during these collisions could be fed into (re-)accelerating relativistic particles and amplifying the ICM magnetic field, resulting in cluster-scale diffuse synchrotron radiation. At radio wavelengths we see evidence of this acceleration in the form of radio halos and radio relics (for reviews see \cite{2012A&ARv..20...54F, 2014IJMPD..2330007B}).

Radio halos are large ($\gtrsim$ 1 Mpc) diffuse radio sources located in the centre of a cluster. They have a smooth regular morphology and are unpolarized down to a few percent level. Their emission typically follows the observed thermal X-ray emission and has a steep spectrum (${\alpha}\lesssim{-1}$). Radio relics are irregularly shaped sources of a similar scale to halos ($\sim0.5$--2.0 Mpc), but are located in the cluster periphery and polarized at the 10-60$\%$ level, indicating the presence of ordered magnetic fields. Like halos, they  also have steep synchrotron spectra. A strong correlation between cluster-scale synchrotron emission and cluster mergers is observed due to the fact that these radio sources are preferentially found in dynamically disturbed systems \citep{2010A&A...517A..10C}. The exact origin of the radio emission in these features is still being debated. The main problem is reconciling the relatively short radiative lifetime of the electron ($\sim10^{8}$~yrs) with the Mpc size and $\sim$Gyr age of these sources. Therefore some form of an in-situ acceleration process must be occurring \citep{1977ApJ...212....1J}.

For radio halos the current theories are that of turbulent re-acceleration \citep[\textit{primary models,}][]{2001MNRAS.320..365B, 2001ApJ...557..560P}, and the production of secondary electrons during collisions between thermal ICM protons and cosmic ray protons trapped in the ICM \citep[\textit{secondary models,}][]{1980ApJ...239L..93D}.

Radio relics are  subdivided into three categories: giant radio relics, AGN relics, and radio phoenices \citep{2004rcfg.proc..335K}. Giant radio relics are Mpc-size arc-like sources of synchrotron radiation. The leading theory behind their formation is that of shock acceleration \citep{1998A&A...332..395E,1983RPPh...46..973D} of either thermal or pre-accelerated fossil electrons from AGN or other radio galaxy activity \citep[e.g.][]{2005ApJ...627..733M, 2008A&A...486..347G, 2011ApJ...734...18K, 2013MNRAS.435.1061P,2017NatAs...1E...5V}, although other models exist \citep{2015ApJ...815..116F}.

Radio phoenices are also believed to be caused by shocks. However, in this case the shocks only adiabatically compresses fossil relativistic plasma from AGN \citep{2001A&A...366...26E, 2011A&A...527A.114V, 2015MNRAS.448.2197D}. AGN relics on the other hand are regions/lobes of fossil plasma from radio galaxies without any re-acceleration/compression having occurred.

In this paper we present a radio spectral analysis of the merging cluster Abell~2744 from Jansky Very Large Array (VLA) observations at 1--4~GHz. With these  observations we study the nature of the diffuse radio emission in this cluster. In Section~\ref{sec:introa2744} we provide an overview of Abell~2744, describing the results of previous X-ray, radio, and optical studies.  In Section~\ref{sec:data} we explain the calibration and reduction processes used for both the X-ray and radio data. In Section~\ref{sec:results} we present the radio maps constructed at various resolutions along with the associated spectral index maps. Integrated radio spectra and a polarization vector map are also shown. In Section~\ref{sec:discussion} we discuss the application of our results to the current proposed synchrotron acceleration theories. We extract radial profiles of the spectral index across both the radio relic and halo, and perform an analysis of the spectral index variation across the radio halo. Three previously unobserved diffuse sources are discussed in more detail. The results are summarized in Section~\ref{sec:conclusion}. 

Throughout this paper we assume a flat, $\Lambda$CDM cosmology with $H_{0} = 70$ kms$^{-1}$ Mpc$^{-1}$, matter density $\Omega_{m} = 0.3$ and dark energy density $\Omega_{\Lambda} = 0.7$ \citep{2014ApJ...794..135B}. 
At the cluster's redshift, 1\arcmin~corresponds to a scale of $\sim$ 272~kpc. 

Cluster, foreground and lensed compact radio sources will be discussed in detail in a separate paper.

\section{Abell 2744}
\label{sec:introa2744}
Abell~2744 is a complex merger event located at $z=0.308$. Owing to its virial mass and large area of high magnification, it was chosen as one of the HST Frontier Field clusters \citep{2014AAS...22325401L,2016arXiv160506567L}. 

X-ray observations have revealed several substructures near the centre, including cold and dense remnant gas cores to the north and south, a prominent hot gas cloud between the two main galaxy groups, as well as an additional X-ray luminous structure to the northwest  \citep{2004MNRAS.349..385K}. Kinematic studies also showed a bimodal velocity dispersion in the cluster centre, with a third group of cluster members associated with the Northwestern X-ray peak \citep{2006A&A...449..461B,2009A&A...500..947B}. 

The merger scenario for this cluster is that of a primary, bullet-like {merger event \citep{2002ApJ...567L..27M}} that  took place in the N-S direction with a large line-of-sight (LOS) component to the merger axis, and that a secondary event is occurring with the infall of  a third sub-cluster \citep{2004MNRAS.349..385K}. From X-ray analysis it was suggested that the merger mass ratio of the sub-clusters is roughly equal \citep{2004MNRAS.349..385K} though it was unclear if the dominant massive cluster core was located to the South \citep{2011ApJ...728...27O} or the North \citep{2004MNRAS.349..385K}. The role of the Northwest sub-cluster is also debated; with recent analysis of Chandra data by \cite{2011ApJ...728...27O} suggesting it is actually likely in a post, off-centre, core passage phase heading towards the North/Northeast. 

A  more complicated explanation was  suggested by \cite{2011MNRAS.417..333M} who conducted a detailed gravitational lensing analysis of the cluster. They identify four distinct substructures corresponding to peaks in surface mass density, including a prominent Southern core, a Northern clump, a Northwestern clump of dark matter separated from the known interloper gas cloud, and a Western `Ghost' substructure completely stripped of gas. In order to explain the positional offset of the dark matter peaks and X-ray brightness peaks, they suggest a complex merger scenario in which an initial NE-SW merger caused a `slingshot' effect in a second, almost simultaneous NW-SE merger event \citep[see figure 9 of][]{2011MNRAS.417..333M}. 

More recently, a weak-lensing analysis of A2744 was conducted by \cite{2016ApJ...817...24M} using new Subaru/Suprime-Cam images. They identified four similar substructures as \cite{2011MNRAS.417..333M} but determined a much reduced mass for the Northern Core, and instead identified a more prominent Northeastern Core. They therefore proposed that a major merger event occurred in the East-West direction with another taking place in the North-South direction just East of centre, that pushed the main cluster gas towards the Northwest. Based on multiple N-body simulations they find the `slingshot' scenario offered by \cite{2011MNRAS.417..333M} unlikely. Instead, based on the observation of two associated dark matter peaks, they suggest that the interloper is the result of a third, minor off-axis merger event between two subhalos close to the LOS. During this event the gas from both subhalos was completely stripped, resulting in the observed dark matter-X-ray luminosity peak separations. Interestingly in another recent lensing analysis, \cite{2016MNRAS.463.3876J} detected an additional four substructures, further complicating the picture, see Figure~\ref{fig:xrayoptical}.

\subsection{Previous radio studies of Abell 2744}

Radio emission in Abell~2744  was originally identified by \cite{1999NewA....4..141G}, detecting a peripheral radio relic to the north-east, and a central radio halo. The cluster has since been observed at radio frequencies of  1.4 GHz by \cite{2001A&A...376..803G} and at 325~MHz by \cite{2007A&A...467..943O} and \cite{2013A&A...551A..24V}. From the 1.4 GHz observations, \cite{2001A&A...376..803G} determined the halo to be one of the most powerful known ($P_{\rm{1.4 GHz}}\approx2.6 \times 10^{25}$ W~Hz$^{-1}$). Herein these observations shall be referred to as `O07', `V13' and `G01' respectively.

In the investigation by O07, the radio halo was found to have a patchy spectral index distribution in accordance with primary models \citep[e.g.][]{2008SSRv..134...93F}. A spatial correlation was also suggested between the halo regions with flattest spectral indices and those with the highest X-ray temperature. Hints of a spectral index gradient across the relic in A2744 were identified by O07, with spectral steepening occurring inwards in the direction of the cluster center. Such gradients across relics are common and are thought to be the result of energy losses in the downstream post-shock region \cite[e.g.,][]{2010Sci...330..347V}. \cite{2014A&A...562A..11I} found evidence of an X-ray shock in the form a temperature jump, co-located with the known radio relic (R1, see Figure~\ref{fig:xrayoptical}) in A2744. More recently, using \textit{XMM-Newton} and Suzaku data, \cite{2016arXiv160302272E} reported the detection of a surface-brightness and temperature jump at the Eastern edge of the R1 radio relic, corresponding to a weak shock with a Mach number of $\mathcal{M}=1.7^{+0.5}_{-0.3}$ ($1\sigma$ uncertainties).

\vspace{5mm}
\section{Observations and Data Reduction}
\label{sec:data}
\subsection{VLA observations}
A2744 was observed with the VLA in the 1--2 GHz L- and 2--4~GHz S-band in the DnC-, CnB-, and BnA-array configurations. An overview of the observations is given in Tables~\ref{tab:observationsL} and \ref{tab:observationsS}. The data were recorded with the default wide-band setup of 16 spectral windows spanning the entire bandwidth, with each window having 64 channels.

\begin{table*}
\begin{center}
\caption{L-band Observations \label{tbl-1}}
\hfill \break
\hfill\break
\begin{tabular}{llll}
& DnC array & CnB array & BnA array \\
\hline
\hline
Observation dates & 22 Sep 2014 & 10 Jan 2015 & 2 June 2015 \\
Total on-target observing time (hrs) & 2.3 & 5 & 5 \\
Frequency range (GHz) & 1--2 & 1--2 & 1--2 \\
Correlations & full stokes & full stokes & full stokes \\
Largest angular scale (arcsec) & 970 & 970 & 120 \\
Channel width (MHz) & 1 & 1 & 1 \\
\hline
\label{tab:observationsL}
\end{tabular}
\end{center}
\end{table*}

\begin{table*}
\begin{center}
\caption{S-band Observations \label{tbl-2}}
\hfill \break
\hfill\break
\begin{tabular}{llll}
& DnC array & CnB array & BnA array \\
\hline
\hline
Observation dates & 20 Sep 2014 & \parbox{3cm}{ Run 1: 11 Jan 2015\\Run 2: 13 Jan 2015} & \parbox{3cm}{ Run 1: 22 May 2015\\Run 2: 23 May 2015} \\
Total on-target observing time (hrs) & 2.3 & 4 (2+2) & 10 (5+5) \\
Frequency range (GHz) & 2--4 & 2--4 & 2--4 \\
Correlations & full stokes & full stokes & full stokes \\
Largest angular scale (arcsec) & 490 & 490 & 58 \\
Channel width (MHz) & 2 & 2 & 2 \\
\hline
\label{tab:observationsS}
\end{tabular}
\end{center}
\vspace{10mm}
\end{table*}

The data have been calibrated using the Common Astronomy Software Applications  \citep[CASA;][]{2007ASPC..376..127M} package version 4.4.0. Firstly the data was Hanning smoothed and data affected by RFI (radio frequency interference) and other sources such as antenna shadowing were flagged. The pre-determined elevation dependent gain tables and antenna offset positions were also applied. 
An initial set of gain solutions was determined for the primary calibrator sources 3C147 and 3C138 over a small window of channels in the centre of the bandpass where the phase variations per channel are small. This removes any time dependent phase variation effects in each channel. We then calibrated the delays and bandpass in conjunction with the gain solutions. Next, the cross-hand delays were calibrated using the polarized calibrator 3C138. The gain solutions for the phase calibrator (J0011--2612) were determined along with the polarization leakages. The gains for all the calibrators were then combined solving for the J0011--2612 flux density. We used the \cite{2013ApJS..204...19P} flux-scale. As a final step all relevant calibration solutions were  applied to the target field data (A2744). After initial calibration and flagging, several rounds of self-calibration were performed to further refine the calibration for each individual dataset. All images were  made employing the W-projection algorithm in CASA \citep{2008ISTSP...2..793C}. Clean masks were used for each image step. These masks were made using the PyBDSM source detection package \citep{2015ascl.soft02007M}. The spectral index and curvature was taken into account during the deconvolution  \citep[{\tt nterms=3};][]{2011A&A...532A..71R}.

\begin{figure*}[h!]
\centering
\includegraphics[angle=180,width=0.7\textwidth]{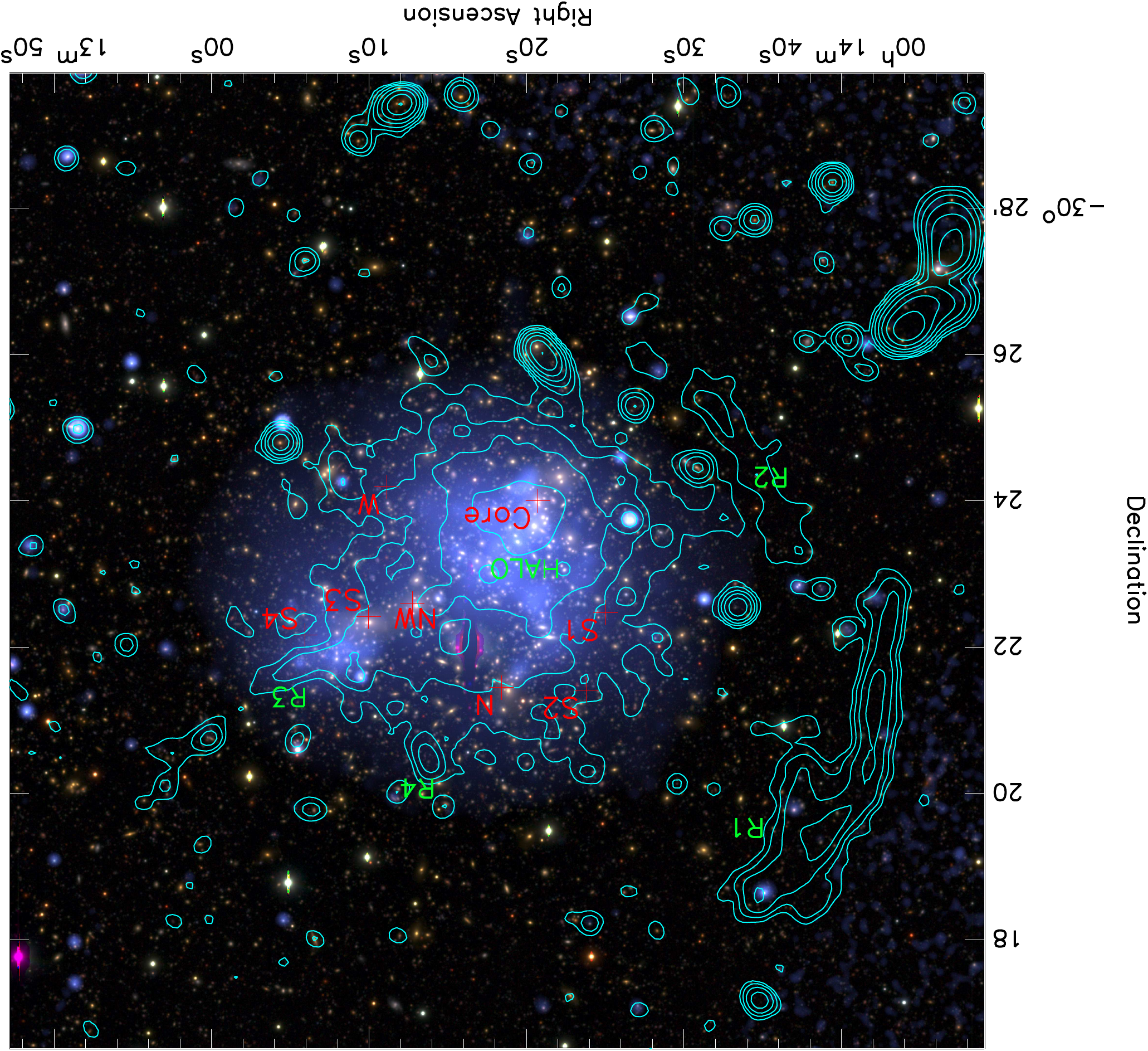}
\caption{A2744 optical, X-ray and radio overlay. The optical BRz Subaru color image comes from \cite{2016ApJ...817...24M}.  The Chandra 0.5--2.0~keV  image is shown in blue. Overlaid in cyan are radio contours from the 1--2~GHz wide-band 15$\arcsec$ uv-tapered map. 
The contour levels are drawn at  $[1,2,4,\ldots] \times 5\sigma_{\rm{rms}}$. Cluster substructures are labelled with red crosses following \cite{2016MNRAS.463.3876J}. Diffuse extended cluster radio sources are indicated with green labels, see also Figure~\ref{fig:labels}.
\vspace{3mm}}
\label{fig:xrayoptical}
\end{figure*}

After self-calibration, the datasets were combined (in each relative frequency band). We imaged the combined datasets using WSClean \citep{2014MNRAS.444..606O}, using Briggs weighting \citep{briggs_phd} with robust = 0.0 and employing the wide-band and multiscale algorithms and using various uv-tapers. Again, clean masks were constructed as mentioned above. The images were corrected for the primary beam attenuation using primary beam images created in CASA.

In order to create spectral index maps we imaged the combined datasets in CASA with an inner u-v range cut (corresponding to the shortest baselines in the S-band data),  uniform weighting, Gaussian taper, and multi-scale clean \citep{2011A&A...532A..71R}. Each dataset was imaged four times, each tapered to a different resolution in order to resolve diffuse emission on varying spatial scales. The final images were then corrected for primary beam attenuation.

To create the deepest possible radio maps we imaged the L- and S-band datasets together using WSClean. These images were only used as visual guides and no flux density values quoted in this paper have been extracted from them. A summary of the final image properties is given in Table~\ref{tbl:images}. The S-band polarization image will be discussed in Section~\ref{sec:pol}.

\subsection{Chandra observations}
We used 125~ks of archival \textit{Chandra} ACIS observations (ObsID: 7712, 2212, 7915, 8477, 8557).
As described in \cite{2005ApJ...628..655V}, the data were calibrated using the {\tt chav} package, applying the most recent calibration files\footnote{We used CIAO v4.8 and CALDB v4.7.2}. The calibration process  involves filtering all counts from bad pixels and counts with recomputed ASCA grades of 1, 5, or~7. We then correct for position-dependent charge transfer inefficiency and applied gain maps to calibrate photon energies. Periods of high background during the observation time were also filtered out by examining the count rate in the  6--12~keV band and removing periods with a flux 1.2 times above the mean. The total filtered clean exposure time was 126~ks. Standard blank sky background files were used for background subtraction. We used a pixel binning factor of 4. For more details about the data reduction the reader is referred to \cite{2005ApJ...628..655V}.

\begin{table*}
\begin{center}
\caption{Image properties \label{tbl:images}}
\hfill \break
\hfill\break
\begin{tabular}{llll}
Image & Weighting & Resolution (arcsec $\times$ arcsec) & rms noise level ($\mu$Jy~beam$^{-1}$) \\
\hline
\hline
L-band uv taper 30$\arcsec$ & Uniform & $30 \times 30$ & $31$ \\
L-band uv taper 15$\arcsec$ & Uniform & $15 \times15$ & $19$ \\
L-band uv taper 10$\arcsec$ & Uniform & $10 \times 10$ & $16$ \\
L-band uv taper 5$\arcsec$ & Uniform & $5 \times 5$ & $15$ \\
L-band & Briggs, robust = 0.0 & $4.15 \times 2.83$ & $10$ \\
S-band uv taper 30$\arcsec$ & Uniform & $30 \times 30$ & $43$ \\
S-band uv taper 15$\arcsec$ & Uniform & $15 \times 15$ & $15$ \\
S-band uv taper 10$\arcsec$ & Uniform & $10 \times 10$ & $10$ \\
S-band uv taper 5$\arcsec$ & Uniform & $5 \times 5$ & $7.1$ \\
S-band & Briggs, robust = 0.0 & $1.65 \times 1.40$ & $4.1$ \\
1--4 GHz wide-band uv taper 30$\arcsec$ & Briggs, robust = 0.0 & $30.5 \times 29.8$ & $-$ \\
1--4 GHz wide-band uv taper 15$\arcsec$ & Briggs, robust = 0.0 & $15.6 \times 15.3$ & $-$ \\
1--4 GHz wide-band uv taper 10$\arcsec$ & Briggs, robust = 0.0 & $10.9 \times 10.4$ & $-$ \\
1--4 GHz wide-band uv taper 5$\arcsec$ & Briggs, robust = 0.0 & $6.01 \times 5.52$ & $-$ \\
1--4 GHz wide-band & Briggs, robust = 0.0 & $2.29 \times 1.81$ & $-$ \\
\hline
\end{tabular}
\end{center}
\end{table*}

\begin{figure*}[th!]
\centering
\includegraphics[width=0.49\textwidth, angle=0]{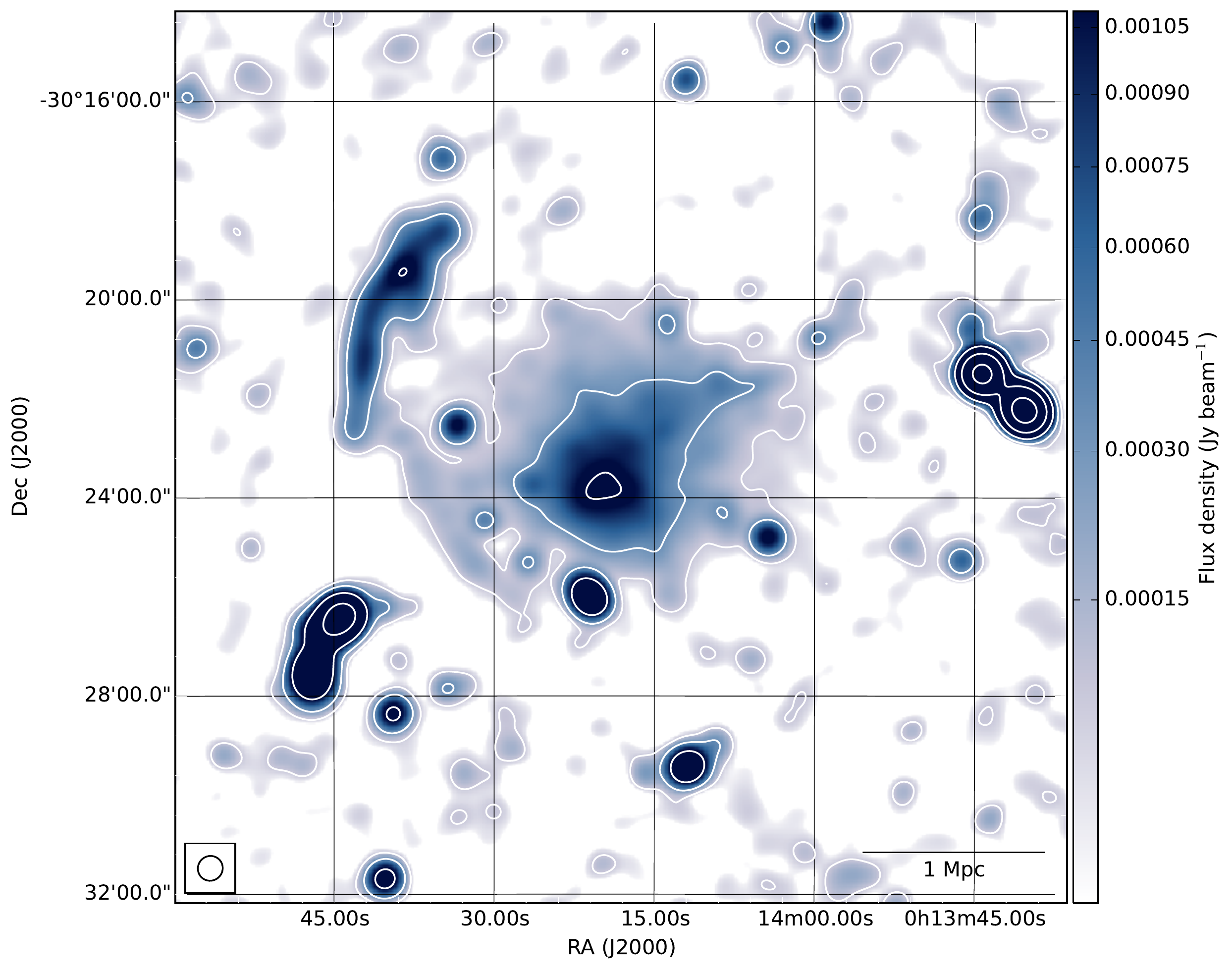}
\includegraphics[width=0.49\textwidth, angle=0]{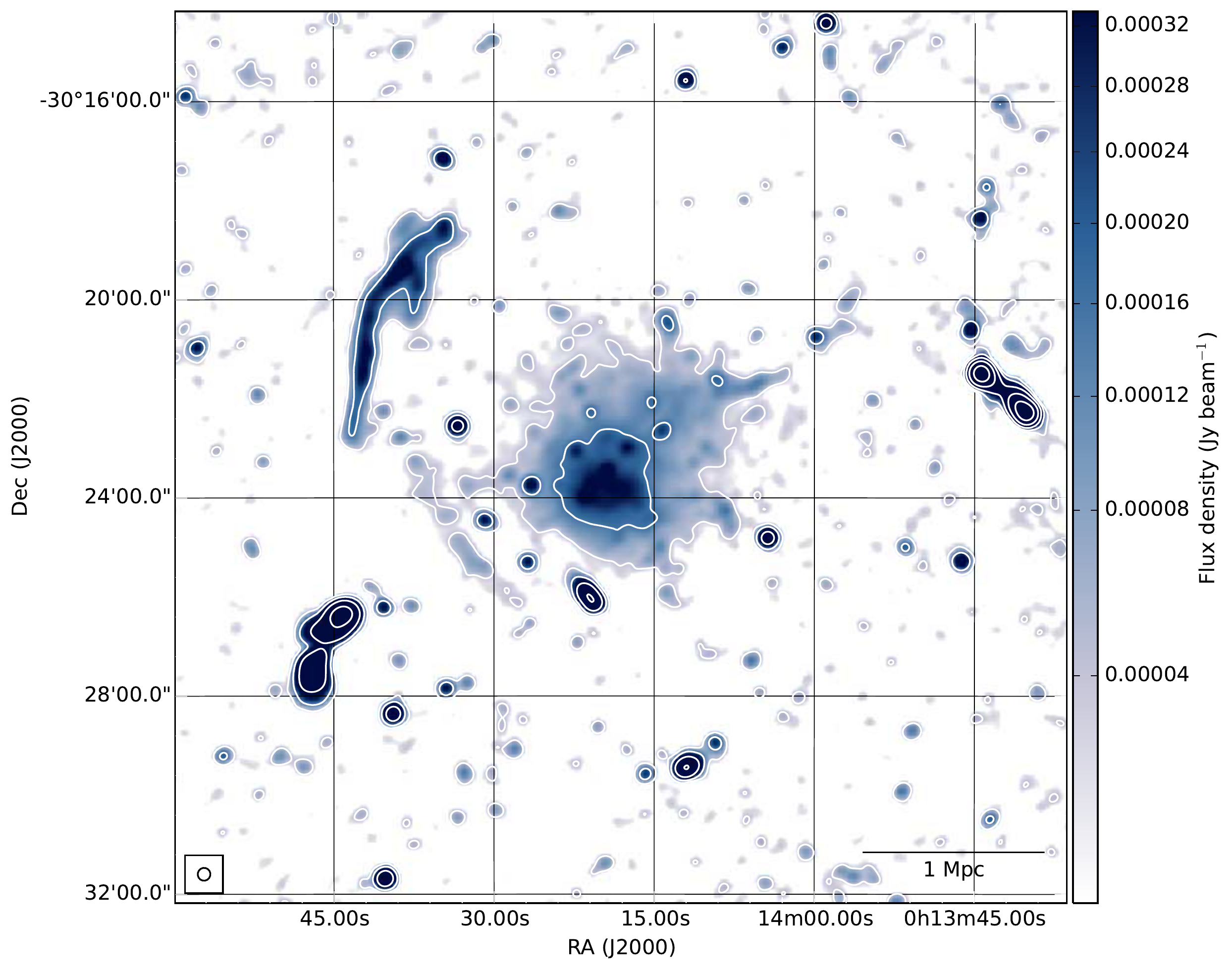}
\caption{VLA combined 1--4~GHz  continuum images of Abell~2744.  \textit{left}: $30\arcsec$ tapered resolution image. \textit{right}: $15\arcsec$ tapered resolution image. In each image the first contour is at the 3.5$\sigma_{\rm{rms}}$ level, with additional contours spaced by factors of 4. }
\label{radio_images}
\end{figure*}

\begin{figure*}[th!]
\centering
\includegraphics[width=0.75\textwidth, angle=0]{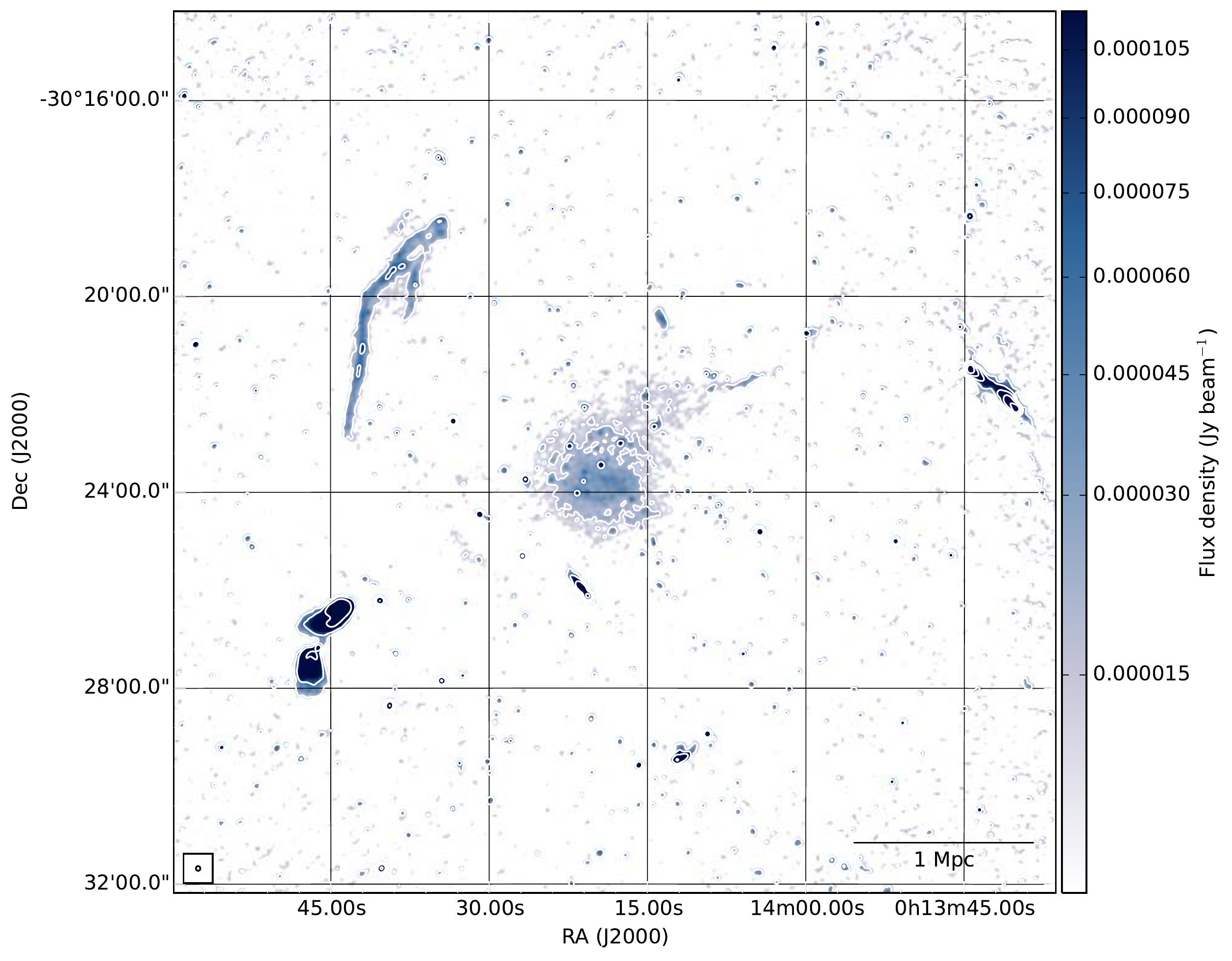}
\caption{VLA combined 1--4~GHz continuum images of Abell~2744 tapered to $5\arcsec$ resolution. The first contour is drawn at the 3.5$\sigma_{\rm{rms}}$ level, with additional contours spaced by factors of 4.}
\label{radio_imageshr}
\end{figure*}

\section{Results}
\label{sec:results}
\begin{figure*}[th!]
\centering
\includegraphics[width=0.95\textwidth, angle=0]{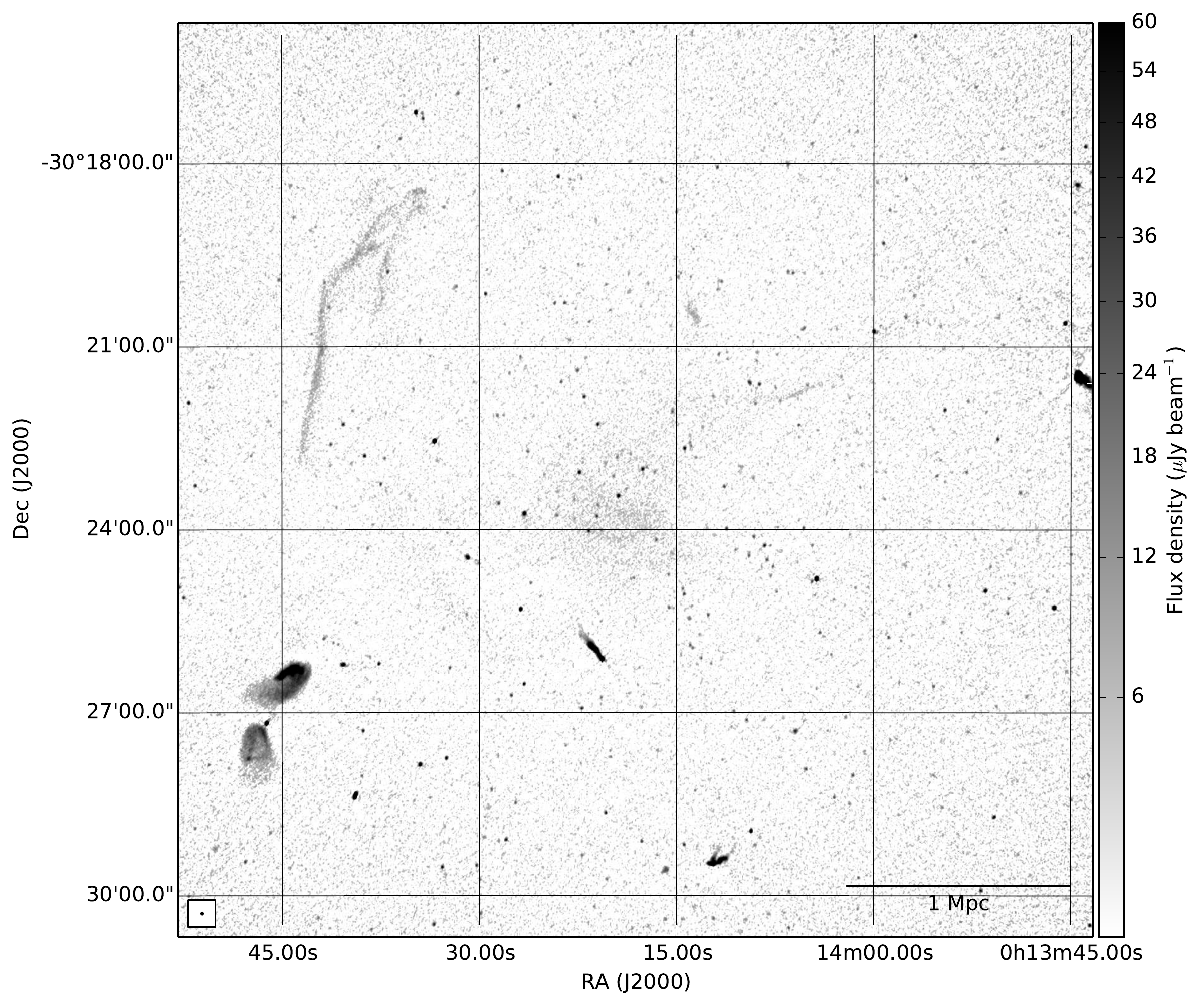}
\caption{VLA combined 1--4~GHz high resolution continuum  image of Abell~2744. The beam size is $2.29\arcsec \times 1.81\arcsec$.}
\label{fig:hi_res_image}
\end{figure*}

\begin{figure*}[h!]
\centering
\includegraphics[width=0.5\textwidth, angle=0]{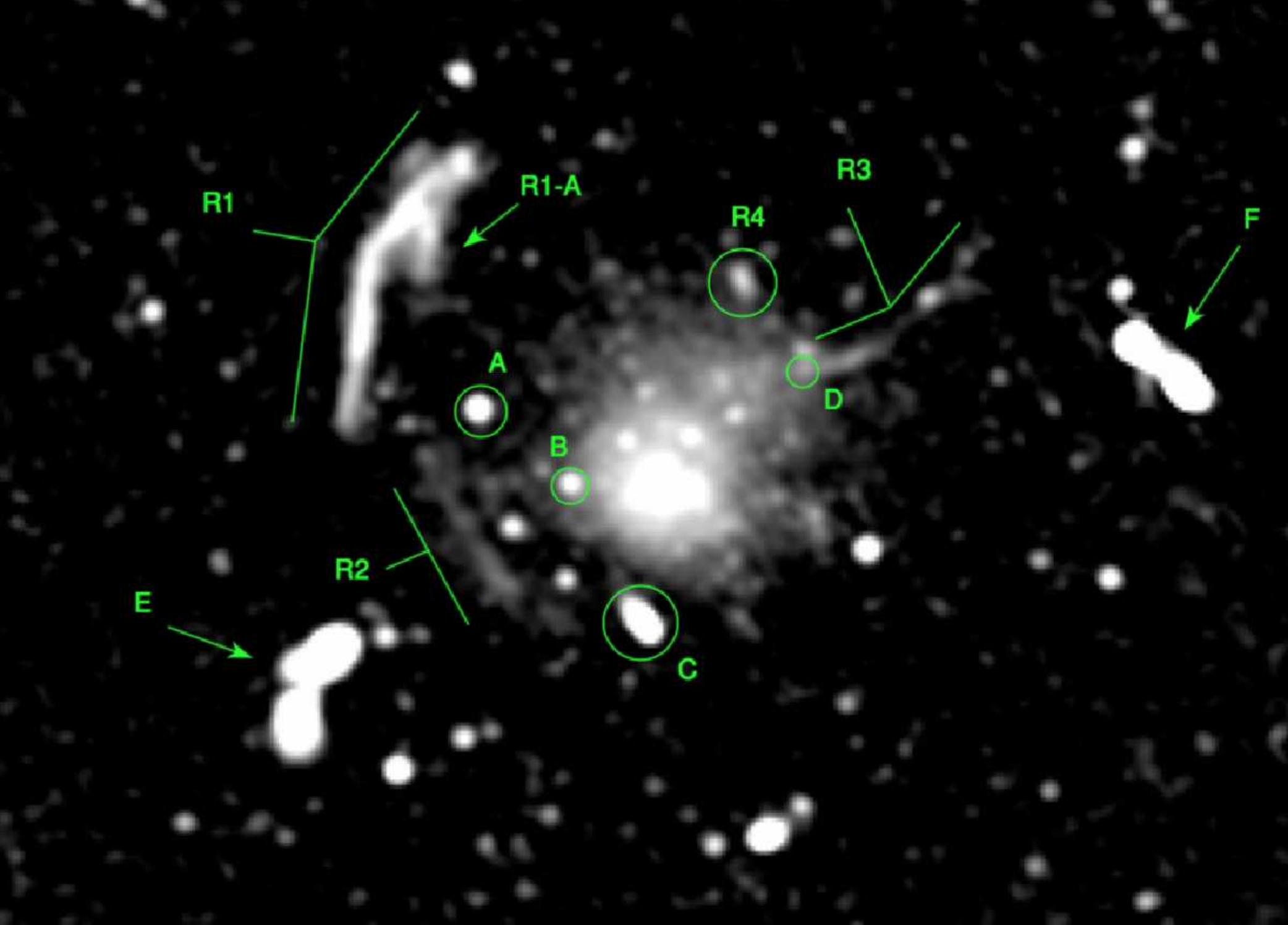}
\includegraphics[width=0.483\textwidth, angle=0, trim={0 2cm 0 0},clip]{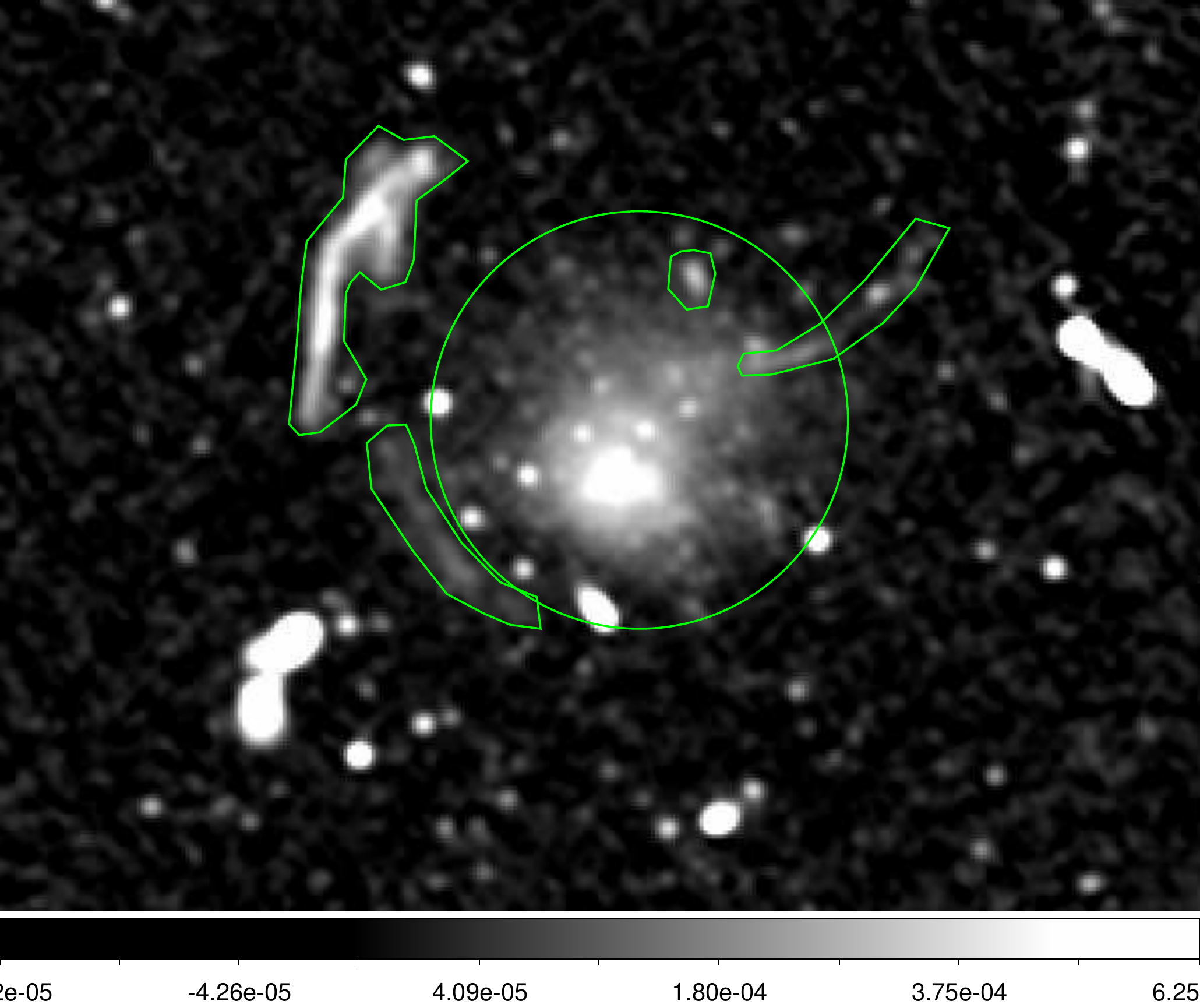}
\caption{\textit{left:} Image showing the regions of interest within the field of view that are discussed in this paper. Shown in greyscale is the L-band u-v tapered 15$\arcsec$ resolution image. \textit{right:} The regions used to calculate the integrated flux densities of the diffuse cluster sources. In greyscale is the L-band u-v tapered $15\arcsec$ resolution image.}
\label{fig:labels}
\end{figure*}

\begin{figure}[h!]
\includegraphics[width=1.0\columnwidth, angle=0]{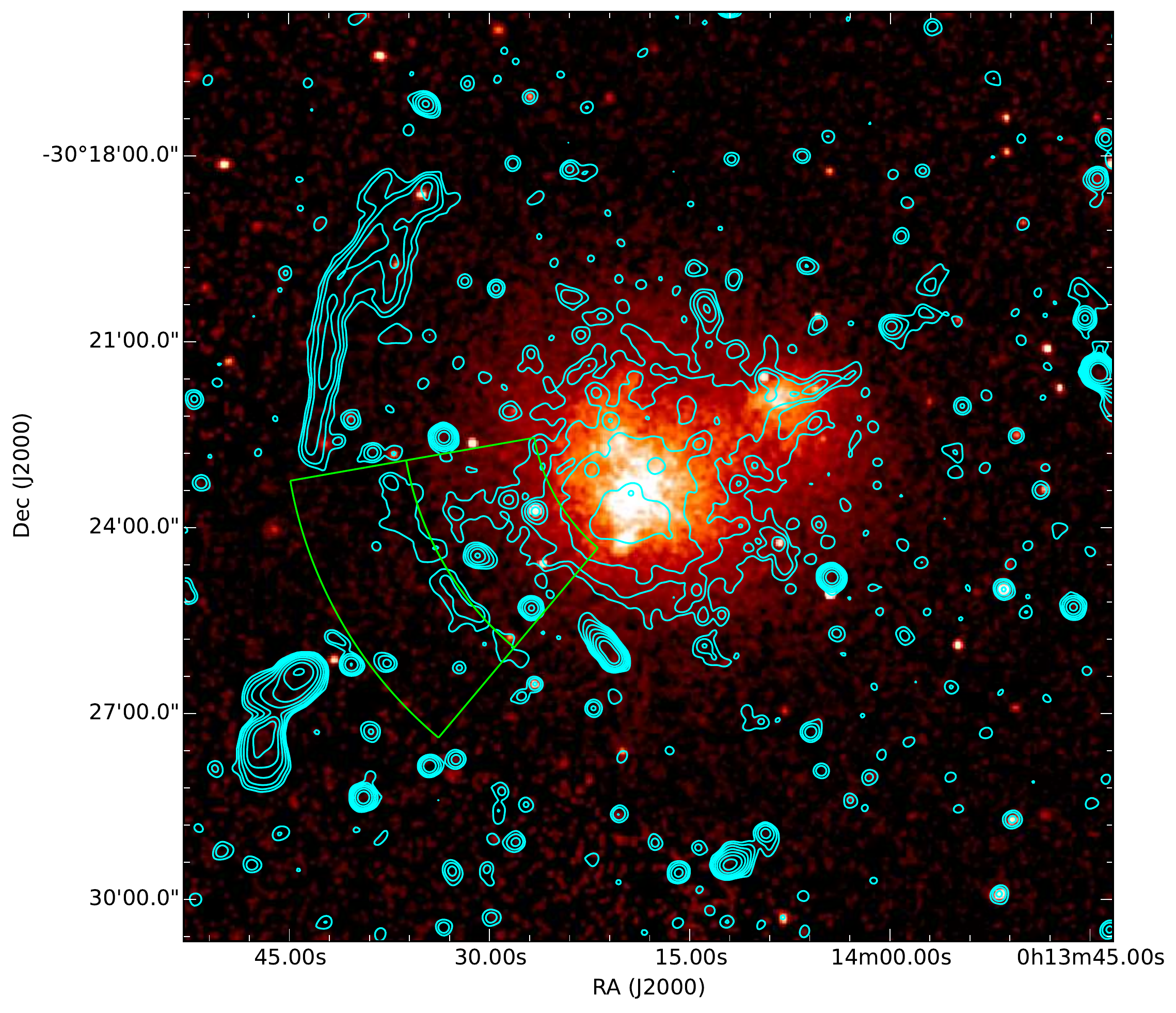}
\caption{A2744 radio and X-ray overlay: The background-subtracted, exposure-corrected Chandra 0.5--2.0~keV  image is shown in orange. Overlaid in cyan are radio contours from the 1--4 GHz wide-band 10$\arcsec$ uv-tapered radio map. The contour levels correspond to the 3.5$\sigma_{\rm{rms}}$ level and above, with scaled spacings of factors of 4. The sector used to extract a surface brightness profile from the X-ray image is displayed in green, see Section~\ref{sec:R2}. The central line within the sector indicates the location of the detected density jump; at a distance of $r_{\rm{edge}} =5.15\arcmin^{+0.16}_{-0.13}$ from the sector centre.\vspace{3mm}}
\label{fig:chandra}
\end{figure}

In Figures~\ref{radio_images}, \ref{radio_imageshr}, and \ref{fig:hi_res_image} we present the combined L- and S-band total intensity radio maps at different resolutions. For the labelling of sources see Figure~\ref{fig:labels}. At all resolutions we detect both the radio halo and the relic (R1), along with several other radio sources. In Figure~\ref{fig:chandra} we display a radio and X-ray overlay of the cluster. A combined X-ray and optical image, marked with the subcluster components found by \cite{2016MNRAS.463.3876J}, is displayed in Figure~\ref{fig:xrayoptical}.

From the low-resolution VLA images we find that the halo has an extent of $\sim 2.1 $~Mpc and the halo roughly follows the X-ray emission from the ICM (see Figure~\ref{fig:chandra}). The peak radio emission in the central part of the halo also appears to be co-located with the peak X-ray emission. The morphology of the halo is similar to that observed by O07, V13 and G01, with a slight elongation in the NW direction and an asymmetric brightness distribution. The largest linear size we detect of 2.1 Mpc appears to be a good average of the 1.6, 1.9 and 2.34 Mpc values obtained by O07, V13 and G01. 

The morphology of the relic (R1) is similar to that observed in previous observations but with some notable new features. The high resolution images reveal that the relic does not appear to have the same smoothly curved shape as previously seen, instead we find it has a relatively straight morphology with a distinct `kink' in it. Interestingly, the upper portion of the relic appears to be further complicated by a second linear component of diffuse emission extending away from the shock front (labelled R1-A in Figure~\ref{fig:labels}).

Diffuse elongated source R2, located to the south of R1, covers an area of $1.15\times0.25$~Mpc and is located approximately 0.9~Mpc from the cluster centre. It is best visible in the low-resolution L-band images, but also appears in the low-resolution S-band images.
In addition, we observe a filamentary thin 1.1~Mpc elongated source (R3) to the NW of the radio halo. The source seems to protrude directly outward from  the Northwest sub-cluster which is also covered by emission from the radio halo. However, due to our line of sight it is unclear whether or not it is projected on top of the halo or if it is connected to it. Finally, we identify a patch of diffuse emission north of the radio halo (labelled R4 in Figure~\ref{fig:labels}, see also Figure~\ref{fig:hi_res_image}), $\sim 200$~kpc from the cluster centre. It covers a much smaller area than the other diffuse sources ($50\times30$~kpc). We do not find an optical counterpart to the source in the images from \cite{2016ApJ...817...24M}.

\subsection{Other individual sources}
In Figure~\ref{fig:labels} we mark several compact radio sources within the field of view. Source~A is located roughly halfway between the relic and the cluster centre (see section~\ref{sec:R1}). It is identified in V13 as source \object{ABELL 2744: [VGD2013] S2}. Source~B is the ``Jellyfish'' galaxy \object{F0083} identified by \cite{2012ApJ...750L..23O}. These galaxies are characterized by trailing knots of star formation caused by extreme ram-pressure stripping \citep{2007IAUS..235..198C,2014ApJ...781L..40E}. Source C, known as \object{NVSS J001421-302558}, is an example of a ``head-tail'' radio galaxy. Source~D is briefly discussed in Section~\ref{subsec:halo}.

The  foreground \citep[$z=0.1966$;][]{2003astro.ph..6581C} source~E is a radio galaxy, identified in NVSS as three separate radio components \object{NVSS J001444-302635}, \object{NVSS J001446-302722}, and \object{NVSS J001445-302644} due to its extended nature (together making up a single radio source).  In a similar fashion, source~F is also identified as two individual components \object{NVSS J001340-302212} and \object{NVSS 001344-302130}.

\subsection{Integrated Fluxes}
In order to obtain accurate measurements of the integrated radio spectra of the halo and the relic, the flux densities of all compact sources within the regions of interest need to be carefully subtracted from the total diffuse emission. Figure~\ref{fig:labels} (right panel) shows the regions used to calculate the integrated flux densities for the diffuse sources. Note, unless stated otherwise, all flux density values mentioned in this section are obtained from the 15$\arcsec$ resolution radio maps where we find the highest signal-to-noise ratio. 

The uncertainties ($\sigma_{S}$) in the flux density measurements ($S$), were taken as:

\begin{equation}
\sigma_{S}=\sqrt{(0.05S)^{2}+\sigma_{RS}^{2}} \hspace{0.5cm},
\end{equation}

where:
\begin{itemize}
\item $\sigma_{RS}=\sqrt{\sigma_{R}^{2}+\sigma_{CS1}^{2}+\sigma_{CS2}^{2}+\ldots}$, is the statistical error on the diffuse source flux density. $\sigma_{R}=\sigma_{\rm{rms}}\times\sqrt{N_{\rm{beams}}}$, with $\sigma_{\rm{rms}}$ being the image noise level and $N_{\rm{beams}}$ being the number of beams covered by the diffuse source. $\sigma_{CS}$ is the statistical error on the flux densities of the individual compact sources (where applicable), as reported by the PyBDSM software package.
\item $0.05S$ is the error due to calibration uncertainties, taken as 5$\%$. 
\end{itemize}

The measured source flux densities, spectral index values and observed physical characteristics of the relic (R1), halo and diffuse sources R2, R3 and R4 are summarized in Table~\ref{tbl-fluxes}. More details are given in the subsections below.

\begin{table*}[t!]
\begin{center}
\caption{Diffuse radio source properties \label{tbl-fluxes}}
\hfill \break
\begin{tabular}{cccccccccc}
Source & D$_{\perp}$ & LLS & $S_{\rm{1.5 GHz}}$ & $S_{\rm{3 GHz}}$ & P$_{\rm{1.4 GHz}}$ & $\alpha$ & $\alpha_{\rm{inj}}$ & $\mathcal{M}^{\dagger}$ & $\bar{P}^{\dagger\dagger}$ \\
& Mpc & Mpc & mJy & mJy & (10$^{24}$ W Hz$^{-1})$ & & & & \\
\hline
\hline
Halo & $-$ & 2.1 & $45.1 \pm 2.3$ & $17.0 \pm 1.0$ & $17.4 \pm 0.9$ & $-1.43 \pm$ 0.11 & $-$ & $-$ & $-$\\
Relic (R1) & 1.3 & 1.5 & $11.74 \pm 0.62$ & $4.78 \pm 0.14$ & $4.37 \pm 0.23$ & $-1.32 \pm 0.09$ & $-1.12 \pm 0.19$ & 2.05$^{+0.31}_{-0.19}$ & 27 $\%$ \\
R2 & 0.9 & 1.15 & $2.18 \pm 0.17$ & $0.64 \pm 0.10$ & $0.96 \pm 0.07$ & $-1.81 \pm 0.26$ & $-1.31 \pm 0.26^{*}$ & 1.86$^{+0.29}_{-0.17}$ & 43 $\%$ \\
R3 & $-$ & $\gtrsim1.1$ & $1.46 \pm 0.14$ & $0.95 \pm 0.10$ & $0.47 \pm 0.04$ & $-0.63 \pm 0.21$ & $-$ & $-$ & 30 $\%$ \\
R4 & 0.2 & 0.05 & $0.88 \pm 0.07$ & $0.35 \pm 0.05$ & $0.30 \pm 0.03$ & $-1.34 \pm 0.23$ & $-0.84\pm0.23^{*}$ & $2.62^{+1.76}_{-0.50}$ & 30 $\%$ \\
\hline
\end{tabular}
\end{center}
{${\dagger}$ derived from $\alpha_{\rm{inj}}$ and Equation~\ref{eq:mach}}\\
${\dagger\dagger}$ emission weighted mean\\
{$^{*}$ $\alpha_{\rm{inj}}$ was not directly measured, but calculated using Equation~\ref{eq:inject}}
\end{table*}

\subsubsection{Halo}
\label{subsec:halo}

For the halo, we subtracted all point sources by imaging the field using an inner u-v range cut of 3.2 $~k\lambda$ (robust=0), corresponding to elimination of spatial scales of $\gtrsim~300$~kpc. This model was then subtracted from the uv-data. The data was then re-imaged with uniform weighting and a 30\arcsec~uv-taper. After this process, sources C, D and R4 (see Figure~\ref{fig:labels}) remained visible as discrete sources, predominantly in the S-band. The fluxes from these sources were manually measured and then subtracted. 

After accounting for primary beam attenuation and source subtraction, we find the radio halo has integrated flux densities of $S_{\rm{1.5 GHz}} = 45.14 \pm 2.34$~mJy in the L-band and $S_{\rm{3 GHz}} =17.03 \pm 0.99$~mJy in the S-band. Using these values we find the integrated spectral index value of the halo to be $\alpha= -1.43 \pm 0.11$. 

Based upon the L-band integrated flux density measurements, the monochromatic radio power is calculated to be $P_{\rm{1.4 GHz}}$=(1.74 $\pm$ 0.09) $\times$ 10$^{25}$ WHz$^{-1}$, using the relation:

\begin{equation}
P_{\rm{1.4 GHz}}=4{\pi}D_{\rm{L}}^{2}S_{\nu_{0}}{\nu_{0}}(1+z)^{-\left(\alpha+1\right)} \hspace{0.5cm},
\label{eq:power}
\end{equation}

where $D_{L}$ is the luminosity distance to the source, $s_{\nu_{0}}$ is the integrated flux density at observing frequency $\nu_{0}$ (1.5~GHz  in our case), $z$ is the redshift of the source and $\alpha$ is the spectral index used in the $k$-correction, taken  as the $-1.43$. The radio power is in reasonable agreement with previous measurements.

\subsubsection{Relic and other diffuse emission}

The same u-v cut could not be applied to remove compact sources from the regions of the relic (R1), R2, R3 and R4 as these sources contained diffuse emission corresponding to physical scales smaller than $\sim~300$~kpc. We used the PyBDSM source detection package to identify all compact sources in the high resolution (robust=0.0) images with flux densities above the 5$\sigma$ level. 

The total flux of the relic R1, after source subtraction of compact sources, measures $S_{\rm{1.5 GHz}}= 11.74 \pm 0.62$~mJy in the L-band and $S_{\rm{3 GHz}}= 4.78 \pm 0.14$~mJy in the S-band. This corresponds to an integrated spectral index value of $\alpha= -1.32 \pm 0.09$. Using Equation~\ref{eq:power} we estimate the monochromatic radio power of the relic to be $P_{\rm{1.4GHz}}=(4.37 \pm 0.23) \times 10^{24}$~W~Hz$^{-1}$, where we have used our measured integrated spectral index value of $\alpha=-1.32$.  

Diffuse  source R2 has significantly lower flux densities measuring  $S_{\rm{1.5 GHz}}=2.18 \pm 0.17$~mJy and $S_{\rm{3 GHz}}= 0.64 \pm 0.10$~mJy. The average spectral index value is steeper than R1 with $\alpha= -1.81 \pm 0.26$. In contrast, source R3 to the NW of the halo has a flatter  spectrum. The integrated flux density values of $S_{\rm{1.5 GHz}}= 1.46 \pm 0.14$~mJy and $S_{\rm{3 GHz}}= 0.95 \pm 0.10$~mJy correspond to a spectral index value of $\alpha= -0.63 \pm 0.21$. For R4, we find flux densities of $S_{\rm{1.5 GHz}}= 0.88 \pm 0.07$~mJy and $S_{\rm{3 GHz}}= 0.35 \pm 0.05$~mJy which correspond to a spectral index of $\alpha= -1.34 \pm 0.23$.

\subsection{Integrated radio spectrum using P-band data}

\begin{figure*}[th!]
\includegraphics[width=0.5\textwidth, angle=0]{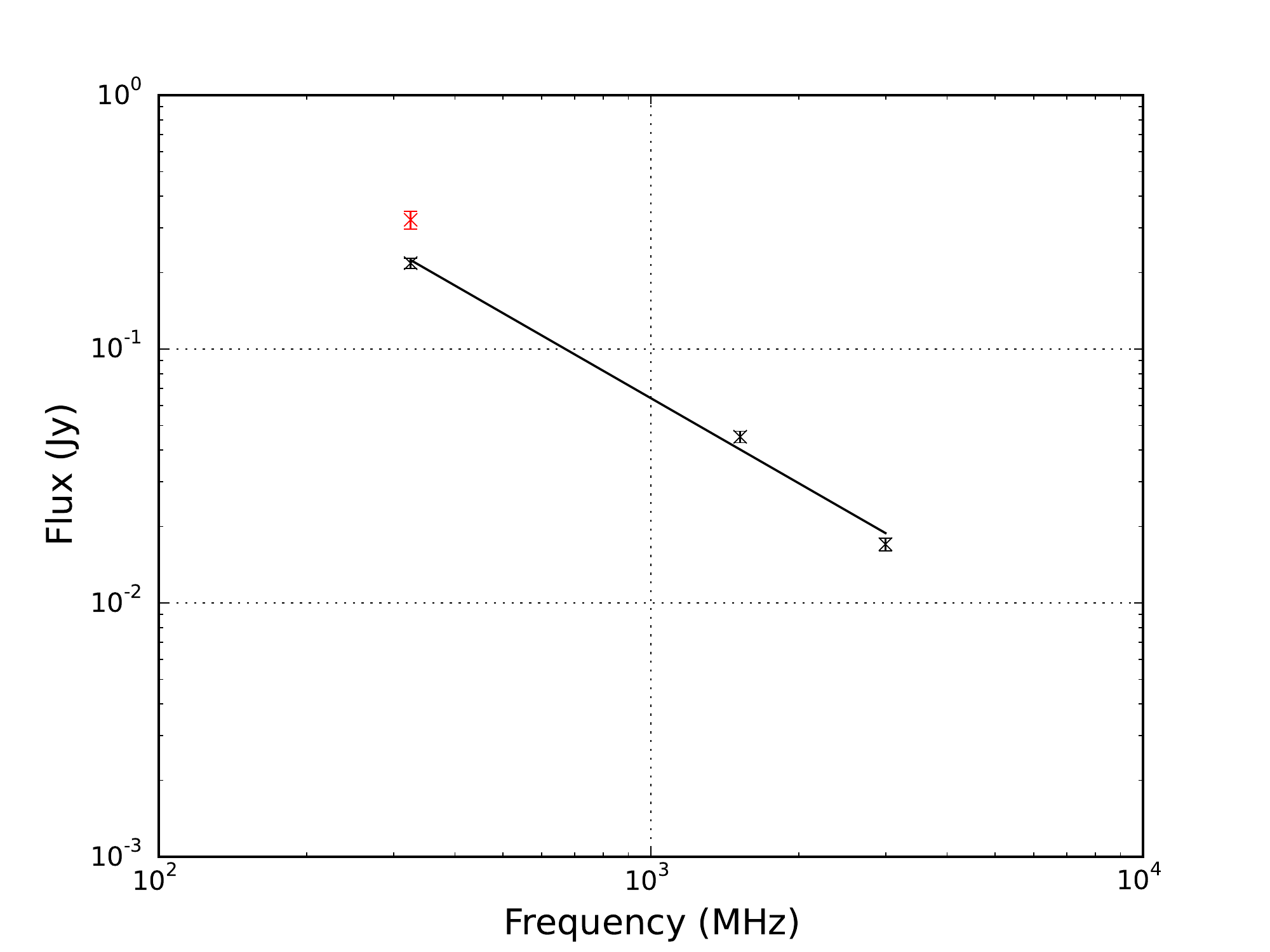}
\includegraphics[width=0.5\textwidth, angle=0]{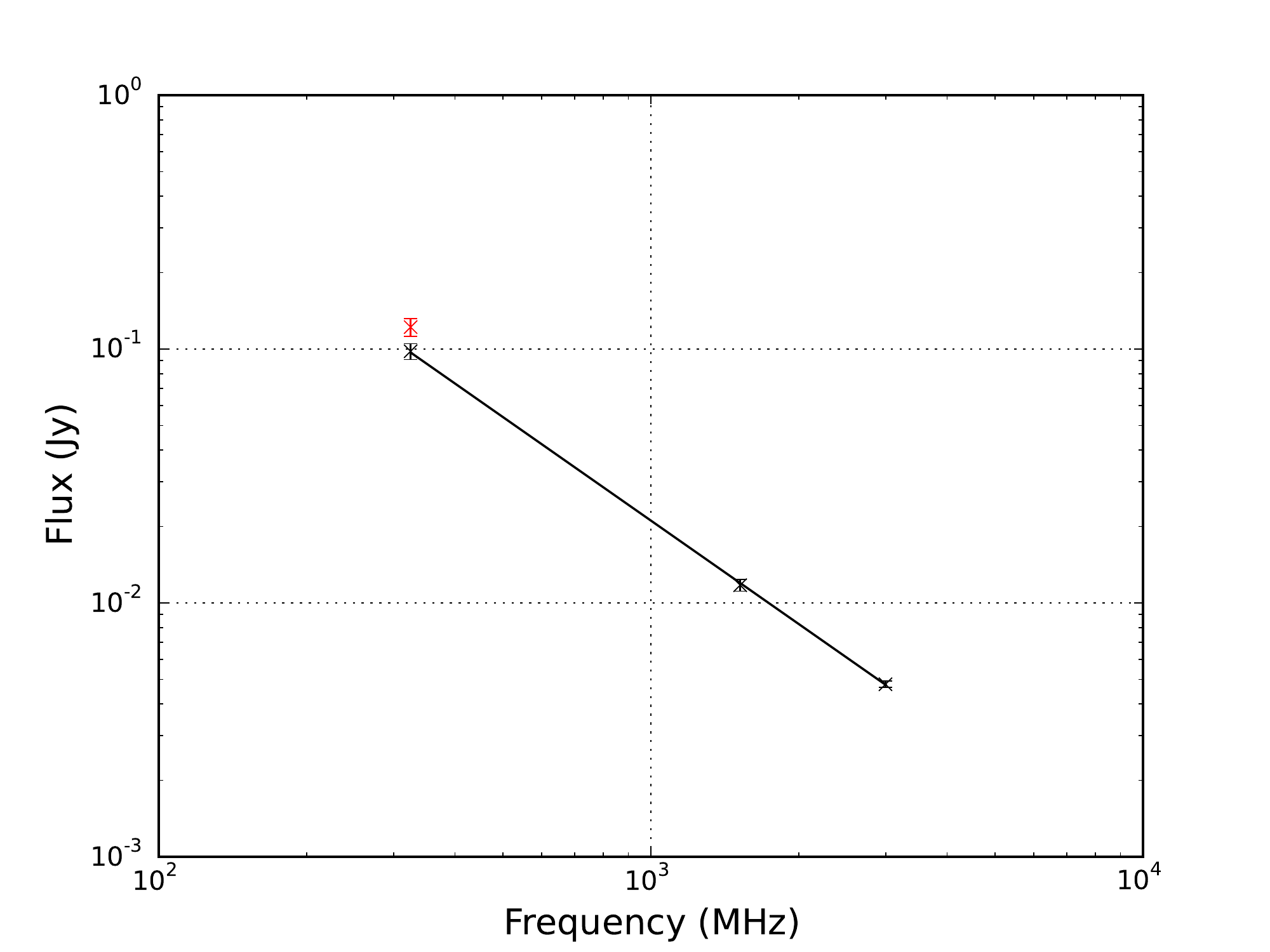}
\caption{Integrated flux densities of the halo (\textit{left}) and relic R1 (\textit{right}) between 325~MHz and 3.0 GHz. The 1.5 GHz and 3.0 GHz flux densities are taken from the $30\arcsec$ radio map. The 325~MHz points are taken from O07 (black) and V13 (red). The black line show a power-law fit through the black data points.}
\label{fig:int_spix_diag}
\end{figure*}

Using the observed flux densities, we produced integrated radio spectra for the relic (R1) and the halo by combining our measured flux densities at both frequencies (1.5~GHz and 3.0~GHz) with those obtained  in the P-band at 325~MHz by O07. The results are shown in Figure~\ref{fig:int_spix_diag}. The integrated radio halo emission, between 325~MHz and 1.5 GHz, has a spectral index of $\alpha_{325}^{1500}= -1.02\pm 0.04$. This value then steepens between 1.5~GHz and 3.0~GHz to $\alpha_{1500}^{3000}= -1.43 \pm 0.11$. 
If we instead take the  325~MHz measurement from V13, the halo spectral index is well described by a single power-law with index -1.32 $\pm$ 0.14. Therefore it remains unclear whether the radio halo spectrum steepens at higher frequencies because of the different P-band flux densities reported by  O07 and V13.

Using the flux density from O07, we find that the integrated radio spectrum from the relic R1 (right panel of Figure~\ref{fig:int_spix_diag}) between the full range of 325~MHz to 3~GHz is well described  by a single power law spectrum, with a fitted spectral index of  $\alpha_{325}^{3000} = -1.36 \pm 0.11$. If we take the flux density from V13 for the power-law fit, we find a consistent result with $\alpha_{325}^{3000} = -1.44 \pm 0.13$.

\subsection{Spectral index maps}
\label{sec:spixmaps}

\begin{figure*}[h!]
\centering
\includegraphics[ angle=180,width=0.49\textwidth]{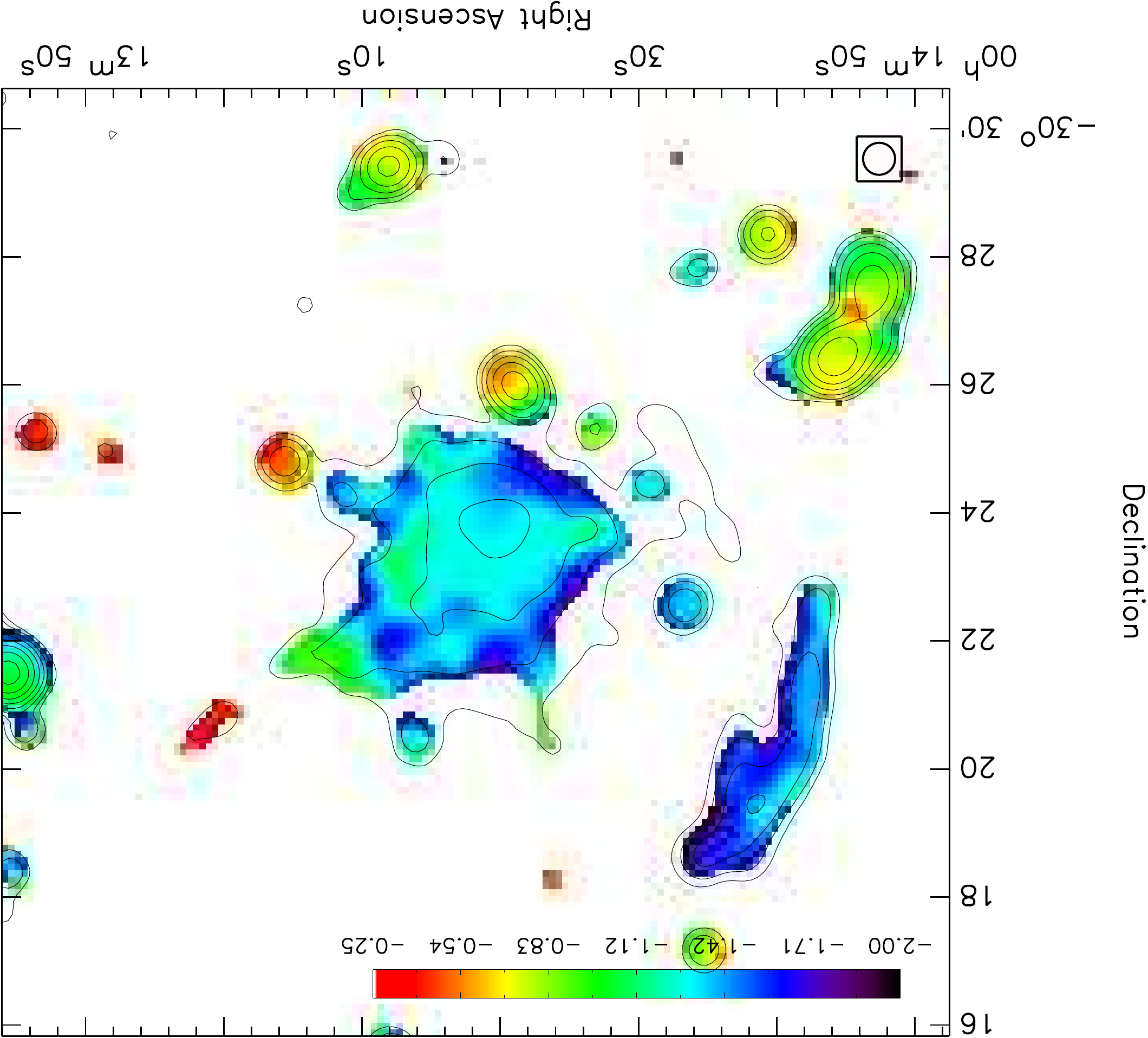}
\includegraphics[ angle=180,width=0.49\textwidth]{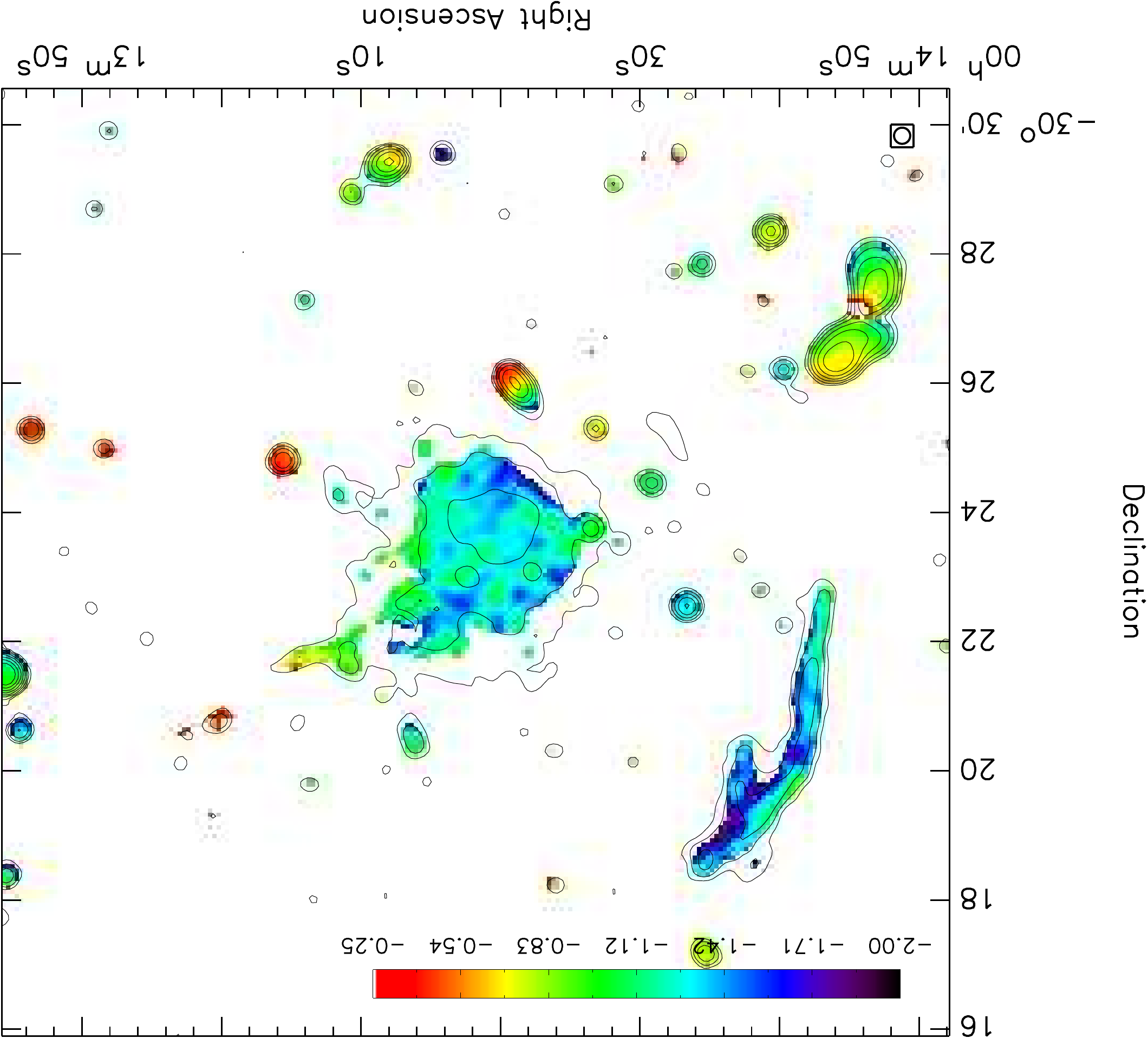}
\caption{Spectral Index map of A2744 between 1.5--3.0~GHz, tapered to a resolution of $30\arcsec$ (\textit{left}) and $15\arcsec$ (\textit{right}). Contour levels are obtained from the 1.5~GHz image and placed at levels of $[1,2,4,8,\ldots] \times n\sigma_{\rm{rms}}$, with $n=3$ for the left image and $n=4$ for the right image.}
\label{spix_map_30}
\end{figure*}

To construct spectral index maps we created L- and S-band images with uniform weighting and an inner uv-range cut corresponding to the S-band DnC-array data. The 30$\arcsec$ and 15$\arcsec$ resolution maps are shown in Figure~\ref{spix_map_30}. The corresponding error maps along with the 10$\arcsec$ and 5$\arcsec$ spectral index maps are displayed in Appendix~\ref{sec:appA}. Only those pixels where the flux values at both frequencies exceed 3$\sigma_{\rm{rms}}$ are displayed. 

We observe a relatively constant spectral index across the radio halo, with patches of steeper values at the radio halo outer boundary. A spectral index gradient across the relic (R1) is evident with values varying from $\sim -0.9$ at the Eastern edge to $\sim -2.5$ at the Western edge. The gradient is consistently perpendicular to the N-S orientation of the relic, with steeper spectral index values in the direction of the cluster centre. It is present along the entire length of the relic, though it exhibits some non-uniformity with the flattest values along the Eastern edge ranging from $\sim -0.9$ to $\sim -1.3$.
A spectral index gradient oriented towards the cluster centre is also present in the emission of diffuse source R4, with values varying from $\sim -0.8$ to $\sim -1.5$. The elongated emission feature to the NW of the halo (source R3) shows little variation in spectral index along its length (although the spectral index cannot be traced along the full source extent) and exhibits flatter spectral values than R1 at around $\alpha \sim -0.7$.

\begin{figure}[h!]
\includegraphics[angle=180,width=0.49\textwidth]{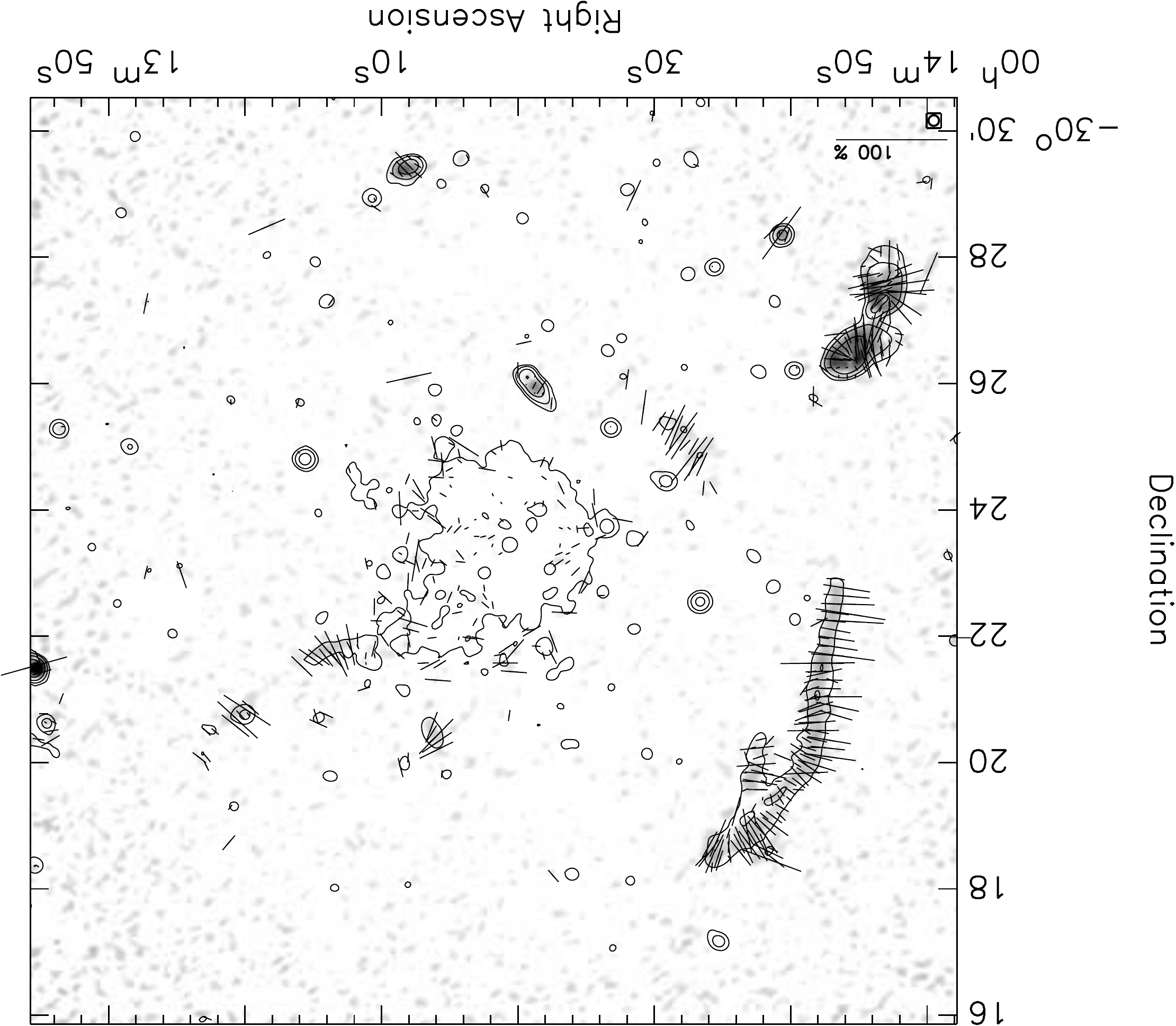}
\caption{S-band vector map showing the magnitude and orientation of the electric field vectors. Vectors are plotted for every 4 pixels. The  linear polarized intensity image is shown in grayscale. The Stokes~I radio contours are from the S-band $10\arcsec$ u-v tapered image and are plotted at levels of $[1,4,16,\ldots] \times 3\sigma_{\rm{rms}}$.}
\label{fig:pol_vec_map}
\end{figure}

\subsection{S-band polarization map}
\label{sec:pol}
S-band images of the Stokes parameters I, Q and U were obtained employing robust=0 weighting and a uv-taper of 10\arcsec. These polarization images serve the purpose of classifying the diffuse extended sources found in the cluster\footnote{A detailed Faraday Rotation Measure analysis, combining all available L- and S-band data will be presented in a future paper.}. These images were corrected for the primary beam attenuation. From the I, Q, and U images the polarization angle ($\phi$) and linear polarized intensity ($p$) can be determined via:
\begin{equation}
p=\sqrt{Q^{2}+U^{2}}
\end{equation}
\begin{equation}
\phi=0.5\arctan{\frac{U}{Q}} \hspace{0.5cm}.
\end{equation}
 A subsequent polarization vector map was constructed from these images, which depicts the magnitude and orientation of the electric field (Figure~\ref{fig:pol_vec_map}).  Polarization vectors are only plotted for pixels above 4$\sigma_{\rm{rms}}$ map noise.
 The polarization vectors shown in Figure~\ref{fig:pol_vec_map} are that of the electric field. 
 
{It is important to note that we did not correct the polarization vectors for the effect of Faraday Rotation. The galactic Rotation Measure (RM) at the location of A2744 is about 6~rad~m$^{-2}$ \citep{2009ApJ...702.1230T}. However, given the relatively high frequency of 2~GHz at the lower end of the S-band and the low galactic RM, corrections for polarization angles would be $< 8\degr$. R1, R2, and R4 are located in the cluster outskirts and therefore the contribution from the ICM to the Faraday Rotation is likely going to be  small as well. For R3, the cluster's RM component might become more important, see Sect.~\ref{sec:R3}.}

From Figure~\ref{fig:pol_vec_map} we find that R1-R4 are highly polarized. The radio halo is mostly unpolarized. The highest polarization fractions ($P$) are found on the Eastern edge of the lower portion of relic R1, with values averaging $\sim$52$\%$. As a whole, R1 has a mean polarization fraction of 27$\%$. The southern portion of the relic appears to be more strongly polarized than the northern part, with mean values of 33$\%$ and 25$\%$ respectively. We find mean polarization fractions of 30$\%$, 43$\%$ and 30$\%$ for sources R4, R2, and R3 respectively.

The alignment of the E-vectors along the length of R1 is consistently perpendicular to relic's major axis. The gradual change in orientation from South to North suggests the lower and upper portions of R1 are indeed part of the same physical structure rather than two independent sources seemingly aligned due to projection effects. For component  R1-A (Figure~\ref{fig:labels}) the E-vectors are aligned in the same direction as the bottom half of the relic. We also find that the E-vectors for R2 and R3 are perpendicular to the source elongation, at least in the region where we have enough S/N to determine the polarization angles.  The interpretation of the polarization map is presented Sections~\ref{sec:R1} to~\ref{sec:R4}.

\section{Discussion}
\label{sec:discussion}
\subsection{Radio halo}

\subsubsection{Spectral index}
\label{sec:spix}


 In the turbulent re-acceleration model, electrons are re-accelerated via magnetohydrodynamical turbulence \citep[e.g.,][]{1987A&A...182...21S,2001MNRAS.320..365B, 2001ApJ...557..560P}. 
In this case,  spectral index variations across radio halos should be related to underlying spatial variations in the turbulent energy and magnetic field strength \citep{2014A&A...561A..52V,2007A&A...467..943O,2004A&A...423..111F}. The most reliable radio halo spectral index map that has been made so far, using LOFAR and VLA data, is the one for the Toothbrush cluster \citep{2016ApJ...818..204V}. For this radio halo, the spectral index variations are remarkably small, with an intrinsic scatter of $\leq 0.04$, suggesting that the turbulent energy does not change significantly across a 0.8~Mpc$^2$ region.

\begin{figure}
\centering
\includegraphics[angle=180, width=0.49\textwidth]{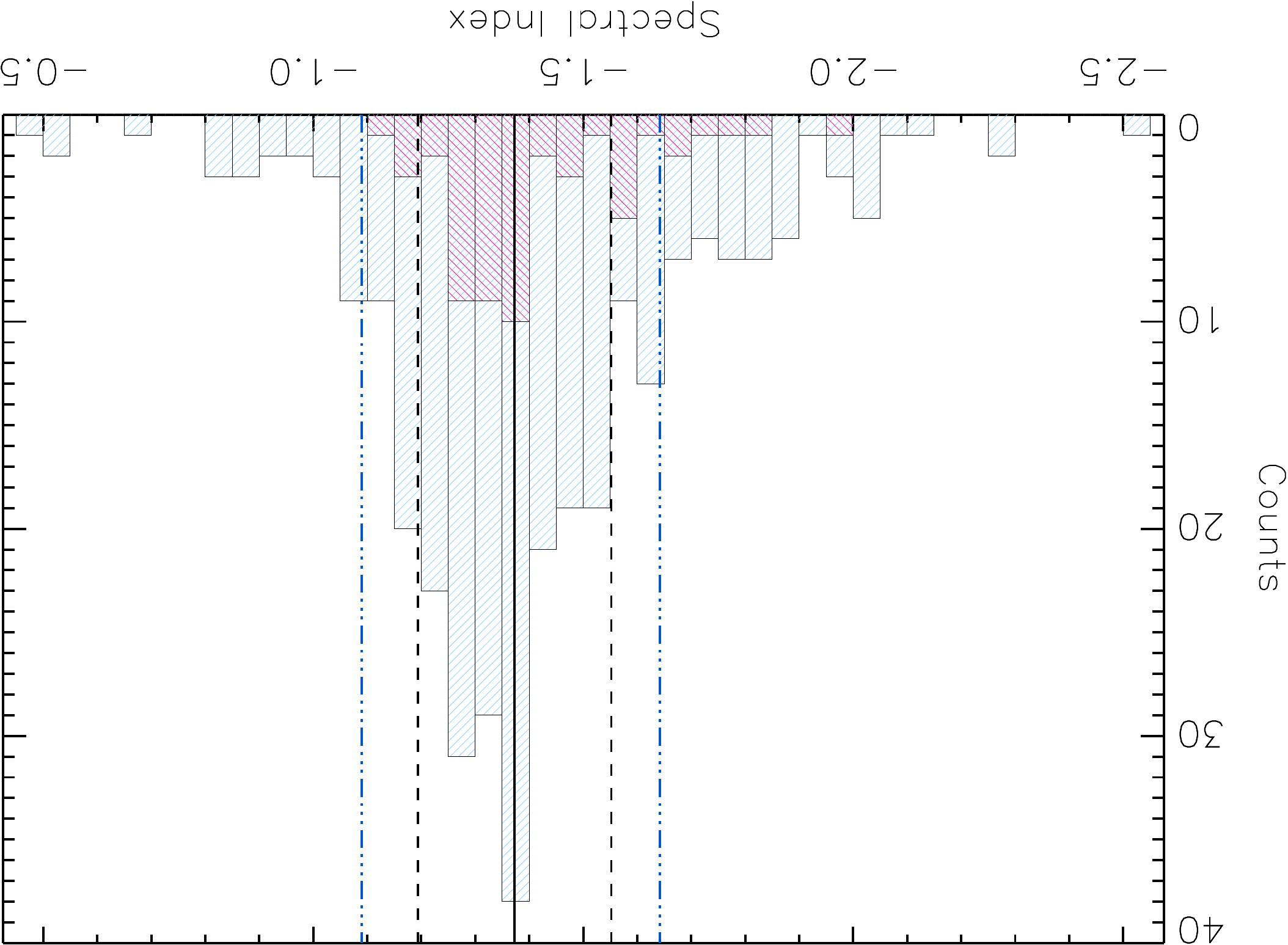}
\caption{Histogram of spectral index values in the radio halo obtained from the 15$\arcsec$ (in blue) and 30$\arcsec$ (in magenta) resolution radio images. The solid line represents the median value of the 30$\arcsec$ distribution. Black dashed lines mark the 30$\arcsec$ standard deviation about the median value, $\sigma_{30\arcsec} =$ 0.18. Blue dashed lines show the 15$\arcsec$ dispersion, $\sigma_{15\arcsec} =$ 0.28.}
\label{fig:hist}
\end{figure}

We determined the spectral variations across the radio halo in a similar way as \cite{2016ApJ...818..204V}.
To test the statistical significance of these fluctuations we carried out a study of the spectral index value distributions for both the 30$\arcsec$ and 15$\arcsec$ resolution radio images. We extracted flux densities, in both the L- and S-bands, in grids of identically sized boxes placed across the halo (see Appendix~\ref{sec:appA}). Spectral index values were then calculated for each individual box. 

Only those boxes with a combined L- and S-band flux of:
\begin{equation}
S_{L}+S_{S} {>} 5(\sigma_{L}^{2}+\sigma_{S}^{2})^{1/2} \hspace{0.5cm},
\end{equation}
where $\sigma$ is the flux density error, are considered. 

The magenta and blue histograms in Figure~\ref{fig:hist} correspond to the 30$\arcsec$ and $15\arcsec$ radio images respectively. The distributions have median values of ${\langle}\alpha_{30\arcsec}\rangle = -1.37$ and ${\langle}\alpha_{15\arcsec}\rangle = -1.37$, and standard deviations of $\sigma_{30\arcsec} =$ 0.18 and $\sigma_{15\arcsec} =$ 0.28. 

To test the level of variation in the values we fit a simple zeroth order polynomial to the data and obtain reduced chi-squared values of $\chi_{30\arcsec}^{2} =$ 1.50 and $\chi_{15\arcsec}^{2} =$ 0.68. The value of 0.68 indicates that the observed fluctuations at $15\arcsec$ are predominantly the result of measurement errors and fall within the bounds of statistical noise. Interestingly though, the 1.50 value indicates that at $30\arcsec$ resolution we do observe some intrinsic complexity in the spectral index distribution. This is confirmed by comparing the median error value of the distributions with the corresponding standard deviations. As described in \cite{2014A&A...561A..52V}, if the variations in spectral index are the result of measurement errors then we expect our median error value to be of comparable size to the standard deviation. At $15\arcsec$, since the median error is almost equal in magnitude to the standard deviation, we conclude that the fluctuations in the halo spectral index are not significant. At $30\arcsec$ resolution,  the errors contribute $\sim 50\%$ to the fluctuations, and they cannot account fully for the level of dispersion. The observed fluctuations at this resolution therefore seem to be statistically significant. This corresponds to variations in the spectral index on spatial scales of $\sim 140$~kpc. The asymmetric distribution of the histogram (positively skewed) indicates that the observed dispersion at $30\arcsec$ is predominantly due to the tail of steep spectral index values. From the spectral index map (Figure~\ref{spix_map_30}) we can see these values tend to reside at the outer edges of the halo. 

\begin{figure}
\includegraphics[angle=180,width=0.49\textwidth]{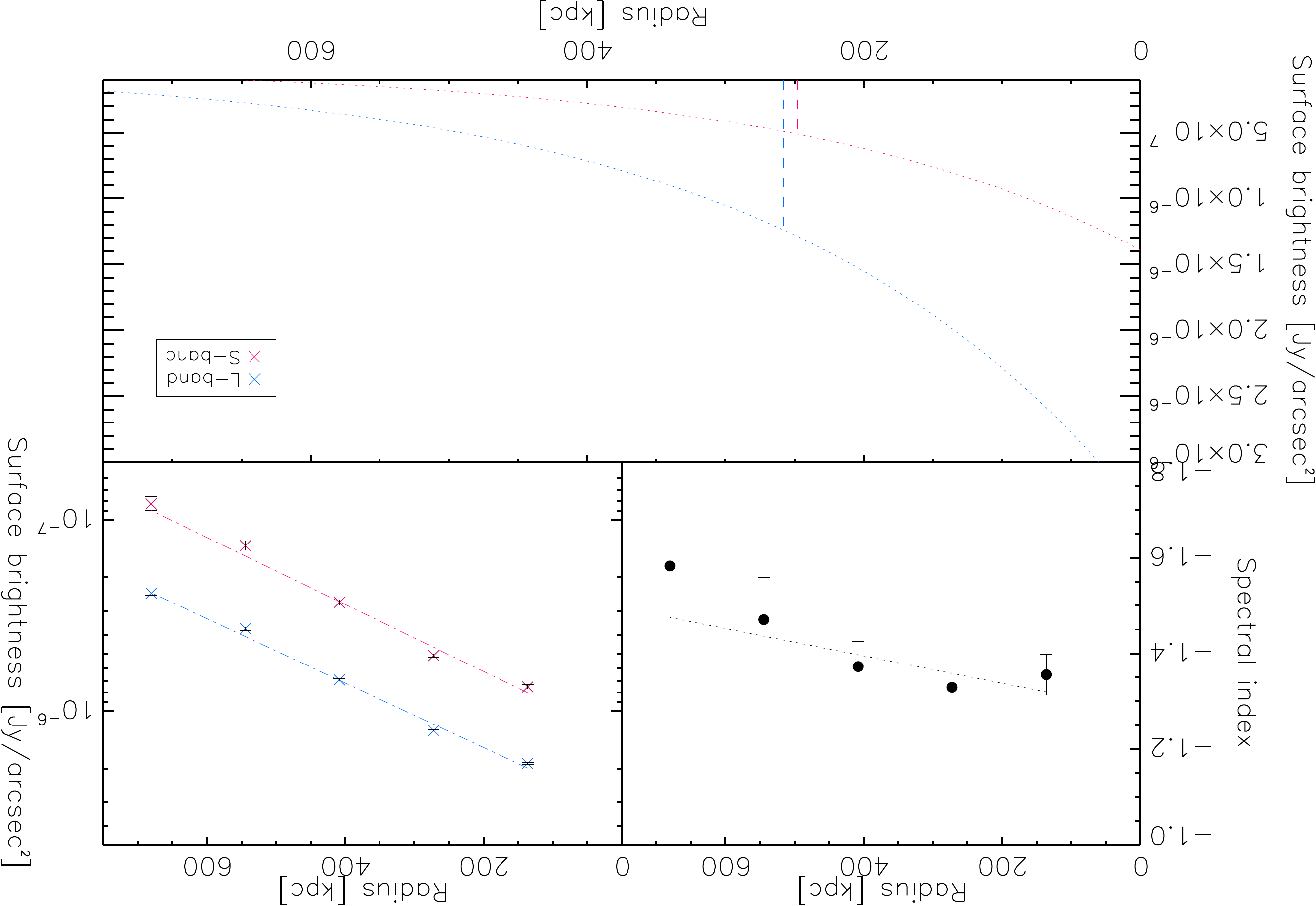}
\caption{\textit{Top left}: Radial profile of the spectral index from the cluster centre. Each value corresponds to the average spectral index in concentric annuli centred on the halo brightness peak and spaced with one 30$\arcsec$ beam width. The dashed line represents the best fit to the data. \textit{Top right}: Azimuthally averaged surface brightness profiles in the L- (blue) and S- (magenta) bands. Values have been extracted from the same concentric annuli as above. Dashed lines represent an exponential fit to the data (Eq.~\ref{eq:RSB}). \textit{Bottom}: Plot of the exponential fit to the surface brightness profiles versus distance to the cluster centre for both frequency bands. Vertical lines represent the corresponding $e$-folding radii ($r_{e}$).}
\label{radial_30}
\end{figure}

We also created a radial profile of the average spectral index value extracted from concentric annuli (see Appendix~\ref{sec:appA}), centered on the point of peak surface brightness and spaced with 1 beam width out to a distance of 150$\arcsec$. The results are shown in Figure~\ref{radial_30}. We fit a first order polynomial to the data and find a slope of gradient $(-2.9 \pm 1.7)\times 10^{-4}$~kpc$^{-1}$. This {suggests} that there is indeed a mild trend of radial steepening in the spectral index value of the radio halo. We speculate that these steeper regions in the outskirts correspond to regions with less efficient turbulent re-acceleration.

In the top right panel of Figure~\ref{radial_30} we present the azimuthally averaged radio brightness profiles of the radio halo at 1.5 and 3.0~GHz. Each data point corresponds to the average surface brightness in the same concentric annuli used for the radial spectral index profile. 

As suggested by \cite{2009A&A...499..679M}, we modeled the brightness profile using a simple exponential law of the form:
\begin{equation}
I(r)=I_{0}e^{-\frac{r}{r_{e}}} \hspace{0.5cm},
\label{eq:RSB}
\end{equation}
where I$_{0}$ is the central radio brightness and r$_{e}$ is the $e$-folding radius. From this we derive values of I$_{\rm{0, 1.5 GHz}} = 3.37 \pm 0.04$~$\mu$Jy/arcsec$^{2}$ and I$_{\rm{0, 3.0 GHz}}= 1.39 \pm 0.05$~$\mu$Jy/arcsec$^{2}$ for the central radio brightness and r$_{\rm{e, 1.5 GHz}} = 257.9 \pm 2.4$~kpc and r$_{\rm{e, 3.0 GHz}} = 248.0 \pm 6.1$~kpc for the $e$-folding radius. The fact that the $e$-folding radius differs in our observations at the L-band and S-band frequencies is consistent with the previous indication of a slight steepening of the spectral index with radial distance.

{To further investigate if the derived spectral indices vary across the radio halo, and are not affected by possible offsets\footnote{e.g, from a non-zero background due to missing short spacings}, we created so called T-T plots \citep{1962MNRAS.124..297T}. These T-T plots compare the flux densities at two frequencies, fitting a straight line through them, and are a useful tool to take into account a non-zero background map level. The resulting reduced $\chi^2$ provides a measure to determine whether the spectral index is constant across the source (reduced $\chi^2\approx1$) or varies (reduced $\chi^2>1$). If the reduced $\chi^2$ is indeed about one, and there are no issues with a non-zero background, the fit offset should be consistent with being zero (i.e., a straight line through the origin).}

{For the radio halo, we created a T-T plot using the flux densities extracted in 30\arcsec~square boxes from the same S- and L-band image used for the spectral index analysis at 30\arcsec~resolution, Figure~\ref{fig:tt} (left panel). These are the same regions used to create the magenta histogram in Figure~\ref{fig:hist}. We then fit a straight line through these datapoints using the {\tt MPFITEXY} routine \citep{2010MNRAS.409.1330W} which utilizes the {\tt MPFIT} package \citep{2009ASPC..411..251M}. For the radio halo we find $\chi^2=71.7$ with $d.o.f.=50$. This does indeed indicate that the spectral index varies slightly across the radio halo.}

\subsubsection{Spatial correlation between radio spectral index and X-ray temperature}

\begin{figure*}
\includegraphics[width=0.9\columnwidth, angle=0]{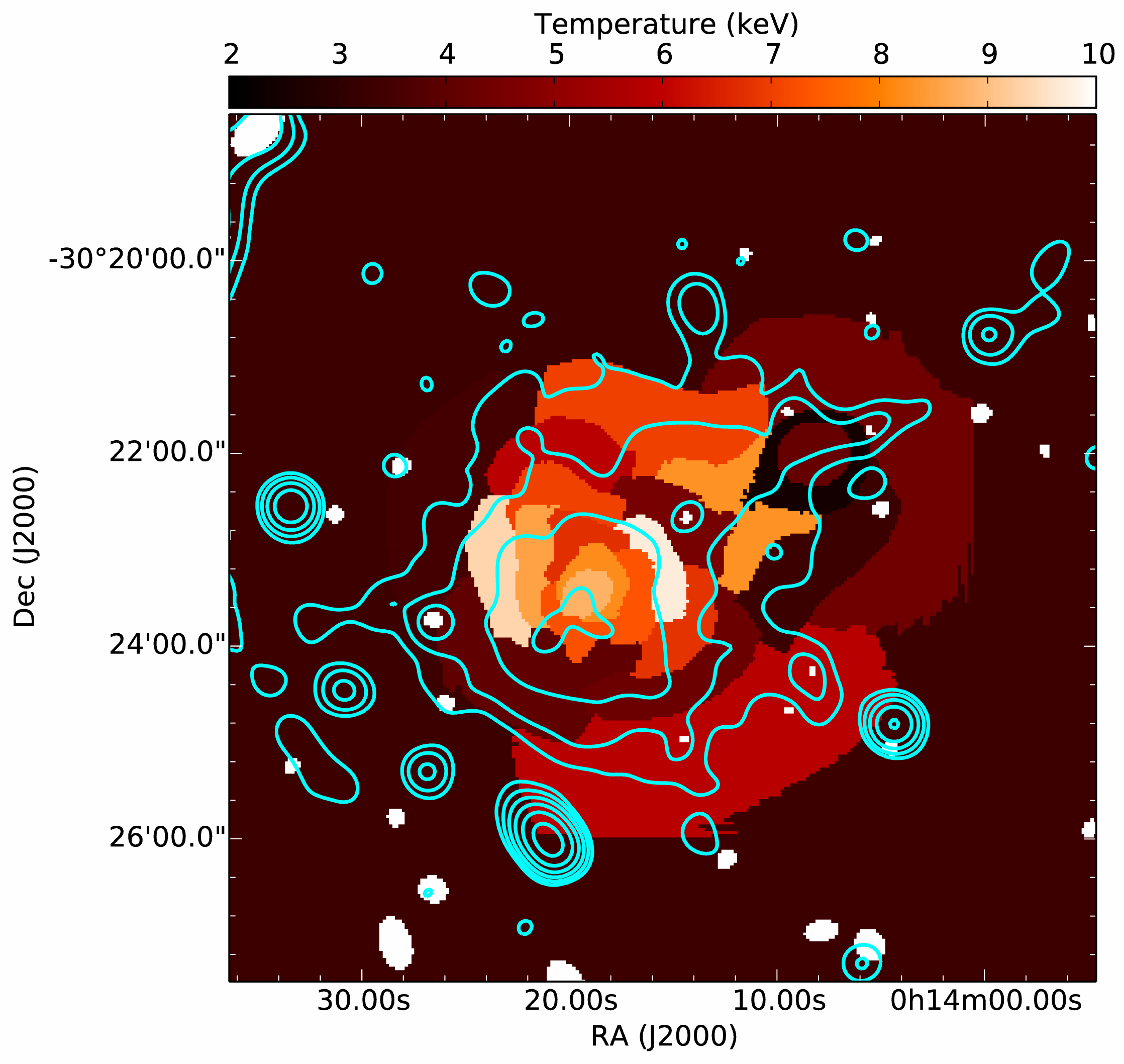}
\includegraphics[trim=0 20cm 0 0cm,width=1.2\columnwidth, angle=180]{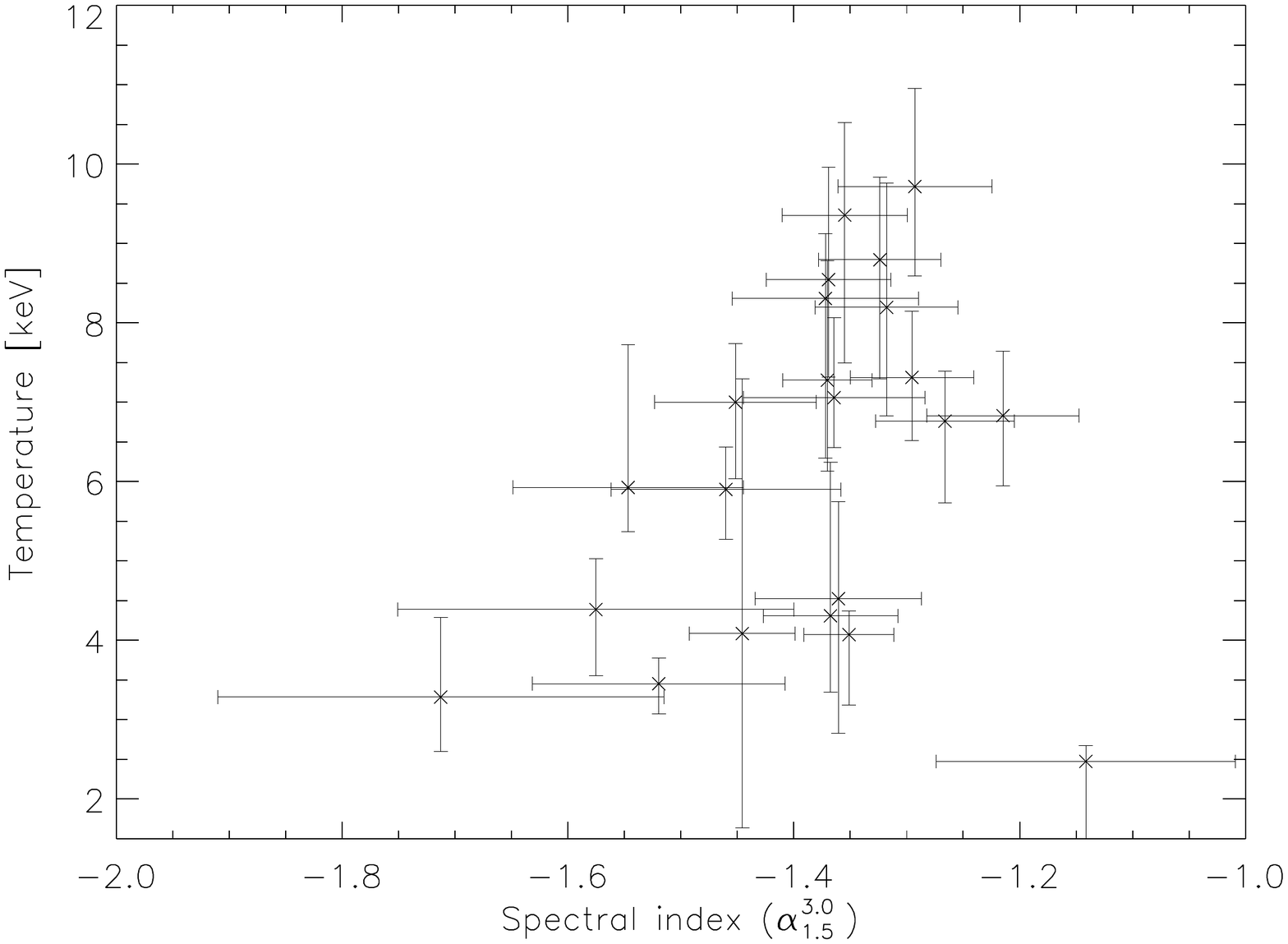}
\caption{\textit{left:} X-ray temperature map. Radio contours from the combined L- and S-band images are overlaid (cyan colors). \textit{right:} Radio halo spectral index plotted against X-ray temperature. The spectral indices were extracted in the same regions as used for the temperature measurements. The Spearman's rank correlation $p$-value is 0.32, i.e., there is a 32\% probability that the data represents an uncorrelated arrangement of points.} 
\label{fig:spixT}
\end{figure*}

\cite{2007A&A...467..943O} reported the presence of a spatial correlation between radio spectral index and ICM temperature in A2744, with the hotter regions of the ICM corresponding to flatter spectral indices. It was argued that this  correlation provided support for the turbulent re-acceleration model. We repeated the investigation from \citeauthor{2007A&A...467..943O} using our new radio data and the deeper Chandra observations.

To create a temperature map, we first divided the cluster into individual regions whose edges follow the X-ray surface brightness using {\tt contbin} \citep{2006MNRAS.371..829S}. We required a signal-to-noise ratio of at least 55 in the 0.5--7.0~keV band for the binning. All compact sources were masked. The extracted spectra were fit with {\tt XSPEC} \citep[v12.8.2,][]{1996ASPC..101...17A}. For the fitting we used an absorbed thermal emission model ({\tt phabs * APEC}). The metallicity was fixed to a value of $0.3$~$Z_\odot$ using the abundance table of \cite{1989GeCoA..53..197A}. The redshift was fixed to $z=0.308$. For the Galactic \rm{H}~\rm{I} column we took a value of $N_{\rm{H}} = 1.38\times 10^{20}$~cm$^{-2}$ \citep[the weighted average $N_{\rm{H}}$ from the Leiden/Argentine/Bonn (LAB) survey,][]{2005A&A...440..775K}. 

The temperature map is shown in Figure~\ref{fig:spixT}. We then extracted radio flux densities in the same regions to compute the spectral indices ($\alpha_{1.5}^{3.0}$). In Figure~\ref{fig:spixT} we plot the spectral indices against the X-ray temperature. We carried out a Spearman's rank correlation test to search for a possible correlation between spectral index and temperature. From this test we find a $p$-value of $0.32$, i.e., the probability that the data represents an uncorrelated arrangement of points\footnote{If we remove the single low-temperature data point in the bottom right corner of Figure~\ref{fig:spixT} (right panel), corresponding to the region surrounding the NW ``interloper", the $p$-value decreases to $0.15$.}. Therefore, we conclude there is no strong evidence for the existence of a correlation between radio spectral index and X-ray temperature. Given our better quality data (both in radio and X-rays), this result should be more reliable than \cite{2007A&A...467..943O}. We note that  \cite{2007A&A...467..943O}  did not provide a statistical measure of the significance of the correlation. Our results are also consistent with recent work by \cite{2014A&A...561A..52V,2016ApJ...818..204V}. We therefore conclude that currently there is no convincing evidence of the existence of a spatial X-ray -- spectral index correlation for individual radio halos.

\subsubsection{Southeastern boundary of the radio halo}  
It is interesting to note that the SE boundary of the radio halo is relatively well defined and aligns with the southern shock reported by \cite{2011ApJ...728...27O}. A very similar configuration is observed for the Toothbrush cluster \citep{2016ApJ...818..204V}, Bullet cluster \citep{2014MNRAS.440.2901S}, Abell~520 \citep{2005ApJ...627..733M,2010arXiv1010.3660M,2014A&A...561A..52V}, the Coma Cluster \citep{2013A&A...554A.140P,2016PASJ...68S..20U}, and Abell~754 \citep{2011ApJ...728...82M}. Therefore this appears to be a common phenomenon and it does suggest there exists some relation between cluster shocks and radio halos. This possibly reflects a change in the ICM's turbulence properties behind the shock front. However, if this turbulence is generated downstream by the shock front, the timescale for the turbulence to decay to the small scales necessary to re-accelerate particles becomes uncomfortably short \citep{2007MNRAS.378..245B,2014IJMPD..2330007B}.

\subsection{Radio relics}

\label{sec:acceleration}
The leading theory for relics is that they trace particles (re)-accelerated at shocks \citep[e.g.,][]{1998A&A...332..395E,2005ApJ...627..733M,2011ApJ...734...18K,2017NatAs...1E...5V}. The  radio spectral index is related to the slope of the underlying electron energy distribution $N(E)dE={\kappa}E^{-\delta}$, with $\alpha=\frac{1-\delta}{2}$. In the case of diffusive shock acceleration \citep[DSA;][]{1983RPPh...46..973D}, the shock Mach number ($\mathcal{M}$) is related to the radio injection spectral index ($\alpha_{\rm{inj}}$) via
\begin{equation}
\mathcal{M}=\sqrt{\frac{2\alpha_{\rm{inj}}-3}{2\alpha_{\rm{inj}}+1}} \hspace{0.5cm}.
\label{eq:mach}
\end{equation}
The observed spectral gradients across relic's width are thought to be indicative of spectral ageing of electrons in the shock downstream region \citep[e.g.,][]{2010Sci...330..347V}. 

For a stationary shock model (i.e, when the radiative lifetime of the synchrotron emitting electrons is much shorter than the timescale on which the shock properties change) with continuous electron injection, the integrated special index ($\alpha_{\rm{int}}$) over the relic can be related to the injection spectral index via the simple relation
\begin{equation}
\alpha_{\rm{int}}=\alpha_{\rm{inj}}-0.5 \mbox{ .}
\label{eq:inject}
\end{equation}

{For a number of relics, with the injection spectral index and integrated spectral index measured independently, Equation 8 seems to be a reasonable approximation \citep[e.g.,][]{ 2008A&A...486..347G,2012MNRAS.426...40B,2014MNRAS.445..330H}. However, a few exceptions have been reported where the integrated spectral index is rather flat, giving an unphysical result for $\alpha_{\rm{inj}}$, i.e.,  $\alpha_{\rm{inj}} > -0.5$ \citep{2012A&A...543A..43V,2015A&A...575A..45T}. Therefore, the usage of Equation~\ref{eq:inject} should be limited to those cases where $\alpha_{\rm{inj}}$ cannot be directly estimated from the data, and any Mach numbers derived from such $\alpha_{\rm{inj}}$ values should be treated with caution.}

Relics are usually highly polarized. Following \cite{1998A&A...332..395E}, the observed polarization fraction of relics can be used to determine a lower limit on the viewing angle of the relic. The alignment of unordered magnetic fields at the shock front should be caused by shock compression, with the degree of alignment dependent on the compression factor. The spectral index ($\alpha$) of a population of electrons in equilibrium, undergoing acceleration and cooling (i.e., the integrated spectral index), is related to the shock compression ratio, $R$, via \citep{1983RPPh...46..973D}:
\begin{equation}
R=\frac{\alpha-1}{\alpha+\frac{1}{2}} \hspace{0.5cm},
\label{eq:compression}
\end{equation}
where we assume the shocked gas to have a polytropic index of $\gamma=5/3$.

The observed polarization fraction also depends on the angle between the line of sight and the B-field.
The resulting polarization therefore depends on the viewing angle $\theta$ of the shock surface and the compression ratio $R$. 

\begin{figure}[h!]
\includegraphics[width=0.49\textwidth, angle=0]{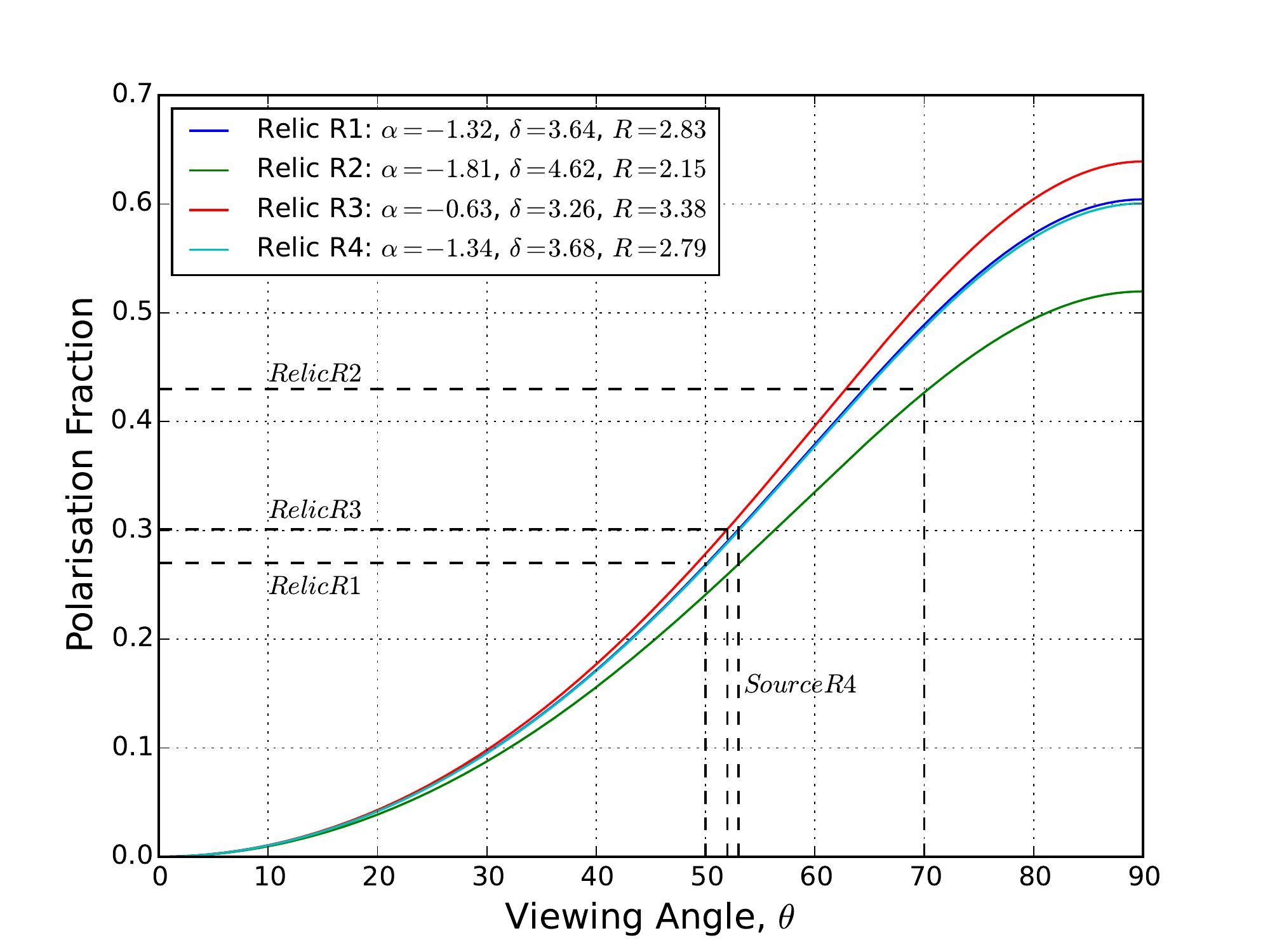}
\caption{A graph showing the polarization fraction of relics R1, R2, R3 and source R4 plotted as a function of the viewing angle, based on Equation~\ref{eq:pol}. Dashed lines correspond to the observed mean polarization fractions.}
\label{fig:pol_frac}
\end{figure}

Assuming a weak magnetic field \citep[i.e., the magnetic pressure of the relic is small compared to the internal gas pressure, see][]{1998A&A...332..395E}, the observed polarization fraction of the relic is related to the viewing angle via the formula:
\begin{equation}
<P_{weak}>=\frac{\delta+1}{\delta+\frac{7}{3}}\frac{sin^{2}(\theta)}{\frac{2R^{2}}{R^{2}-1}-sin^{2}(\theta)} \hspace{0.5cm},
\label{eq:pol}
\end{equation}
where $\delta$ is the electron energy distribution slope (see Section~\ref{sec:acceleration}). Figure~\ref{fig:pol_frac} plots Equation~\ref{eq:pol} for all viewing angles and polarization fractions for our sources R1 to R4.

\subsubsection{Relic R1}
\label{sec:R1}

\begin{figure}
\includegraphics[angle=180,width=0.49\textwidth]{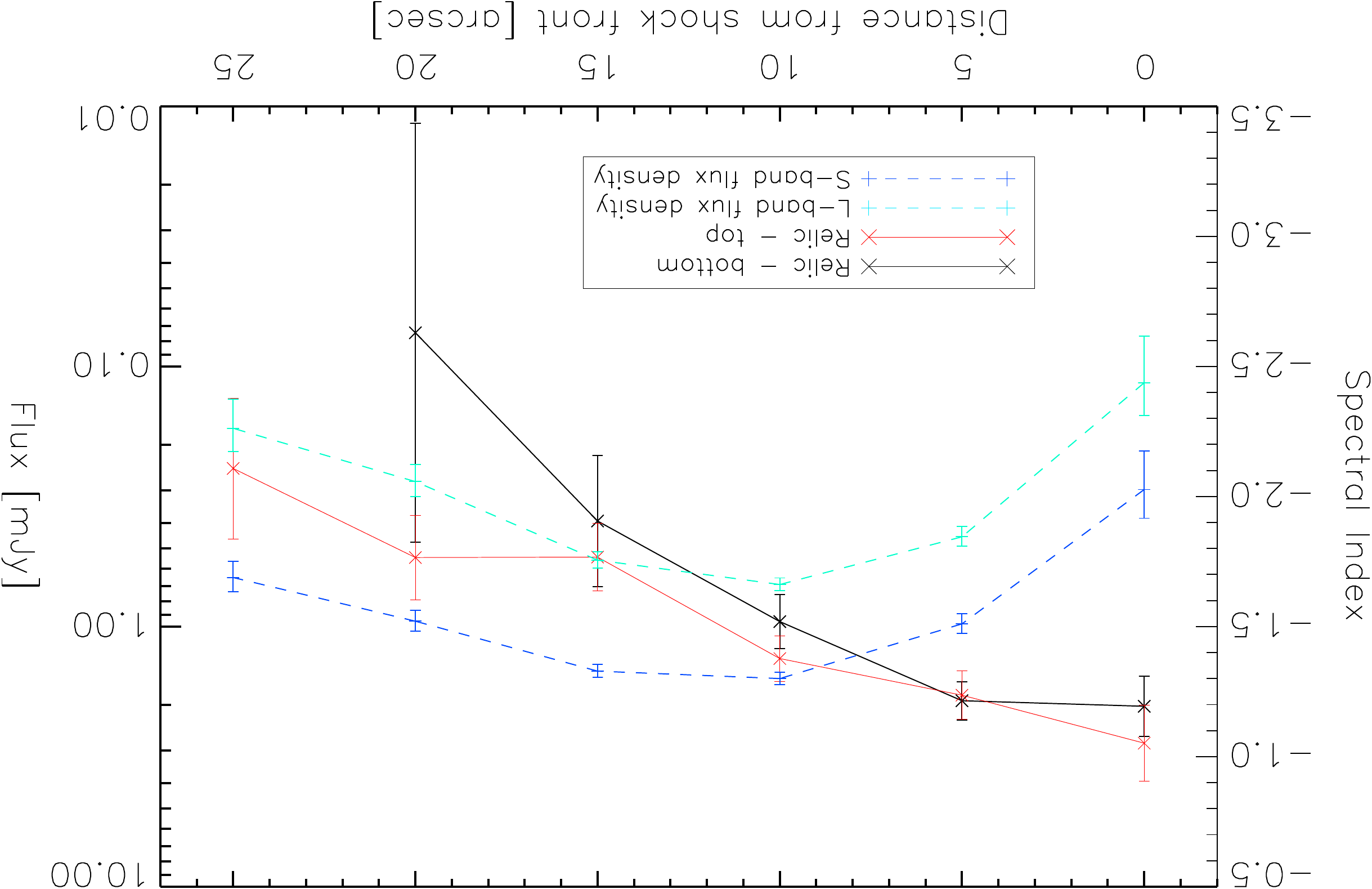}
\caption{Graph showing the integrated spectral index values from $5\arcsec$ wide boxes across the relic.  The corresponding integrated flux densities used to calculate the spectral indices are shown with dashed lines.} 
\label{spix_gradients}
\end{figure}

 In Figure~\ref{spix_gradients} we plot the values of the spectral index, along with their associated flux density measurements, as a function of distance from the shock front. This was done by extracting the flux densities in several $5\arcsec$ wide box regions\footnote{$5\arcsec$ matches the beam size of the highest resolution spectral index map.}.  Given the distinct `wedge-like' shape of the shock front we perform separate measurements across the top and bottom portions of the relic R1, see Figure~\ref{fig:annuliimage}.

Figure~\ref{spix_gradients} displays a clear spectral gradient for both portions of the relic, with the spectral index values steepening away from the relic's outer edge down to values of $\sim -2.5$. From these plots we obtain a flattest spectral index of $\alpha =-1.12\pm0.19$ on the eastern side of the relic, averaging the top and bottom profiles. {We also created a T-T plot using the flux densities extracted from the 5\arcsec resolution L- and S-band maps in the region shown in Figure~\ref{fig:inj_regions}. The T-T plot is shown in Figure~\ref{fig:tt} (right panel). Performing the same procedure as outlined in Section~\ref{sec:spix}, we find $\chi^2=37.3$ with $d.o.f.=11$, clearly confirming the spectral index varies across the relic.\footnote{The T-T plot does show a relatively large offset. However, an offset might be expected because of the large reduced  $\chi^2$, indicating that the data is not well described by a straight line. Therefore, the offset cannot be directly interpreted as a non-zero background level.}
}

{The $\alpha_{\rm{inj}}=-1.12\pm0.19$ value seems somewhat low compared to the $\alpha_{\rm{int}}= -1.32 \pm 0.09$, considering Equation~\ref{eq:inject}. However, we note that this $\alpha_{\rm{inj}}$ value is measured from a map with finite spatial resolution and that projection effects are not included. Both effects lead to an underestimation of $\alpha_{\rm{inj}}$ \citep{2012A&A...546A.124V}. Combining this with the statistical uncertainties of both measurements, the current data do not allow us to draw a firm conclusion on the validity of the assumption of stationary shock conditions for R1.}

Using $\alpha_{\rm{inj}}=-1.12\pm0.19$, we derive a shock Mach number of $\mathcal{M}=2.05^{+0.31}_{-0.19}$ for relic R1. Such a Mach number is consistent with weak merger shocks. Recently, \cite{2016arXiv160302272E} reported evidence of a shock at the relic's Eastern edge using \textit{XMM-Newton} and \textit{Suzaku} X-ray data. Using the Rankine-Hugoniot jump conditions they derived a shock Mach number of $\mathcal{M}=1.7^{+0.5}_{-0.3}$. Our value is in agreement with this within errors. 

\cite{2016arXiv160302272E} also derived an estimate for the acceleration efficiency required of electrons at the shock corresponding to R1 in order to reproduce the observed radio power of the relic. They find that the required efficiency for the shock is 10 to~$10^{3}$ times higher than that predicted by DSA for thermal electrons. It was suggested that DSA re-acceleration of a pool of mildly relativistic fossil electrons, already present in the cluster volume, can reconcile such discrepancies. Given that a variety of sources in the ICM (such as supernovae, AGN etc.) are able to supply relativistic electrons with a broad range of energies, this scenario could be viable in the context of weak shock acceleration. Indeed, observational evidence does exist for a connection between tailed radio galaxies and relics \citep[][]{1991A&A...252..528G,2014ApJ...785....1B,2015MNRAS.449.1486S,2016MNRAS.460L..84B,2017NatAs...1E...5V,2017ApJ...835..197V}. 

\begin{figure}[h!]
\includegraphics[angle=180, width=1.0\columnwidth]{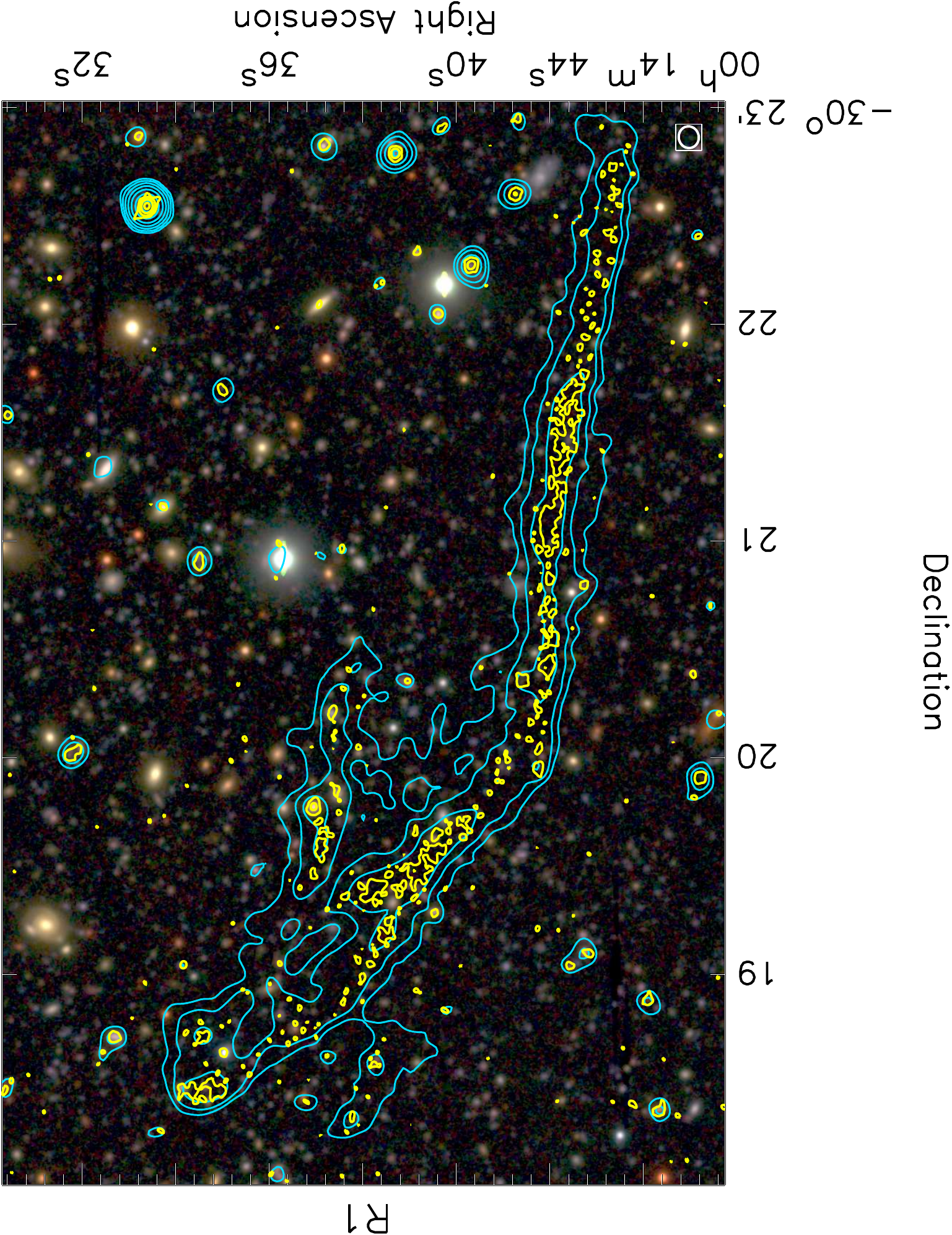}
\caption{Subaru optical BRz  color image \citep{2016ApJ...817...24M} ] {around relic R1} with radio contours overlaid. The blue and red radio contours are from the 5\arcsec~uv-tapered and robust=0 1--4~GHz wide-band images, respectively. Contour levels are drawn at $[1,2,4,\ldots] \times 3\sigma_{\rm{rms}}$ (blue) and $[1,4,8,\ldots] \times 4\sigma_{\rm{rms}}$ (red).}
\label{fig:opticalR1}
\end{figure}

The morphology of R1 with its peculiar extension R1-A could have resulted from the re-acceleration of a remnant patch of fossil AGN plasma.
In A2744, we do observe a number of compact radio sources embedded within the R1 relic region, although presently we cannot establish a direct link between any of these sources and the relic, see Figure~\ref{fig:opticalR1}. One compact radio source is embedded within R1-A (RA=3.6538408\degr, DEC=$-30.3296082\degr$), but this seems to be a background source with $z_{\rm{phot}} = 0.59 \pm 0.07$ \citep{2016ApJ...817...24M}.  

The observed mean polarization fraction of $27\%$ of R1 is typical for radio relics and suggests a high degree of ordering of the magnetic fields ($B$), aligning them within the shock plane \citep[e.g.,][]{2006AJ....131.2900C,2009A&A...494..429B}. It is worth noting that we observe large local fluctuations in this value, particularly towards the southern part of relic R1 where values reach up to a polarization fraction of $\sim60-70\%$, which is close to the maximum value for synchrotron radiation  \citep[e.g.,][]{1986rpa..book.....R}. Using the mean polarization fraction, we determine the viewing angle of relic R1 to be $\theta\gtrsim50^{\circ}$ (Figure~\ref{fig:pol_frac}). This is only a lower limit, with the actual value likely to be higher due to depolarization. With a viewing angle of $\theta\gtrsim50^{\circ}$ we can constrain the geometry of the primary NE-SW merger axis to be within $\sim\pm40^{\circ}$ of the plane of the sky.

\subsubsection{Diffuse Source R2}
\label{sec:R2}
R2 has not been identified as an individual source in any observations to date. However, upon closer inspection of the radio maps presented by V13, hints of the sources are visible. Part of this emission was classified as a radio bridge by V13. Our deeper images suggest that R2 is not a radio bridge but a separate elongated diffuse source with a length of $1.15$~Mpc and located $\sim0.9$~Mpc southeast from the cluster core.

The relatively low flux density, particularly in the S-band, prevents the source from showing up in the spectral index maps. However, using the integrated flux densities we obtain an integrated spectral index value of $\alpha = -1.81 \pm 0.26$.  We estimate the radio power of the source to be $P_{\rm{1.4 GHz}}=(9.59 \pm 0.73) \times 10^{23}$~W~Hz$^{-1}$.

Most importantly, in the E-vector polarization map (Figure~\ref{fig:pol_vec_map}) we find the source to be highly polarized, with a mean value of $43\%$. Given the peripheral location of the source, its Mpc-size, steep  spectrum and high polarization fraction, we conclude that this source is a new radio relic. Henceforth we will refer to source R2 as relic R2. The polarization vector map shows the E-vectors to be aligned perpendicular to the SE edge of the relic. Based on this and the relic's location and orientation, we suggest that the relic is the product of a shock front traveling in the SE direction. Using Equation~\ref{eq:pol}, we derive a viewing angle of $\theta\gtrsim70^{\circ}$ (see Figure~\ref{fig:pol_frac}), meaning we are seeing the relic close to edge-on.

\begin{figure}[h!]
\includegraphics[width=0.49\textwidth, angle=0]{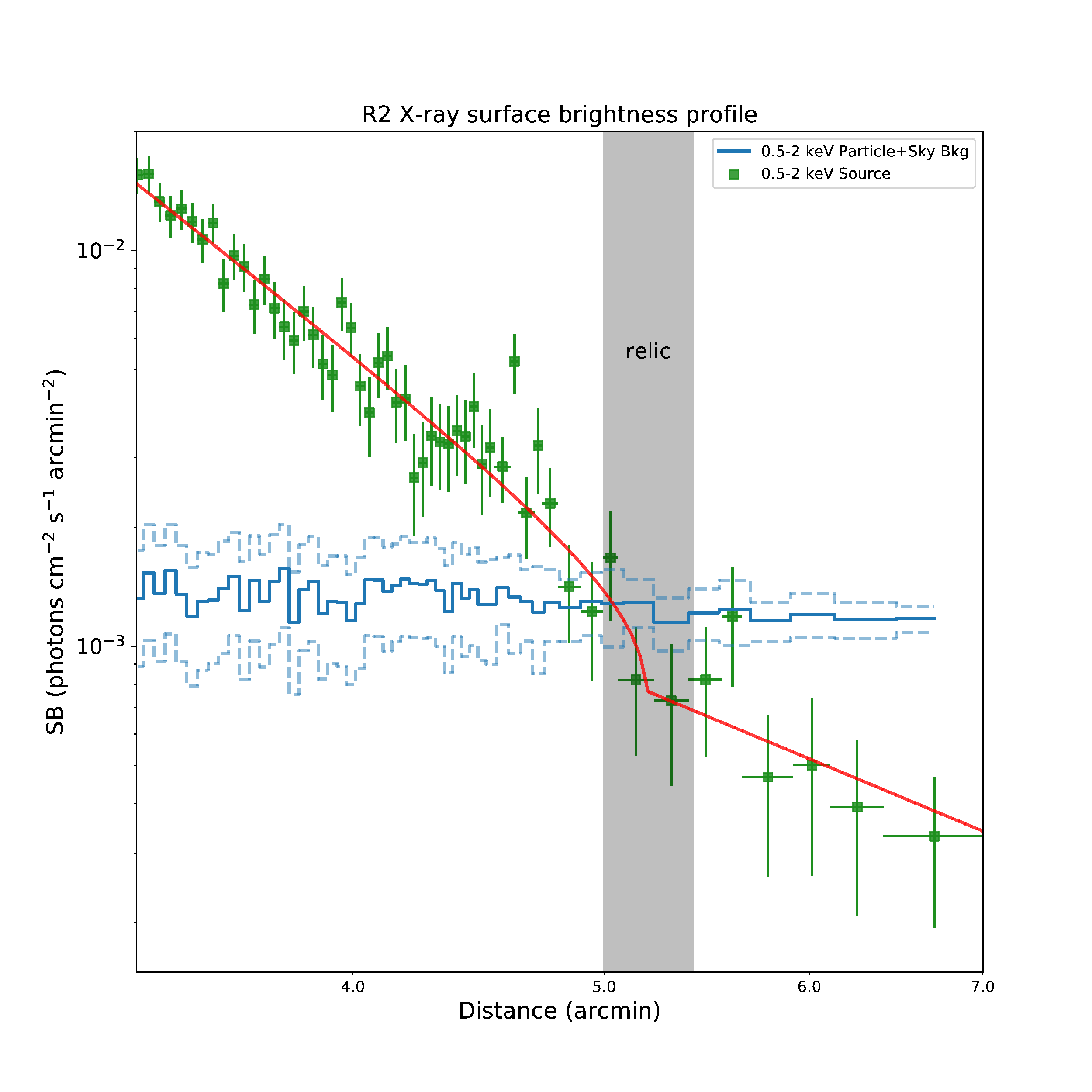}
\caption{X-ray surface brightness profile across relic R2 in the sector shown in Figure~\ref{fig:chandra}, with the distance increasing outwards to the SE. {The red line shows the best-fitting broken power-law density model (Eq.~\ref{eq:pyxel}). The relic location is indicated by the grey-shaded region. The blue and blue dashed lines show the total (sky and particle) background and corresponding uncertainty, respectively.}}
\label{fig:xray_profile}
\end{figure}

Due to the relatively low flux density of R2, it is difficult to identify any specific region of the relic associated with the point of particle injection at the shock front. However, if we make the assumption that the R2-shock follows a simple continuous injection model with $\alpha_{\rm{int}}=\alpha_{\rm{inj}}-0.5$, we obtain an estimate for the injection spectral index of $\alpha_{\rm{inj}}=-1.31 \pm 0.26$. This corresponds to a shock with Mach number $\mathcal{M}=1.86^{+0.29}_{-0.17}$.

To search for the presence of a shock at the location of R2, we extracted a surface brightness profile in the 0.5--2.0~keV band {with {\tt PyXel} \citep{2016AAS...22831709O,2017AAS...22943808O}} in a sector centered on RA $=3.553916\degr$ DEC $= -30.36725\degr$. We used opening angles between 190\degr and 230\degr. This sector was chosen to match the radius of curvature of the relic R2, see Figure~\ref{fig:chandra}. Regions with compact X-ray sources were excluded. The instrumental and sky backgrounds were subtracted.

We fitted a broken power-law density model to the surface brightness (see Eq.~\ref{eq:pyxel}) {with {\tt PyXel}}, assuming that the emissivity is proportional to the density squared 
\begin{equation}
  n(r)=%
  \begin{cases}
    \left(n_{2}/n_{1}\right) n_0 \left(\frac{r}{r_{\rm{edge}}} \right)^{a_2}  \mbox{ ,} &\text{$r \le r_{\rm{edge}}$} \\
    \\
    n_0 \left(\frac{r}{r_{\rm{edge}}} \right)^{a_1}  \mbox{ ,}&\text{$r > r_{\rm{edge}}$}   \mbox{ .}
  \end{cases}
  \label{eq:pyxel}
\end{equation}
In the above equation the subscripts 1 and 2 denote the upstream and downstream regions, respectively.
The parameter $n_{2}/n_{1} \equiv R $ is the electron density jump (i.e., compression ratio).  The parameters $a_1$ and $a_2$ define the slopes of the power-laws, $n_0$ is a normalization factor, and $r_{\rm{edge}}$ is the position of the jump. The model is then integrated along the line of sight assuming spherical symmetry. {The uncertainties on the best-fitting parameters are determined using a Markov chain Monte Carlo (MCMC) method \citep{2013PASP..125..306F}.}

For a shock,  $n_{2}/n_{1}$ is related to the Mach ($\mathcal{M}$) number via the Rankine-Hugoniot  condition \citep{1959flme.book.....L}
\begin{equation}
{\cal M}=\left[\frac{2 R}{\gamma + 1 - R(\gamma -1)}\right]^{1/2}   \mbox{ ,}
\label{eq:machne}
\end{equation}
where we take $5/3$ for the adiabatic index of the gas ($\gamma$).

From the surface brightness profile we see some evidence for a  change in the slope at a radius of about 5$\arcmin$ (see Figure~\ref{fig:xray_profile}). The best-fitting double power-law model finds the presence of a density jump with $R=1.39^{+0.34}_{-0.22}$ at $r_{\rm{edge}} =5.15\arcmin^{+0.16}_{-0.13}$. {All uncertainties are quoted at the $1\sigma$ level. The corresponding MCMC ``corner plot'' for the distribution of the uncertainties in the fitted parameters is shown in Appendix~\ref{fig:corner}. Although the jump is significant given the $1\sigma$ uncertainties, we cannot completely rule out the possibility of a statistical fluctuation (a $\sim10\%$ chance).}
If we assume that this {possible} density jump traces a shock it would correspond to a shock Mach number of $\mathcal{M}=1.26^{+0.25}_{-0.15}$, with the shock traveling to the SE. If the relic traces an outward traveling shock it must be located on the SE side of the relic. {The best fit location of the density jump is slightly closer to the NW side of the relic. However, the relic's outer SE edge is still located within the 90$\%$ confidence limits of $r_{\rm{edge}}$.}

We conclude that R2 likely traces a shock given that (1) it {probably} traces a density jump, (2) the source is located in the cluster outskirts, and (3) the observed radio properties (Mpc extent and polarization properties). We note that the surface brightness edge we find is not the same one as reported by \cite{2011ApJ...728...27O}. The orientation of the shock reported by \citeauthor{2011ApJ...728...27O} and the one corresponding to R2 are roughly parallel, however they are not directly related given that the separation between the two is $\sim 0.5$~Mpc. Interestingly, the edge is located in a sector just above (further clockwise) the one taken by \cite{2010arXiv1010.3660M}, where a very low-contrast jump was reported. Since the edge is located at approximately the same cluster-centric distance as the edge seen by \cite{2010arXiv1010.3660M}, it is likely an extension of same shock front. Therefore, we conclude that there are two shocks fronts in the SE of the cluster with similar orientations, but separated by $\sim 0.5$~Mpc. The presence of a bullet-like feature detected by \cite{2011ApJ...728...27O} south of the main Core (Figure~\ref{fig:chandra}) does suggest a merger event of the main Core with a smaller substructure seen just after core passage. This merger event likely created the shock front reported by \cite{2011ApJ...728...27O}. Given its location, the shock corresponding to R2 must have formed from an  earlier merger event. The orientation of the R2-shock favors a NW-SE merger axis, along the axis where NW and S3 and S4 components are located (Figure~\ref{fig:xrayoptical}). However, given the complexity of the A2744 cluster with multiple ongoing mergers  it is hard to draw firm conclusions and identify which of these components could have been involved in the merger event that created the R2-shock.

\begin{figure}[h!]
\includegraphics[angle=0, width=1.0\columnwidth]{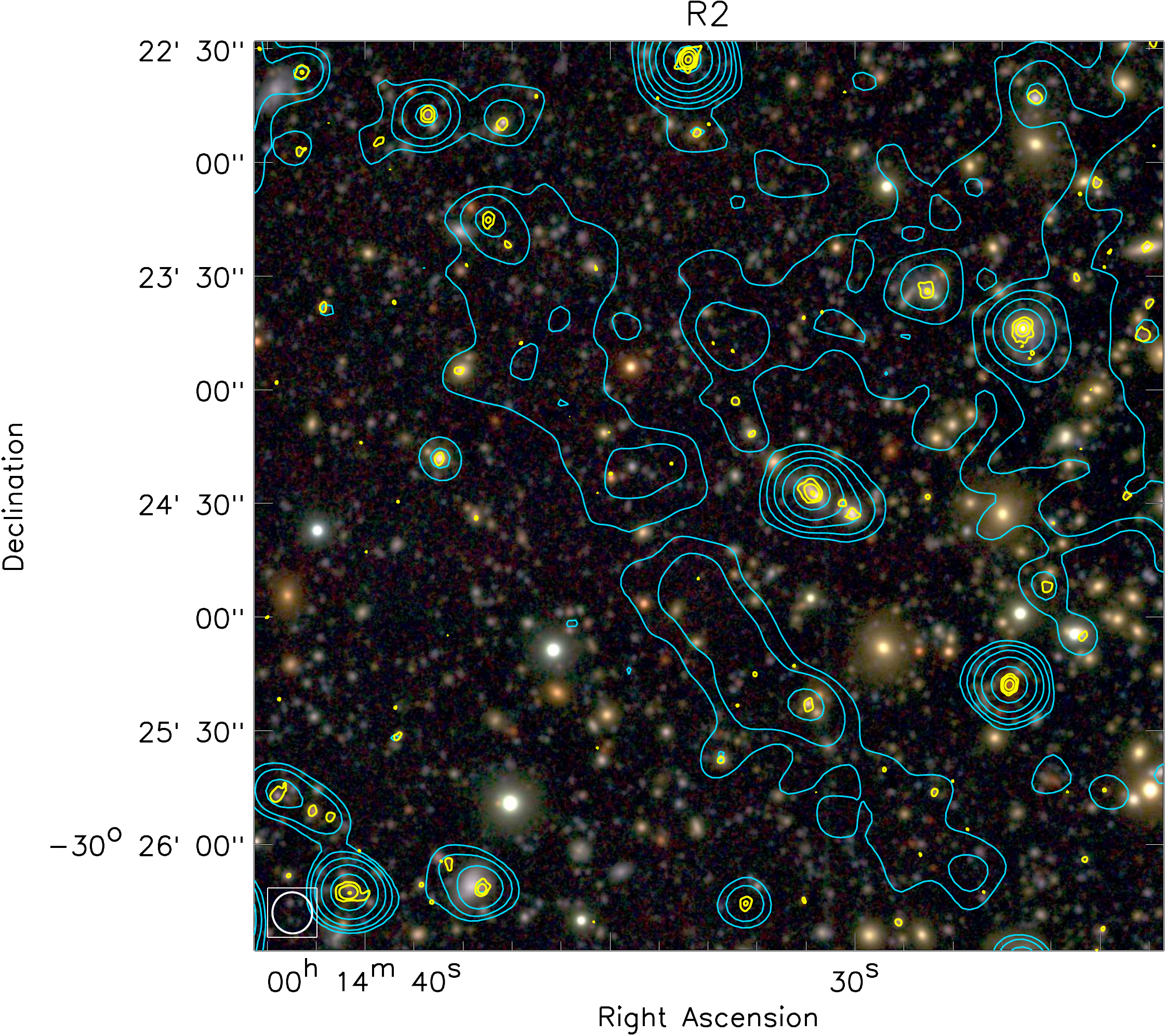}
\caption{Subaru optical BRz  color image \citep{2016ApJ...817...24M}  {around relic R2} with radio contours overlaid. The blue and red radio contours are from the 10\arcsec~uv-tapered and robust=0 1--4~GHz wide-band images, respectively. Contour levels are drawn at $[1,2,4,\ldots] \times 3\sigma_{\rm{rms}}$ (blue) and $[1,4,8,\ldots] \times 4\sigma_{\rm{rms}}$ (red).}
\label{fig:opticalR2}
\end{figure}

For the outer edge corresponding to R2, the X-ray derived Mach number ($\mathcal{M}=1.26^{+0.25}_{-0.15}$) is lower than the radio derived Mach number ($\mathcal{M}=1.86^{+0.29}_{-0.17}$). Similar discrepancies have been seen before \citep[e.g.,][]{2016ApJ...818..204V}, which was invoked as evidence against DSA. Although, the radio derived Mach number was based on the measured integrated spectral index and not on a more reliable measured injection spectral index \citep[see][]{2015JKAS...48....9K}.  {Furthermore, the discrepancy is not very large given the statistical uncertainties and the simplifying assumptions made in the X-ray surface brightness profile fitting.} As such, it is hard to comment on the likely acceleration model for R2 based on the difference between  the radio spectral index and X-ray derived Mach numbers. On the other hand, shocks with   $\mathcal{M} < 2$ should be very inefficient accelerators and therefore a re-acceleration scenario might be preferred. Although several compact, faint radio sources are located in the relic's vicinity, we do not identify a clear candidate for supplying fossil radio plasma for a re-acceleration scenario, see Figure~\ref{fig:opticalR2}.

\subsubsection{Diffuse Source R3}
\label{sec:R3}
Similarly to R2, parts of source R3 were detected in the radio maps produced by V13, though it was not identified as an individual source. In those observations it was deemed to be an elongated extension of the halo and was dubbed the ``North-Western ridge". 

At high resolution ($\gtrsim15\arcsec$) the emission is seen as a long ($\gtrsim 1.1$~Mpc) filament protruding from the NW edge of the halo, out to a distance of $\sim 1.6$~Mpc. The emission appears to arc somewhat to the north as it extends away from the cluster centre. Projection effects make it difficult to determine if the structure is connected to the halo or if it is an independent feature. As such our estimate of a physical size should be considered a lower limit.

We estimate the monochromatic radio power of the source to be $P_{\rm{1.4 GHz}}=(4.72 \pm 0.45) \times 10^{23}$~W~Hz$^{-1}$. However, this is potentially an over-estimate. The relic region was drawn using the polarization intensity map as a guide. As shown in Figure~\ref{fig:labels}, the resulting region extends partway into the halo. The subsequent flux density measurements are likely contaminated with halo flux. 


The source is also highly polarized along its length with a mean polarization fraction of $\sim30\%$. Based on the filament-like morphology, large size, and high polarization fraction we also classify this source as a newly discovered radio relic. Henceforth we shall refer to source R3 as relic R3. Interestingly, the orientation of the aligned E-vectors perpendicular to the relic's orientation (Figure~\ref{fig:pol_vec_map}) suggests a shock propagation direction indicative of an off-centre N-S minor merger. {A caveat is that  R3 is partly projected on top of the NW ``interloper". Therefore, the  polarization vectors might have been affected by Faraday Rotation, depending on the (unknown) relative location of the radio emission with respect to the magnetized ICM.}

The relic possesses one of the flattest spectra observed in radio relics with an integrated spectral index value of $-0.63$. Such a flat spectrum is quite rare and one cannot assume a simple stationary shock model to explain the spectrum. The reason being that in the limit of strong shocks ($\mathcal{M} \gg 1$) the particle energy distribution, $\delta$, approaches an asymptotic flat value of $-2$. This means that the injection spectral index cannot be flatter than $-$0.5, whereas we would obtain a value of $-0.13$ if we would add 0.5 to the value of  $-0.63$ (Eq.~\ref{eq:inject}). If the measured integrated spectral index value is the injection spectrum, we would also be observing a rather strong shock. Based on an injection value of $\alpha_{\rm{inj}}=-0.63\pm0.21$ the shock would have a Mach number measuring $\mathcal{M}=4.04\pm1.4$.

\begin{figure}[h!]
\includegraphics[angle=180, width=1.0\columnwidth]{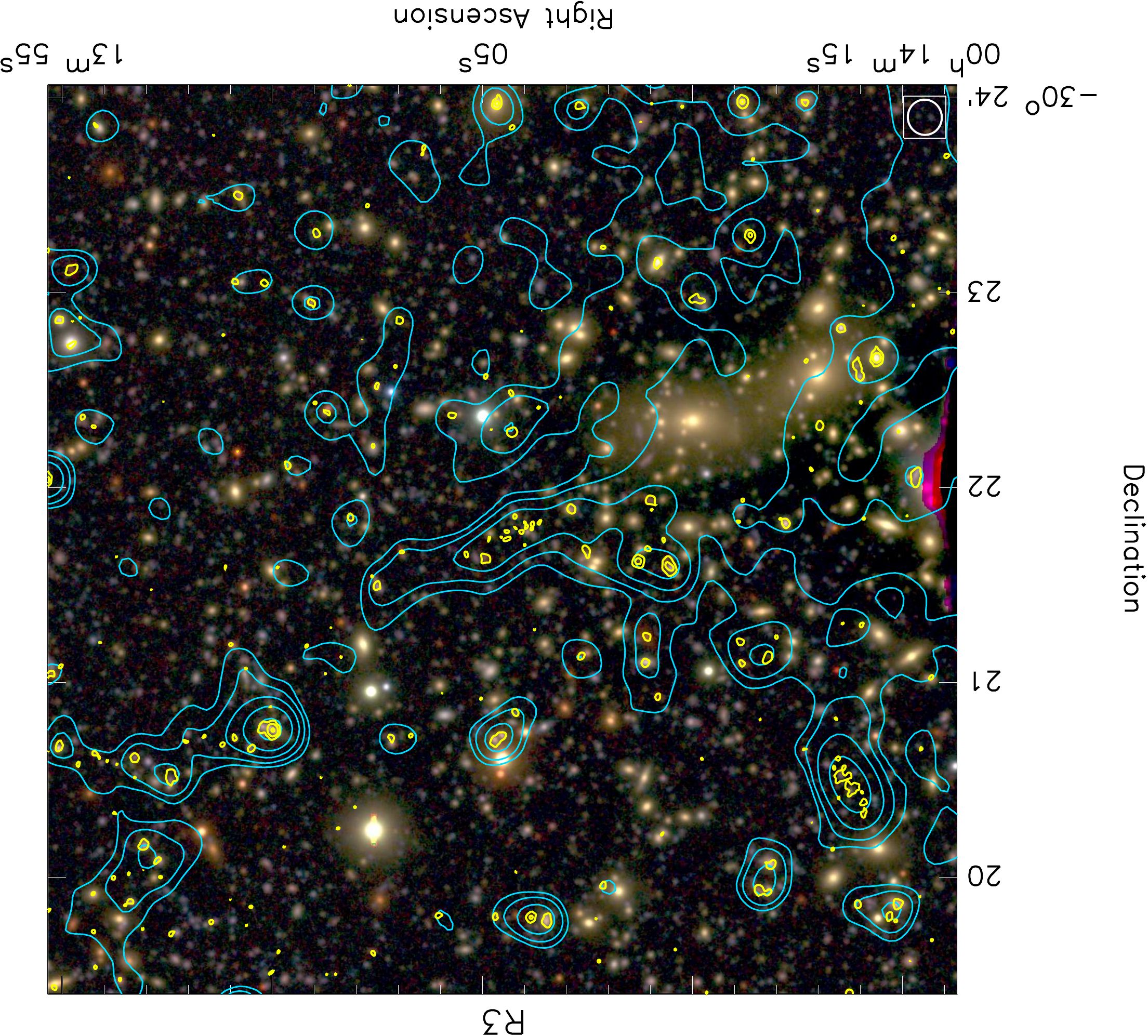}
\caption{Subaru optical BRz  color image \citep{2016ApJ...817...24M}  {around relic R3} with radio contours overlaid. The blue and red radio contours are from the 10\arcsec~uv-tapered and robust=0 1--4~GHz wide-band images, respectively. Contour levels are drawn at $[1,2,4,\ldots] \times 3\sigma_{\rm{rms}}$ (blue) and $[1,4,8,\ldots] \times 4\sigma_{\rm{rms}}$ (red).}
\label{fig:opticalR3}
\end{figure}

It is possible we are observing the integrated spectrum well before the break frequency, whilst the emission is determined simply by the injection spectrum, $S_{\nu}\propto\nu^{\alpha_{\rm{inj}}}$, in which case this relic is  ``young'' and stationary shock conditions do not (yet) apply \citep[e.g.,][]{2012A&A...543A..43V,2015A&A...575A..45T}. Alternatively, the flat spectrum reflects the fossil distribution slope in the case of re-acceleration, or a combination of the two above \citep[re-acceleration + non-stationary shock conditions,][]{2015ApJ...809..186K,2015JKAS...48....9K}. A radio-optical overlay does not reveal any obvious source of fossil radio plasma, Figure~\ref{fig:opticalR3}. The relative bright compact source at RA=3.4999688\degr~DEC=$-30.3459524\degr$ has  $z_{\rm{phot}} = 0.46\pm0.05$ \citep{2016ApJ...817...24M} and is thus likely a background object and not associated with R3.

Using the mean polarization fraction, we derive a viewing angle of $\theta\gtrsim52^{\circ}$ for R3, see Figure~\ref{fig:pol_frac}. Purely from a visual inspection, one would probably expect the relic to be viewed close to perfectly edge-on, considering it measures just 150~kpc across at its widest point. {Interestingly, the relic appears to sit exactly on-top of the NW ``interloper" gas clump/core. Whereas typically, shock fronts should lead gas cores. We see no gas cores on either side of the relic. Therefore, the positioning of the radio shock front/relic perhaps implies that the shock plane is located in front or behind the NW ``interloper"  and just projected on top of it.}

\begin{figure}[h!]
\includegraphics[width=0.49\textwidth, angle=0]{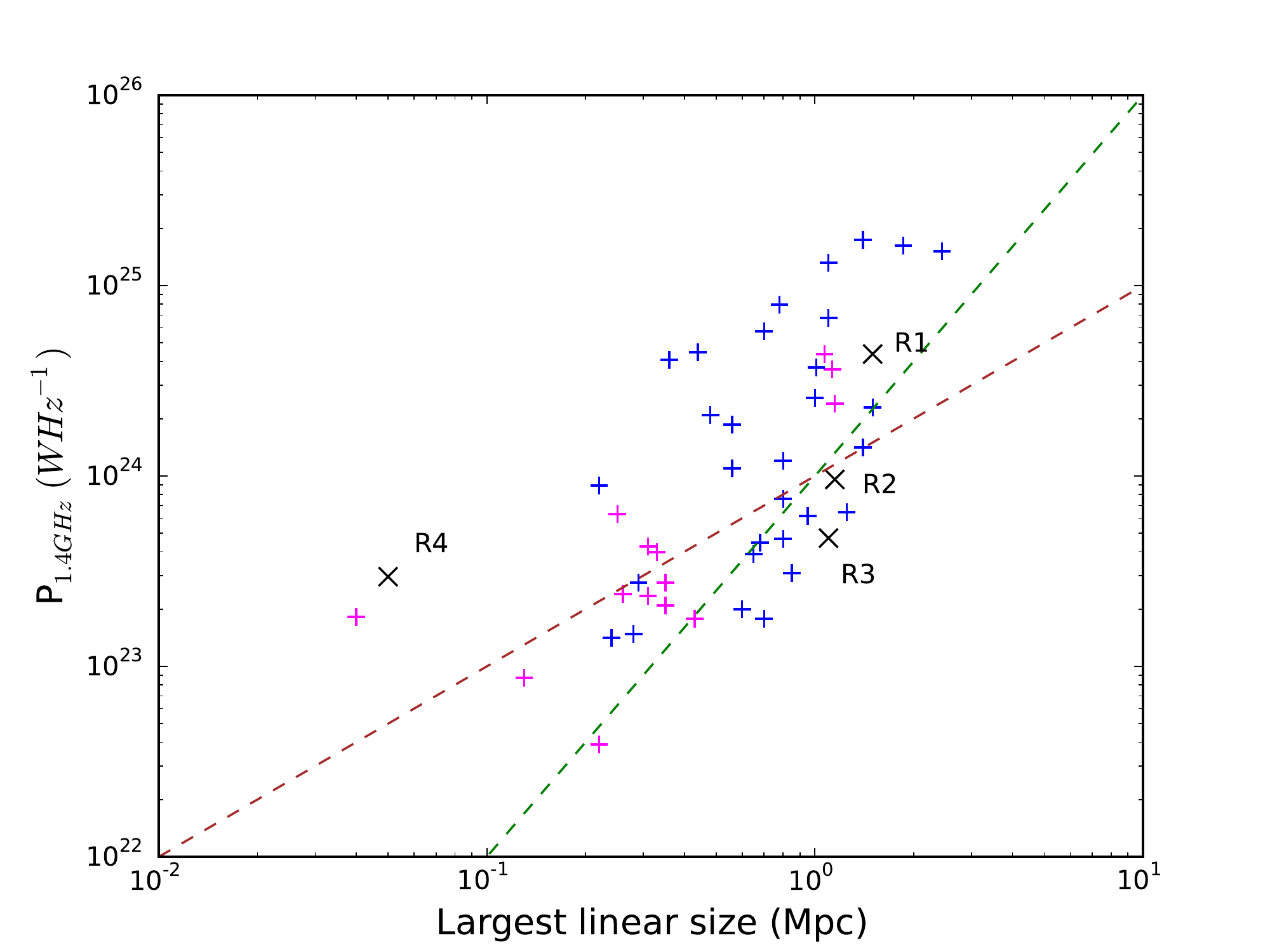}
\caption{1.4~GHz monochromatic radio power versus largest linear size (LLS) for a sample of known radio relics taken from \cite{2012A&ARv..20...54F}. `Elongated' relics are shown in blue. `Roundish' relics are shown in magenta. In black, and labelled with $\times$, are the four candidate radio relics discussed in this paper. The two dashed lines  represent the cases where  $P_{1.4{ \rm{GHz }}} \propto \rm{LLS}$ (red) and $P_{1.4{ \rm{GHz }}} \propto \rm{LLS}^2$ (green). }
\label{P_LLS}
\end{figure}

\subsubsection{Diffuse source R4}
\label{sec:R4}

\begin{figure}[h!]
\includegraphics[angle=180, width=1.0\columnwidth]{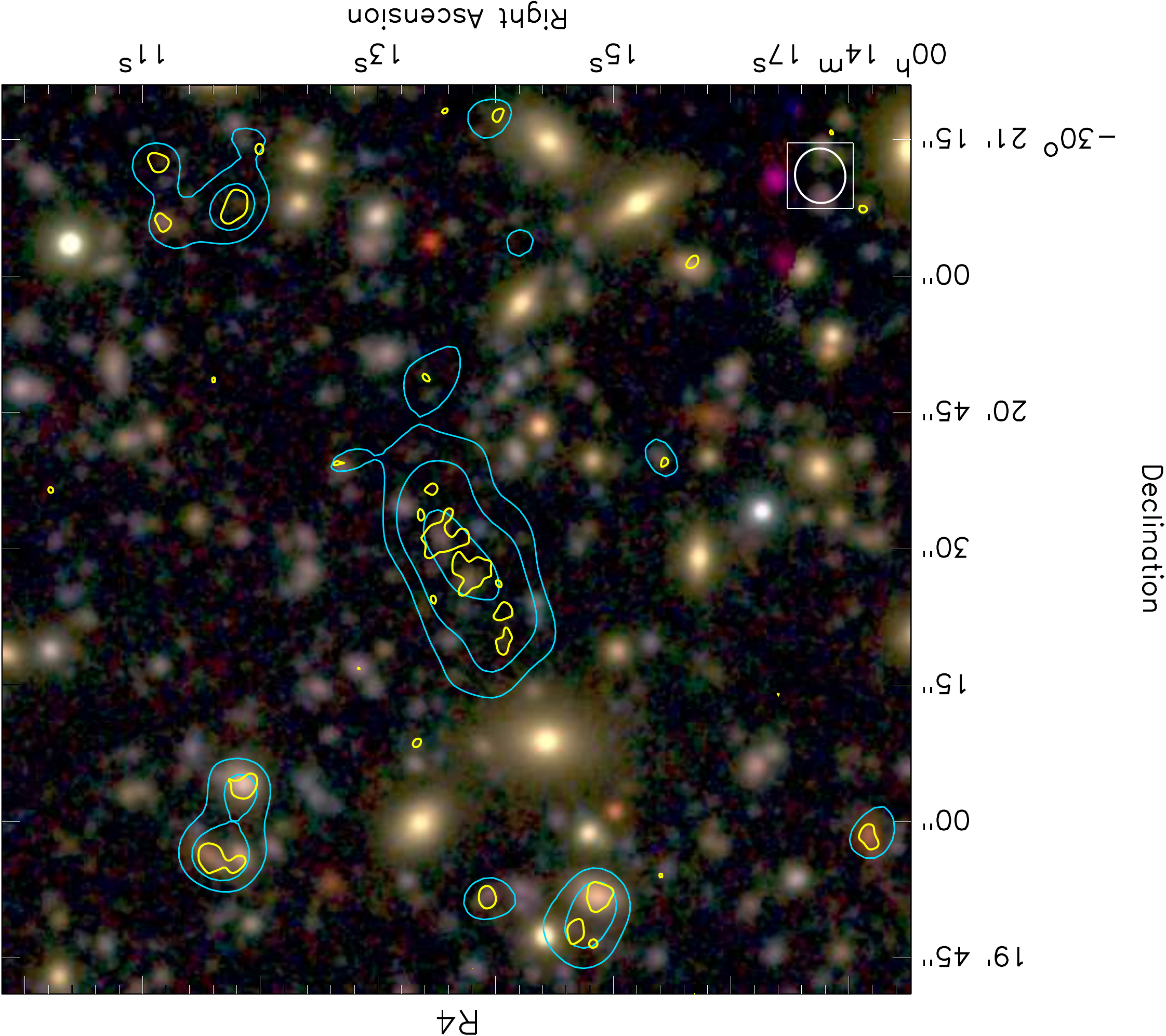}
\caption{Subaru optical BRz  color image \citep{2016ApJ...817...24M}  {around relic R4} with radio contours overlaid. The blue and red radio contours are from the 5\arcsec~uv-tapered and robust=0 1--4~GHz wide-band images, respectively. Contour levels are drawn at $[1,2,4,\ldots] \times 3\sigma_{\rm{rms}}$ (blue) and $[1,4,8,\ldots] \times 4\sigma_{\rm{rms}}$ (red).}
\label{fig:opticalR4}
\end{figure}

Diffuse source R4 consists of a small patch of diffuse emission (Figures~\ref{fig:xrayoptical},~\ref{fig:hi_res_image}, and \ref{fig:opticalR4}). Like R2 and R3, it too appears in the V13 observations, and even in the O07 325~MHz radio map, though it is partly blended with the halo.  

R4 measures just $50 \times 30$~kpc in size and is located $\sim 1$~Mpc north-northwest from the cluster core. We find no obvious optical counterpart in the vicinity, see Figure~\ref{fig:opticalR4}. Most intriguingly, this small patch appears to be highly polarized at $\sim30\%$, with E-vectors aligned perpendicular to the source elongation, just like a typical radio relic. We estimate a monochromatic radio power of  $P_{\rm{1.4 GHz}}=(3.0 \pm 0.3) \times 10^{23}$~W~Hz$^{-1}$ for the source. It is the only diffuse source other than R1 to exhibit a spectral index gradient, with $\alpha$ steepening from $\sim-0.8$ to $\sim-1.5$ in the N-S direction, with an average integrated spectral index of $\alpha = -1.34\pm0.24$. {This integrated spectral index is consistent with the flattest spectral index we observe under the assumption of stationary conditions, i.e., Equation~\ref{eq:inject}. Using the integrated spectral index, Equations~\ref{eq:mach} and \ref{eq:inject} we calculate a shock Mach number of $\mathcal{M}=2.62^{+1.76}_{-0.50}$.}

In the introduction, we mentioned three different types of radio relics: giant radio relics, AGN relics and radio phoenices. These can also be divided more generally, as in \cite{2012A&ARv..20...54F}, into `elongated' and `roundish' relics.  Elongated relics in this case simply correspond to the standard giant radio relics we previously defined. `Roundish' relics consist of diffuse extended radio sources with more regular and roundish morphology, although, some of them display complex filamentary shapes. They too tend to be  located at smaller distances from the cluster centre. The key defining feature though is that of a steep and curved integrated radio spectrum. It is thought they consist of AGN relics and radio phoenices, though other explanations have also been suggested such as buoyant radio bubbles. 

A relation has been shown to exist between the radio power of a relic and its largest linear size (LLS), with larger relics possessing more power. For large relics seen edge-on one would expect that the depth of a relic scales roughly with the LLS and thus $P_{1.4{ \rm{GHz }}} \propto \rm{LLS}^2$. However, the correlation becomes more complicated when `elongated' or `roundish' relics are included. In Figure~\ref{P_LLS} we overlay the results for our four relic candidates onto a plot of the radio power at 1.4~GHz against the LLS of a sample of relics taken from \cite{2012A&ARv..20...54F}. Shown in blue are relics classified as elongated; roundish relics are shown in magenta. 

In general, elongated relics have greater radio power than roundish ones, with the exception of A1664, A2256 and A2345-W, as previously noted by \cite{2012A&ARv..20...54F}. Both types can have low $\sim 1{\times}10^{23}$~W~Hz$^{-1}$ radio powers, but of these, elongated relics tend to have a larger size. Relics R1, R2 and R3 fall nicely into the main elongated relics group. Source R4, on the other hand, appears to fall into the `roundish' relic category, {in particular given its small LLS.}

Based on the morphology of the source and the fact that it possesses all the trademark characteristics of a radio relic we suggest this is likely a new `roundish' relic. {Note that for a `roundish' relic the previous calculation of the shock Mach number is not valid.} However, currently we lack the required high frequency observations to determine if the integrated spectrum of the emission is curved, which is the predominant marker for a roundish relic. We therefore currently refrain from concluding that this is indeed a newly discovered radio relic. 

\section{Conclusions}
\label{sec:conclusion}
In this work, we presented new VLA radio observations of the cluster Abell~2744 at 1--4~GHz. Spectral index maps of the cluster, with resolutions ranging from $30\arcsec$ to $5\arcsec$, were constructed to study the origin of the diffuse emission in the cluster. In our images, we detect the previously known radio relic R1 and radio halo, along with three new diffuse sources, R2, R3, and R4.
Below, we summarize our results:

\begin{itemize}

\item
The spectrum of the radio halo is reasonably well described by a single power-law, with an integrated spectral index of $\alpha_{325}^{3000}=-1.32\pm0.14$, though there is some evidence of a potential steepening at higher frequencies. The spectral index maps reveal the spectral index across the radio halo to be rather uniform, although some steepening seems to occur at the faint outer parts of the halo. We do not find strong evidence for the presence of an spectral index--ICM temperature correlation, as was suggested earlier for the radio halo by \cite{2007A&A...467..943O}. The SE boundary of the radio halo is relatively well defined and aligns with the southern shock reported by \cite{2011ApJ...728...27O}. Similar halo-shock edges have been observed in other well-studied clusters and this appears to be a more common phenomenon, possibly providing important information for our understanding of the formation of radio halos.

\item
For the main relic R1, we measure an integrated spectral index of $\alpha=-1.32\pm0.09$, which is consistent with previous measurements. 
We observe a spectral index gradient across the relic's width which is indicative of acceleration in an outward traveling shock, with radiative losses in the shock-downstream region. We estimate a relic injection spectral index of $\alpha_{\rm{inj}}=-1.12\pm0.19$ from a spectral profile across the relic's width, corresponding to a shock Mach number of $\mathcal{M}=2.05^{+0.31}_{-0.19}$ under the assumption of DSA. This value is found to agree within the uncertainties with the X-ray Mach number derived by \cite{2016arXiv160302272E}. Relic R1 is polarized at the $27\%$ level, with the E-vectors perpendicularly oriented along the relic's major axis.  Using this polarization fraction, we determine the viewing angle of relic R1 to be $\theta\gtrsim50^{\circ}$, from which we constrain the geometry of the primary NE-SW merger axis to be within $\sim\pm40^{\circ}$ of the plane of the sky.

\item The newly discovered relic R2 is located to the SE of the cluster core. Relic R2 is a $\sim 1.2$~Mpc long  elongated source with an integrated spectral index of $\alpha = -1.81 \pm 0.26$. The source is polarized at the $43\%$ level, with the E-vectors perpendicular to the relic's orientation. From a {\chandra} X-ray image, we find evidence for a {possible} density jump at the relic's location. Fitting a broken power-law density model, we determine a density jump of $R=1.39^{+0.34}_{-0.22}$, which would correspond to a Mach number of $\mathcal{M}=1.26^{+0.25}_{-0.15}$. Using the mean polarization fraction we derive a relic viewing angle of $\sim70^{\circ}$, corresponding to a merger axis within $\sim\pm20^{\circ}$ of the plane of the sky.

\item The new relic R3 is located  to the NW of the cluster and measures $\gtrsim1.1$~Mpc in length. This relic is also highly polarized at the $\sim30\%$ level. For the integrated spectral index, we measure a value of $\alpha=-0.63\pm0.21$. The measured spectral index value of $-0.63$ is one of the flattest recorded for radio relics. Assuming what we measure is the injection spectral index, this would correspond to a high shock Mach number of $\mathcal{M}=4.04\pm1.4$ (under the assumption of DSA). Using the mean polarization fraction, we obtain a viewing angle of $\sim 52^{\circ}$. The orientation of the relic is peculiar, as it points northwest away from the Northwest sub-cluster.

\item We found a small patch of diffuse emission $\sim200$~kpc north of the radio halo. The source, R4, measures $50 \times 30$~kpc in size. The spectral index maps reveal a spectral index gradient across R4 and it has an integrated spectral index value of $\alpha=-1.34\pm0.23$. The source is also highly polarized with a mean polarization fraction of $\sim30\%$. The source could be a radio phoenix or AGN relic given its small size.

\item The detection of two new large relics adds yet more complexity to the already complicated merger history of A2744. The location and orientation of the new relic R2 could suggest it traces a shock wave produced when the main cluster core collided with the Northwestern interloper in a SE-NW event. The discovery of relic R3 seems to lend credence to a minor off-axis merger event between two sub-halos suggested by \cite{2016ApJ...817...24M}, although its precise location and orientation remain puzzling. 

\end{itemize}

\acknowledgments
{\it Acknowledgments:}
We thank the anonymous referee for useful comments.
The National Radio Astronomy Observatory is a facility of the National Science Foundation operated under cooperative agreement by Associated Universities, Inc. Support for this work was provided by the National Aeronautics and Space Administration through Chandra Award Number GO5-16133X issued by the Chandra X-ray Observatory Center, which is operated by the Smithsonian Astrophysical Observatory for and on behalf of the National Aeronautics Space Administration under contract NAS8-03060.

R.J.W. is supported by a Clay Fellowship awarded by the Harvard-Smithsonian Center for Astrophysics. W.R.F., C.J., and F.A-S. acknowledge support from the Smithsonian Institution. F.A-S. acknowledges support from Chandra grant GO3-14131X. Basic research in radio astronomy at NRL by T.E.C. is supported by 6.1 Base funding. M.N. acknowledges  PRIN-INAF 2014 1.05.01.94.02. This research made use of APLpy, an open-source plotting package for Python \citep{2012ascl.soft08017R}. This research made use of Astropy, a community-developed core Python package for Astronomy \citep{2013A&A...558A..33A}.

\clearpage
\appendix

\section{Spectral index uncertainty maps, region locations, and T-T plots}
\label{sec:appA}

\begin{figure}[h!]
\centering
\includegraphics[angle=180,width=0.49\columnwidth]{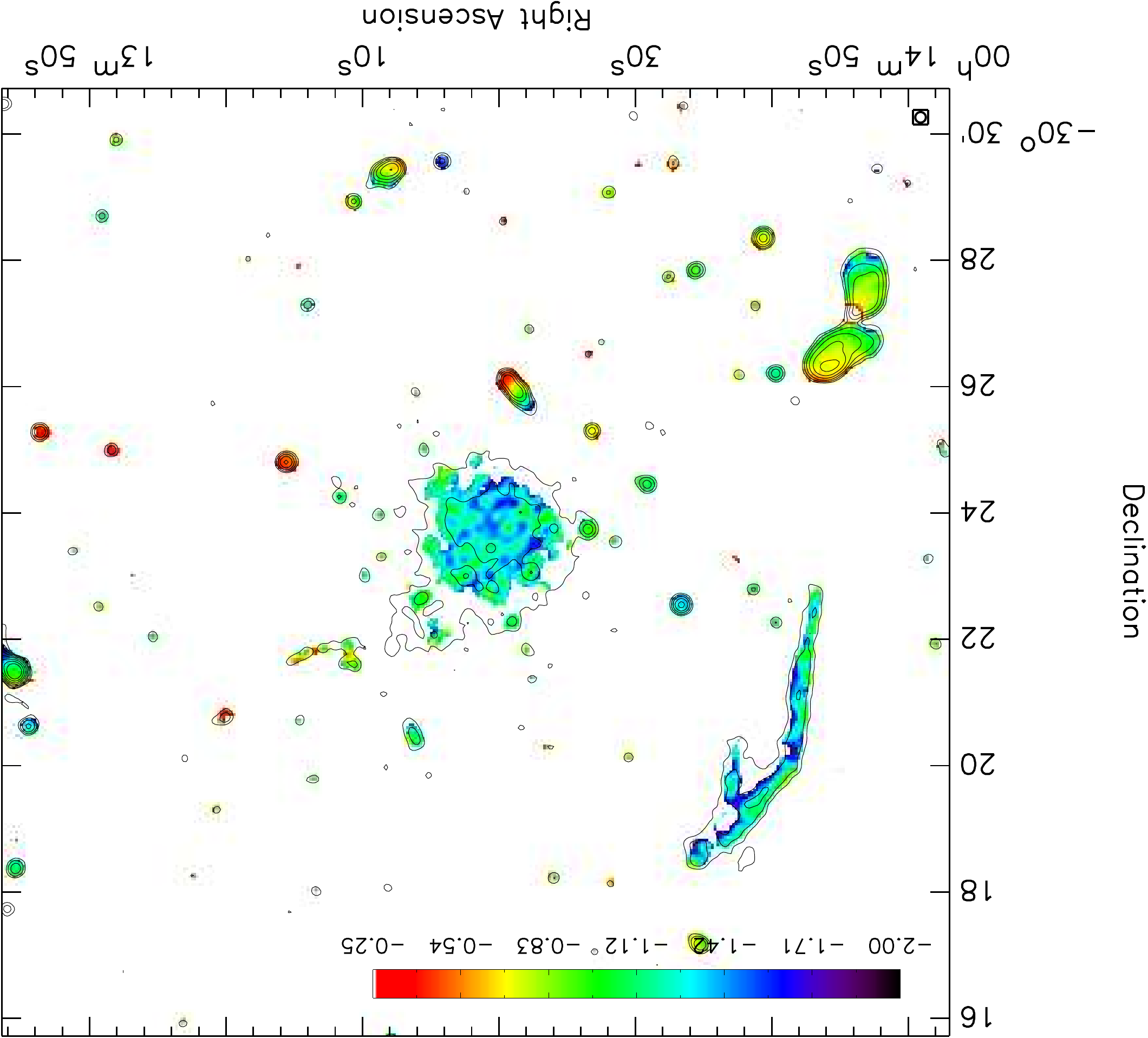}
\includegraphics[angle=180,width=0.49\columnwidth]{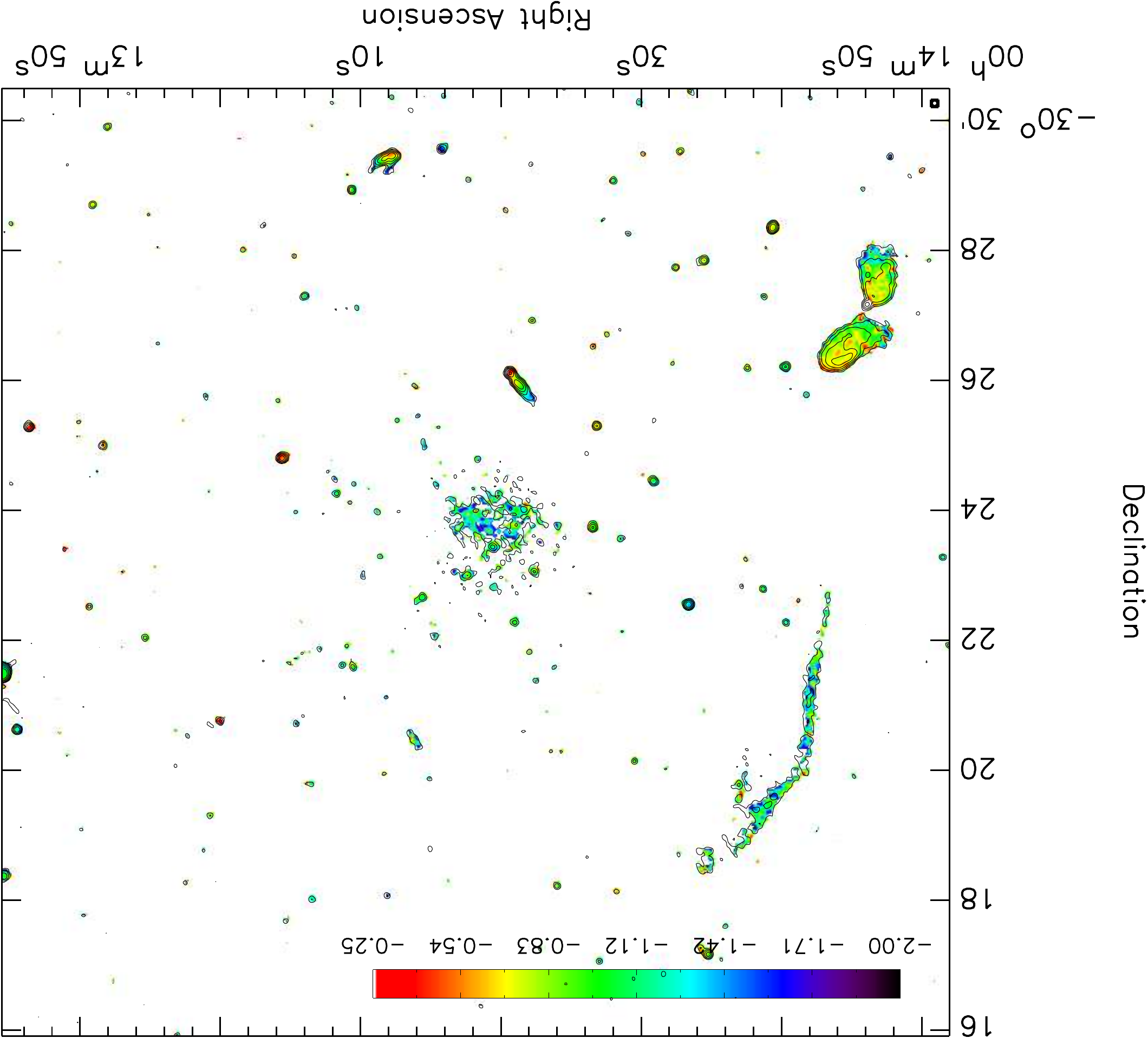}
\caption{Spectral Index maps of A2744 between 1.5--3.0 GHz, tapered to resolutions of $10\arcsec$ (\textit{left}) and 5$\arcsec$ (\textit{right}). Contour levels are obtained from the 1.5 GHz image and placed at the $[1,2,4,8,\ldots] \times 4\sigma_{\rm{rms}}$ and 3$\sigma_{\rm{rms}}$ levels respectively.}  
\label{spix_map_10}
\end{figure}

\begin{figure}[th!]
\centering
\includegraphics[ angle=180,width=0.49\textwidth]{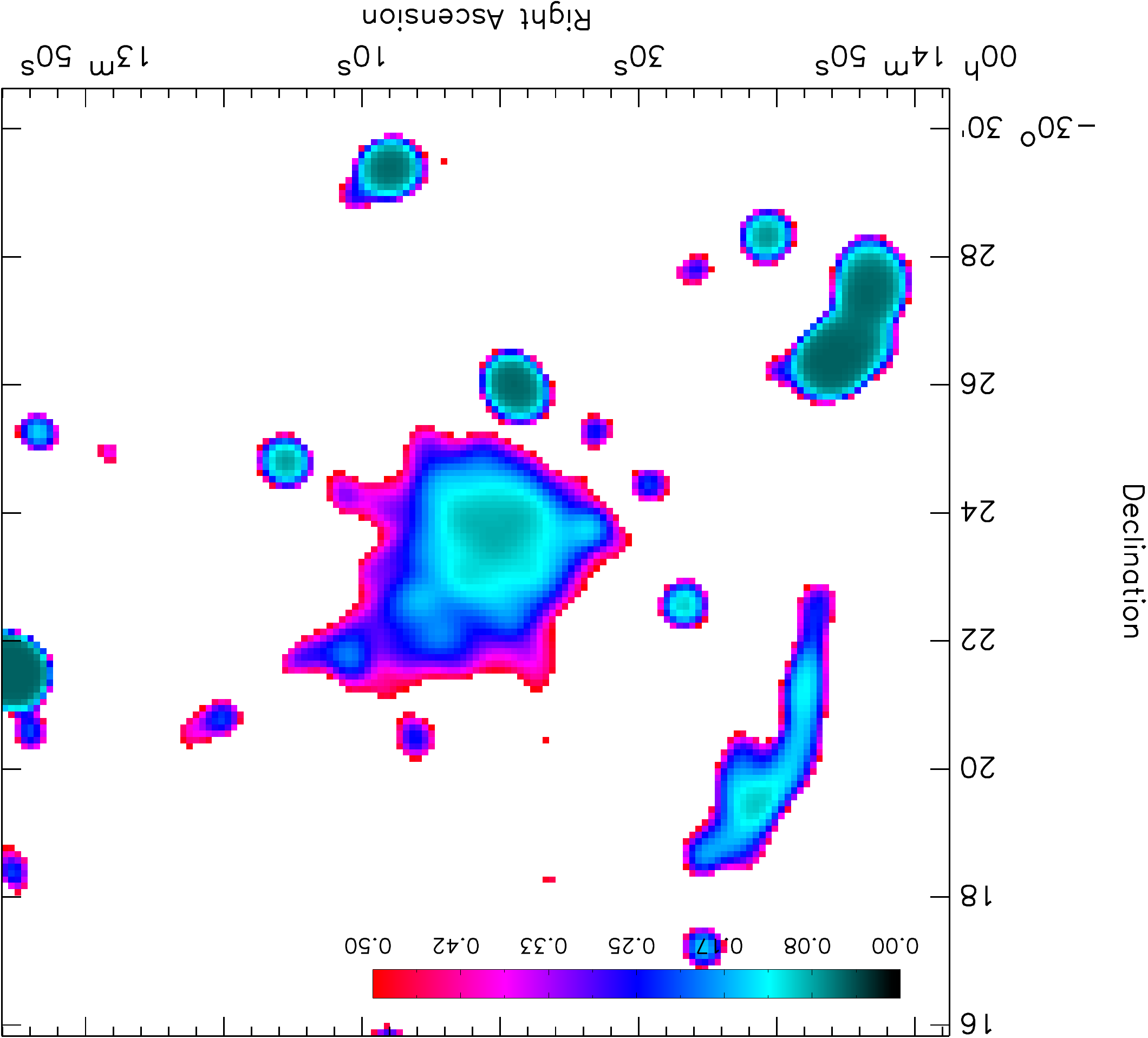}
\includegraphics[ angle=180,width=0.49\textwidth]{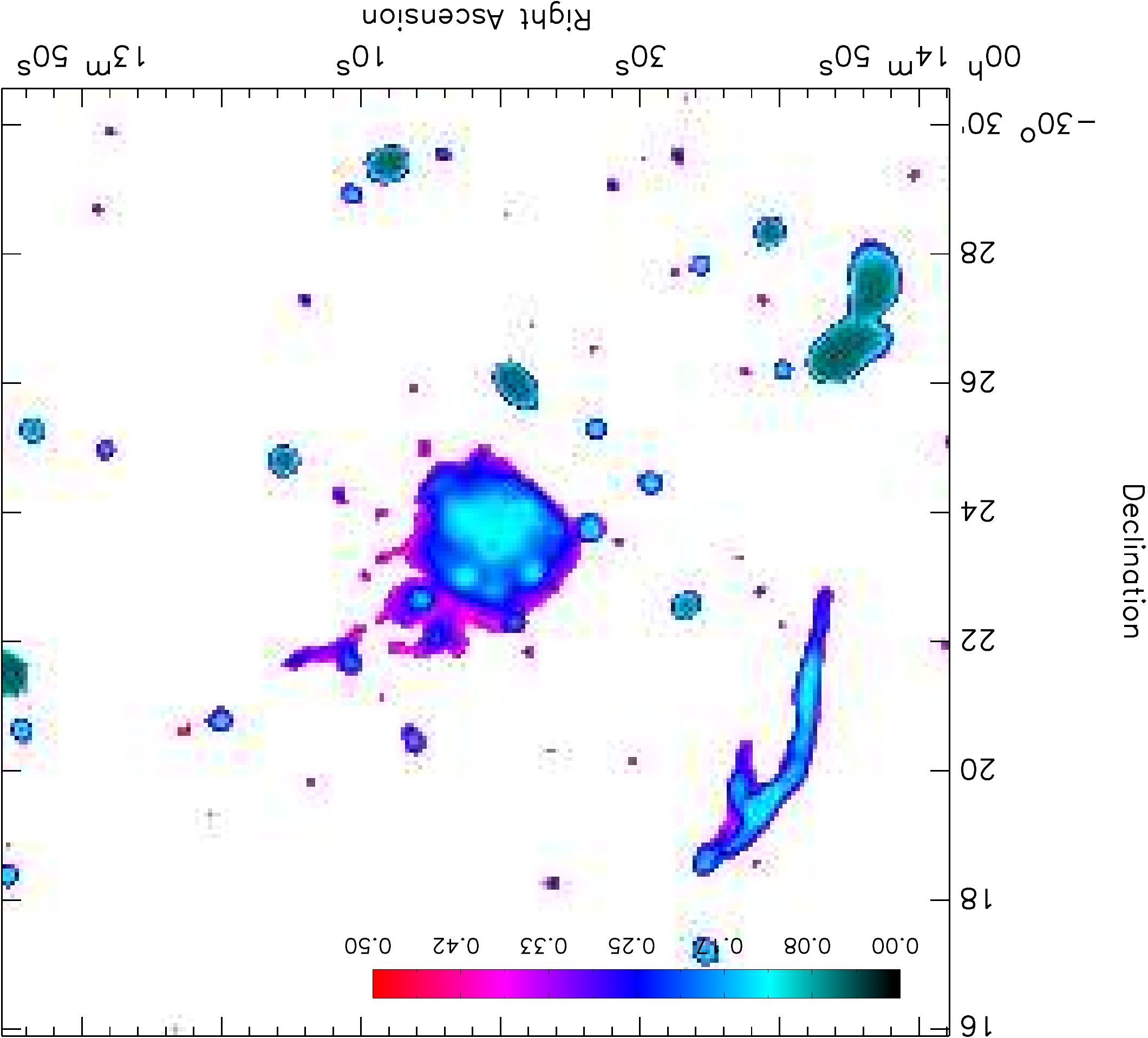}
\includegraphics[ angle=180,width=0.49\textwidth]{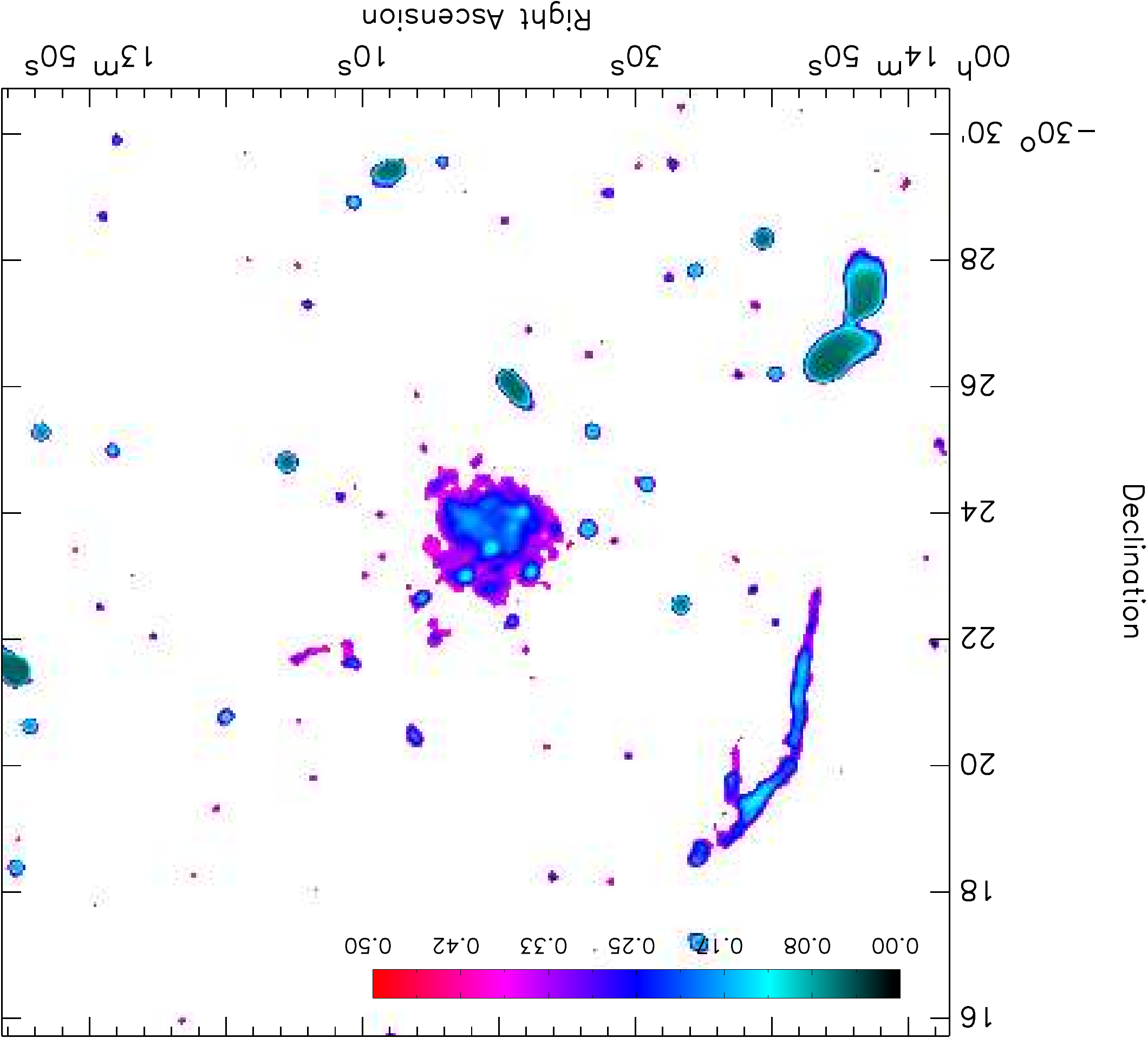}
\includegraphics[ angle=180,width=0.49\textwidth]{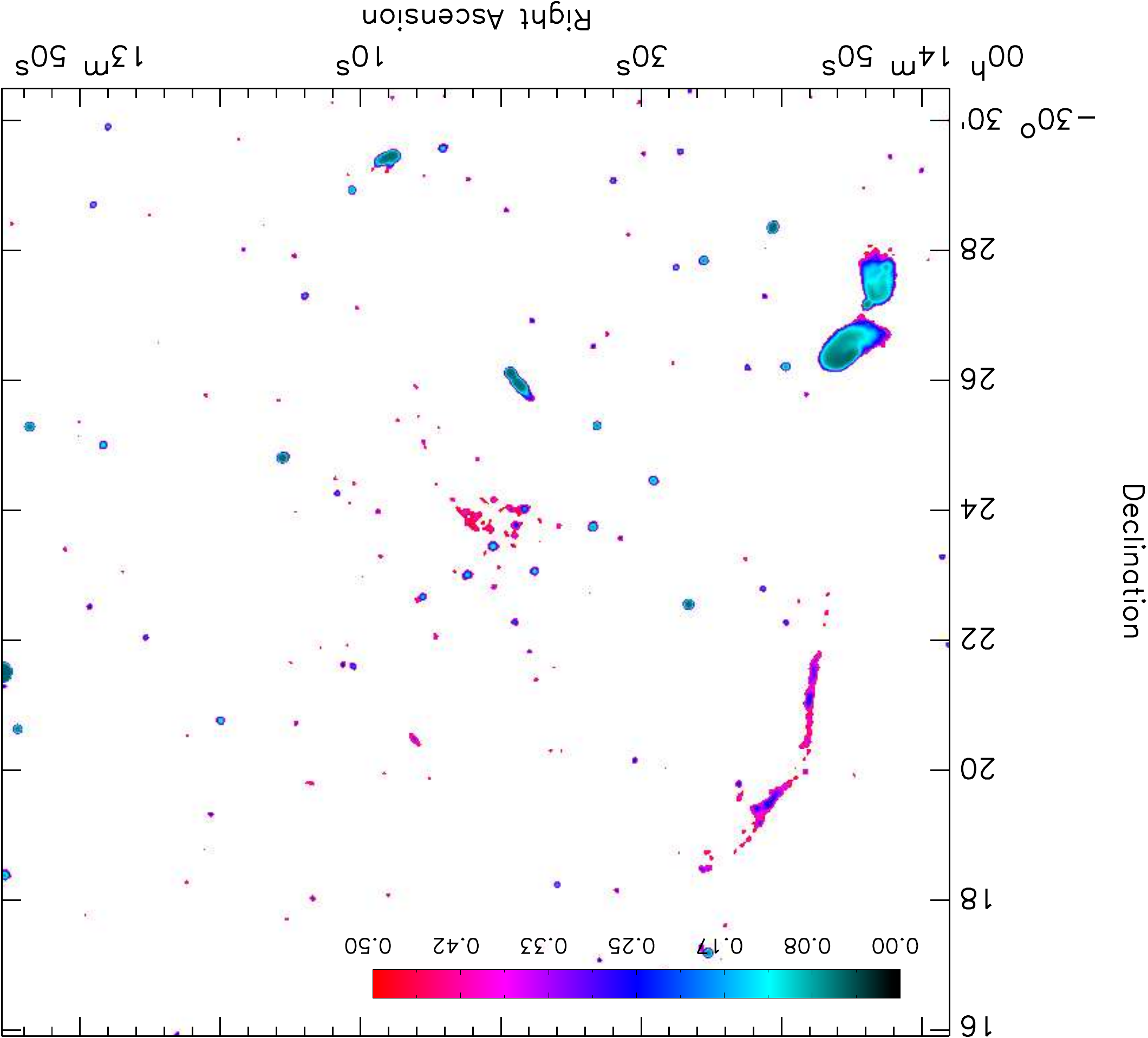}
\caption{Spectral index error maps corresponding to Figures~\ref{spix_map_30} (\textit{top}) and~\ref{spix_map_10} (\textit{bottom}).}
\label{spix_maps_e}
\end{figure}

\begin{figure}[th!]
\centering
\includegraphics[width=0.45\textwidth, angle=0]{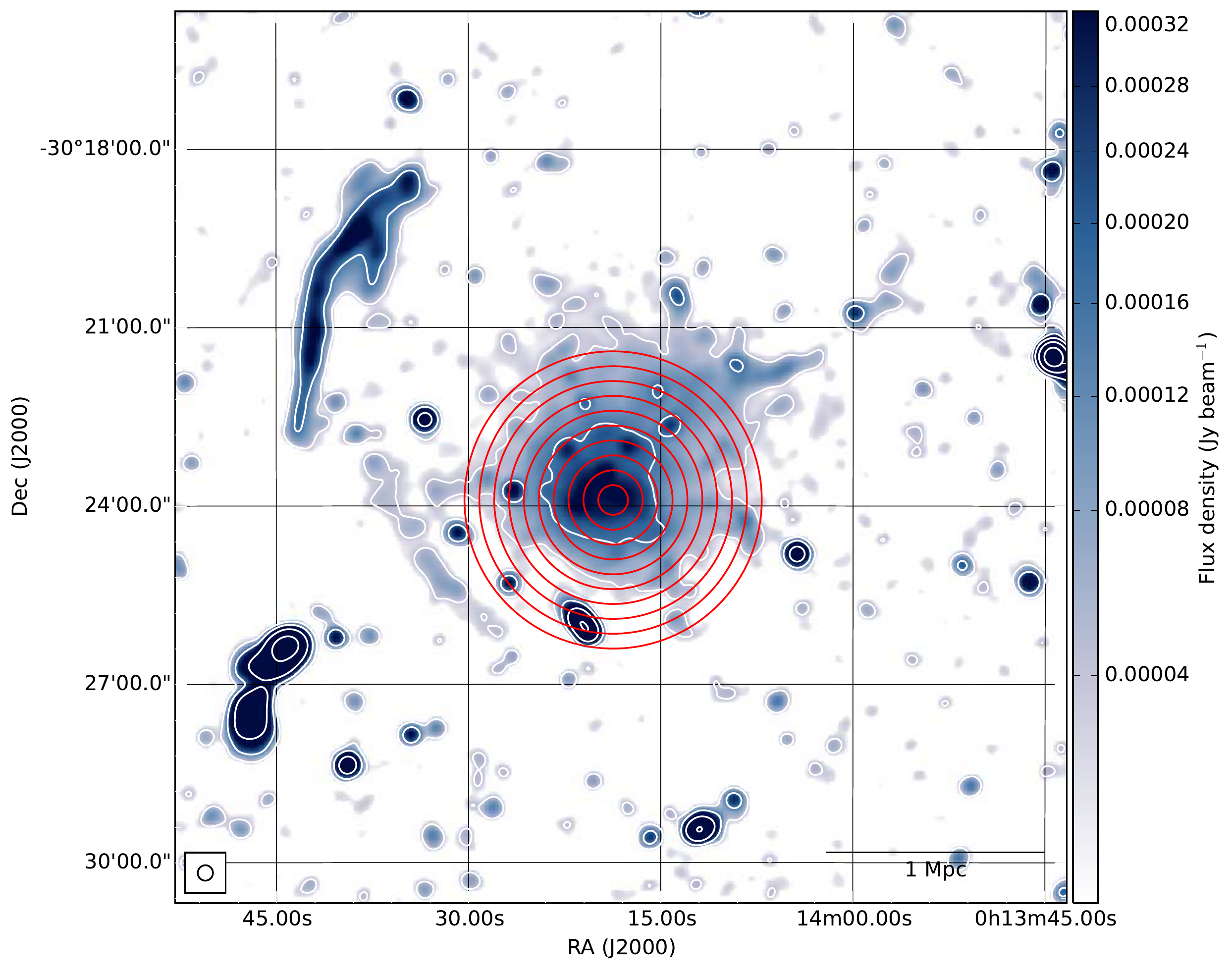}
\includegraphics[width=0.45\textwidth, angle=0]{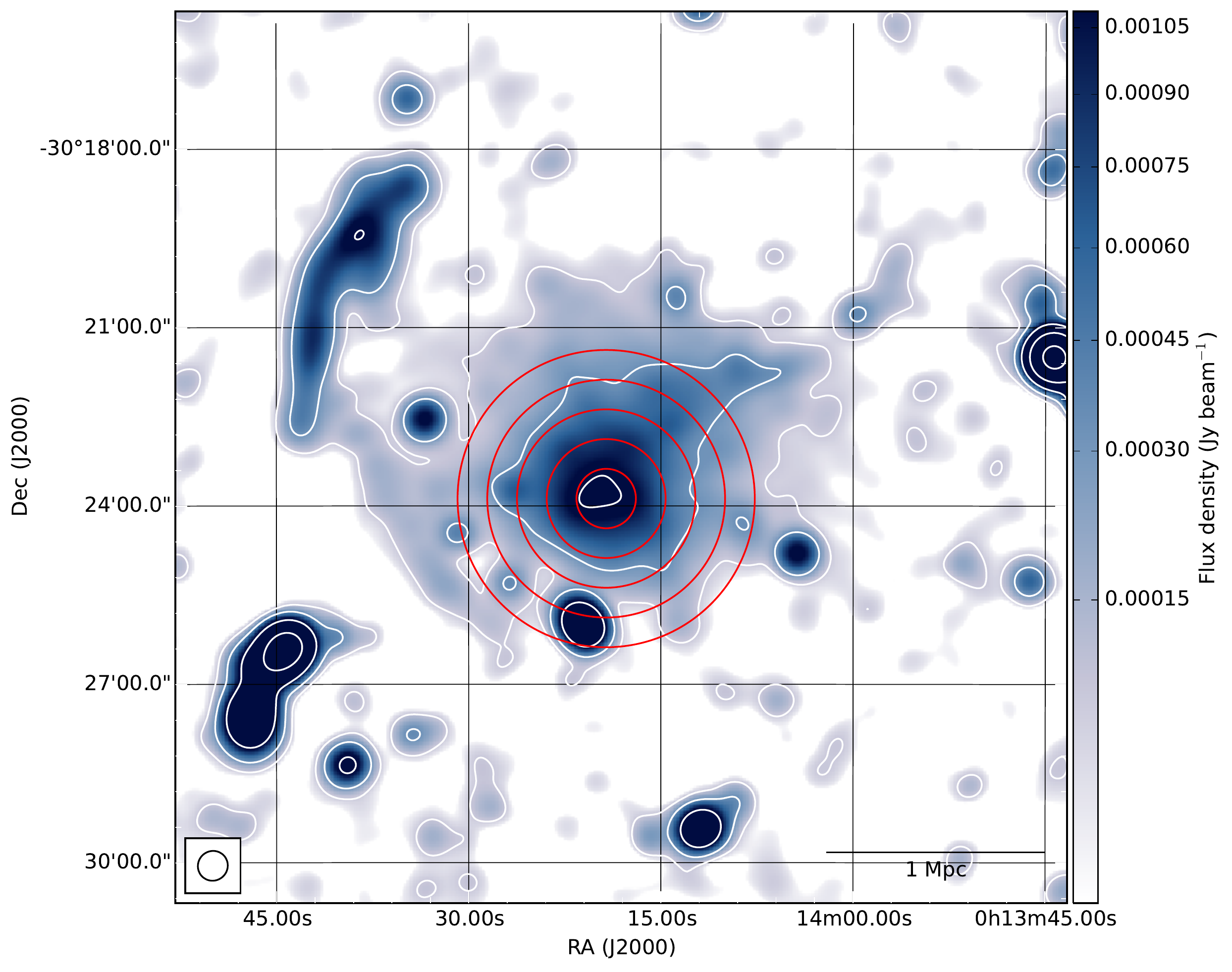}
\includegraphics[width=0.45\textwidth, angle=0]{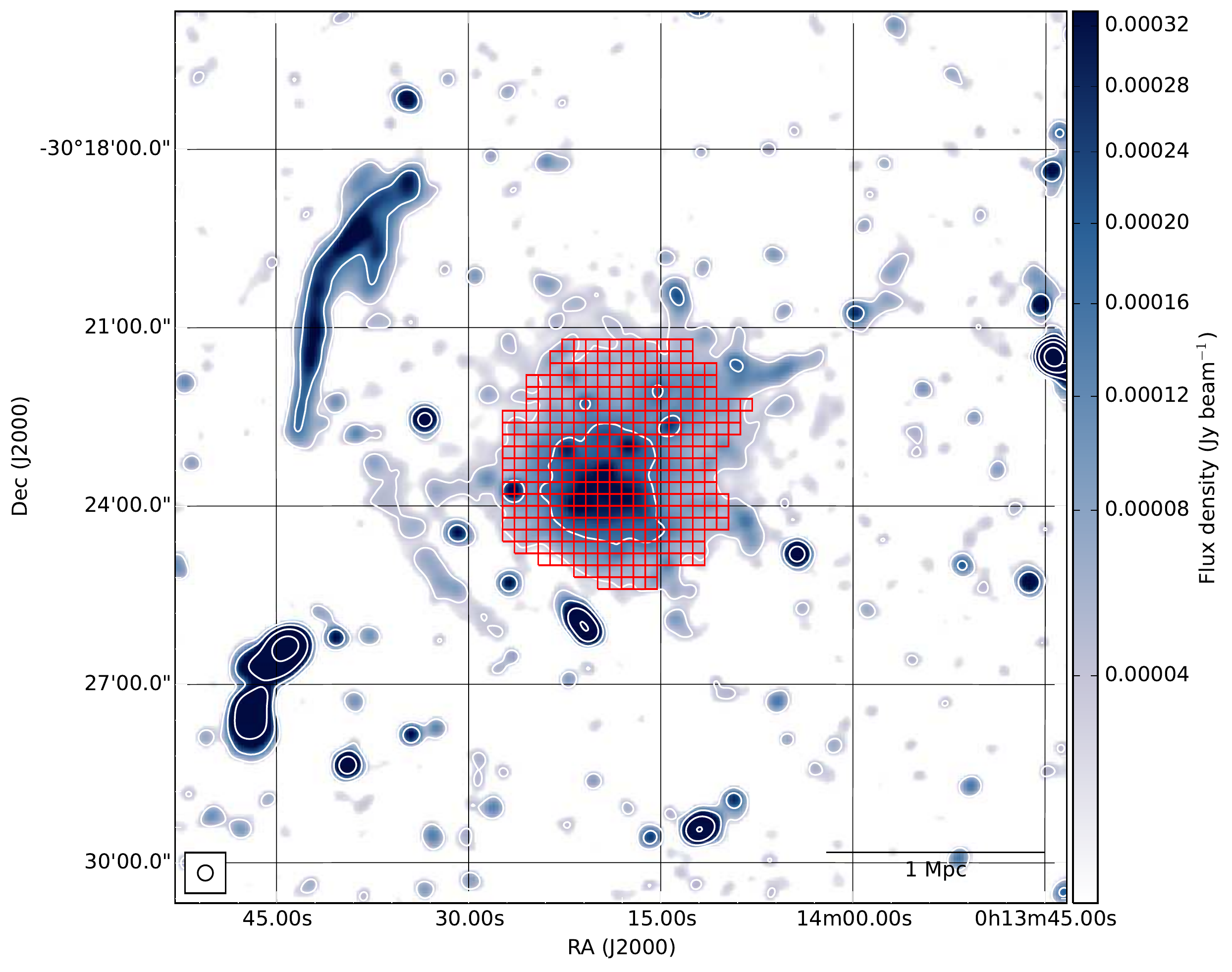}
\includegraphics[width=0.45\textwidth, angle=0]{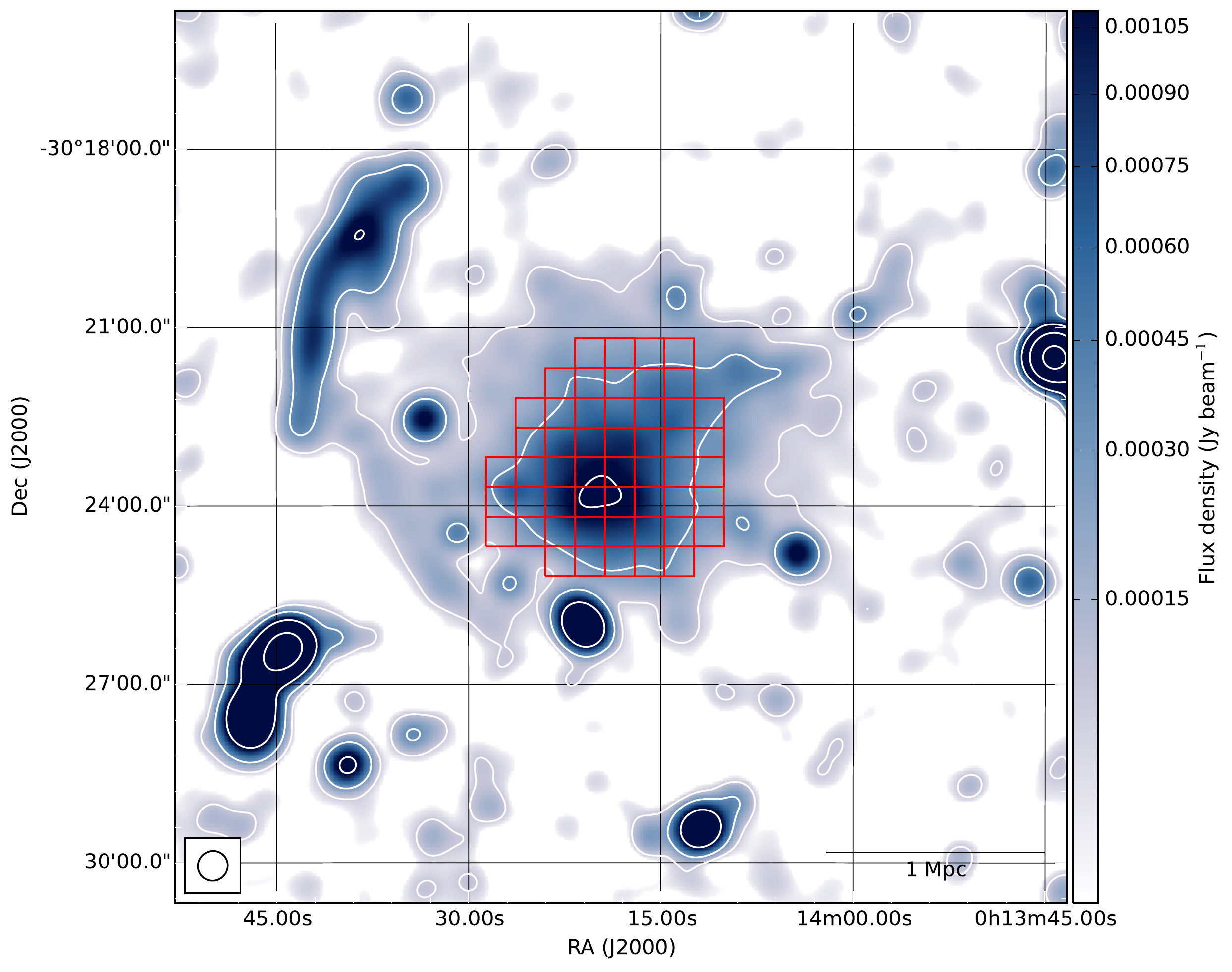}
\caption{Regions where the spectral indices were extracted. Top: Annuli to compute the spectral index radial profile of the halo, see Figure~\ref{radial_30}. These annuli have widths of  $15\arcsec$ (\textit{left}) and $30\arcsec$ (\textit{right}) and are overlaid on the L+S-band combined images tapered to resolutions of $15\arcsec$ and  $30\arcsec$. Bottom: The same images as in the top panel, but in this case, boxes are shown in which we extracted the spectral indices, see Figure~\ref{fig:hist}.}
\label{fig:annuliimage}
\end{figure}

\begin{figure}[h!]
\centering
\includegraphics[width=0.75\textwidth, angle=0, trim={0 2cm 0 0},clip]{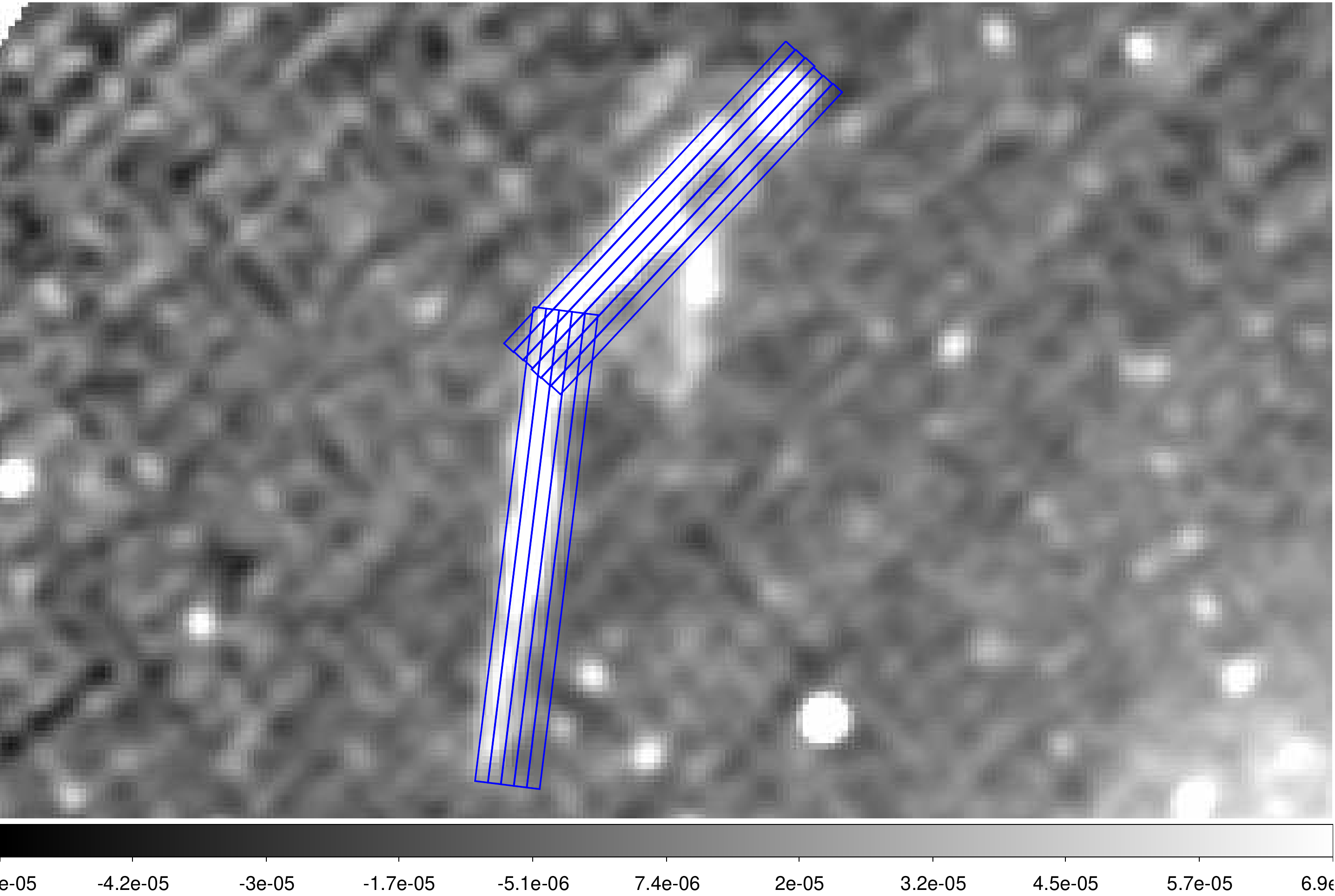}
\caption{Image showing the $5\arcsec$ wide regions used to extract the spectral index gradient across the relic. In greyscale is the S-band $10\arcsec$ resolution image.}
\label{fig:inj_regions}
\end{figure}

\begin{figure}[h!]
\centering
\includegraphics[angle=180, trim={0 0 0 0},width=0.49\textwidth]{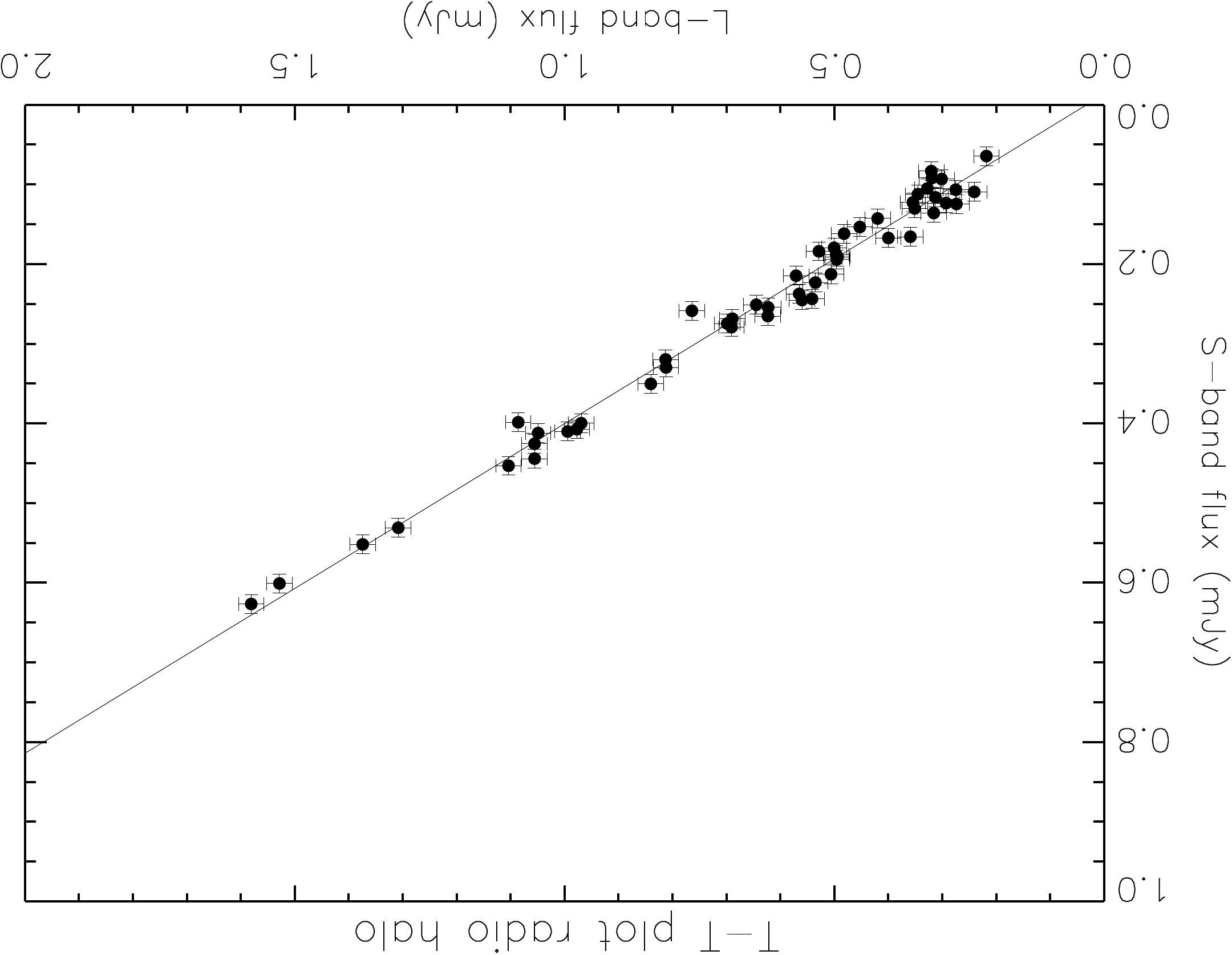}
\includegraphics[angle=180, trim={0 0 0 0},width=0.49\textwidth]{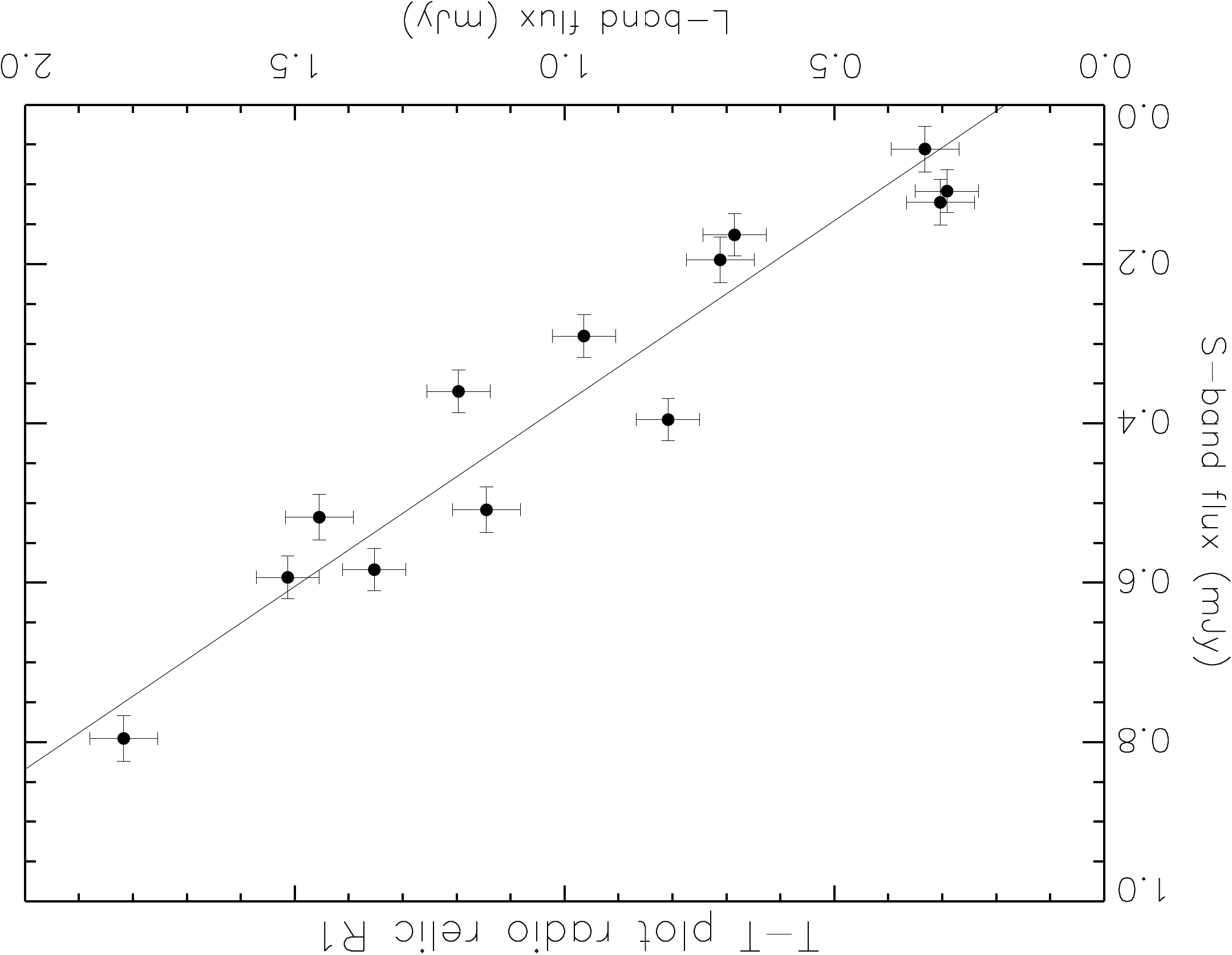}
\caption{{T-T plots  \citep{1962MNRAS.124..297T} for the radio halo (left) and relic (right), see Sections~\ref{sec:spix} and~\ref{sec:R1} for more details.}}
\label{fig:tt}
\end{figure}

\section{MCMC corner plot}
\label{sec:corner}

\begin{figure}[h!]
\centering
\includegraphics[width=1.0\textwidth, angle=0]{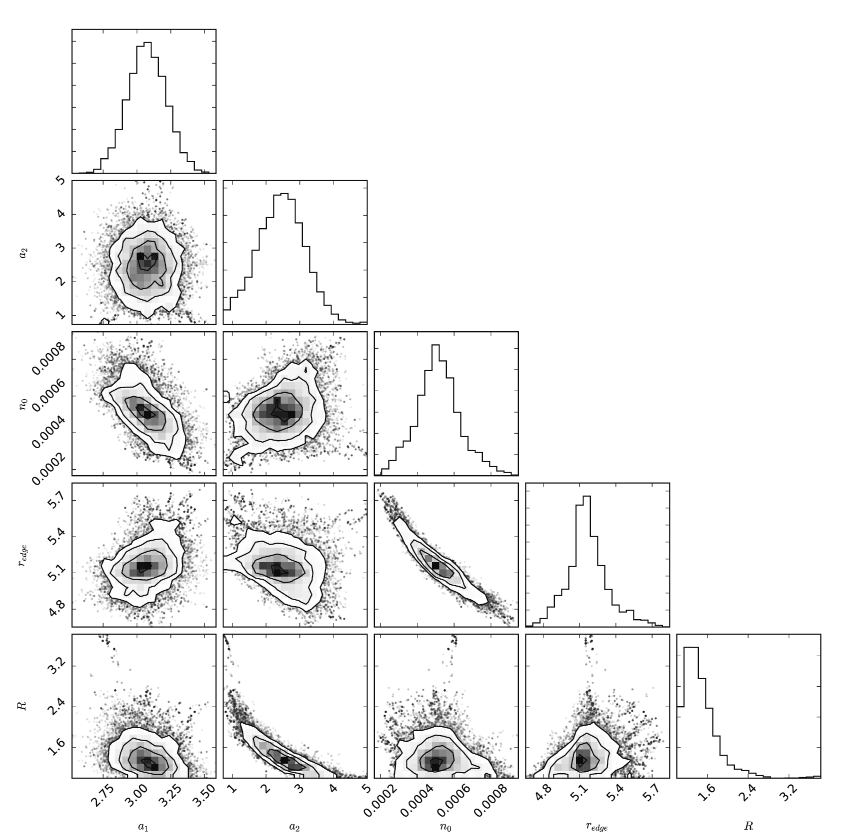}
\caption{{The MCMC ``corner plot'' \citep{corner,2017ascl.soft02002F} for the distribution of the uncertainties in the fitted parameters for the X-ray surface brightness profile across relic R2 (see Section~\ref{sec:R2} and Equation~\ref{eq:pyxel}). Contour levels are drawn at $[0.5,1.0,1.5,2.0]\sigma$.}}
\label{fig:corner}
\end{figure}

\bibliography{ref_filaments.bib}

\begin{thebibliography}{92}
\expandafter\ifx\csname natexlab\endcsname\relax\def\natexlab#1{#1}\fi

\bibitem[{{Anders} \& {Grevesse}(1989)}]{1989GeCoA..53..197A}
{Anders}, E., \& {Grevesse}, N. 1989, \gca, 53, 197

\bibitem[{{Arnaud}(1996)}]{1996ASPC..101...17A}
{Arnaud}, K.~A. 1996, in Astronomical Society of the Pacific Conference Series,
  Vol. 101, Astronomical Data Analysis Software and Systems V, ed. G.~H.
  {Jacoby} \& J.~{Barnes}, 17

\bibitem[{{Astropy Collaboration} {et~al.}(2013){Astropy Collaboration},
  {Robitaille}, {Tollerud}, {Greenfield}, {Droettboom}, {Bray}, {Aldcroft},
  {Davis}, {Ginsburg}, {Price-Whelan}, {Kerzendorf}, {Conley}, {Crighton},
  {Barbary}, {Muna}, {Ferguson}, {Grollier}, {Parikh}, {Nair}, {Unther},
  {Deil}, {Woillez}, {Conseil}, {Kramer}, {Turner}, {Singer}, {Fox}, {Weaver},
  {Zabalza}, {Edwards}, {Azalee Bostroem}, {Burke}, {Casey}, {Crawford},
  {Dencheva}, {Ely}, {Jenness}, {Labrie}, {Lim}, {Pierfederici}, {Pontzen},
  {Ptak}, {Refsdal}, {Servillat}, \& {Streicher}}]{2013A&A...558A..33A}
{Astropy Collaboration}, {Robitaille}, T.~P., {Tollerud}, E.~J., {et~al.} 2013,
  \aap, 558, A33

\bibitem[{{Bennett} {et~al.}(2014){Bennett}, {Larson}, {Weiland}, \&
  {Hinshaw}}]{2014ApJ...794..135B}
{Bennett}, C.~L., {Larson}, D., {Weiland}, J.~L., \& {Hinshaw}, G. 2014, \apj,
  794, 135

\bibitem[{{Bonafede} {et~al.}(2009){Bonafede}, {Giovannini}, {Feretti},
  {Govoni}, \& {Murgia}}]{2009A&A...494..429B}
{Bonafede}, A., {Giovannini}, G., {Feretti}, L., {Govoni}, F., \& {Murgia}, M.
  2009, \aap, 494, 429

\bibitem[{{Bonafede} {et~al.}(2014){Bonafede}, {Intema}, {Br{\"u}ggen},
  {Girardi}, {Nonino}, {Kantharia}, {van Weeren}, \&
  {R{\"o}ttgering}}]{2014ApJ...785....1B}
{Bonafede}, A., {Intema}, H.~T., {Br{\"u}ggen}, M., {et~al.} 2014, \apj, 785, 1

\bibitem[{{Bonafede} {et~al.}(2012){Bonafede}, {Br{\"u}ggen}, {van Weeren},
  {Vazza}, {Giovannini}, {Ebeling}, {Edge}, {Hoeft}, \&
  {Klein}}]{2012MNRAS.426...40B}
{Bonafede}, A., {Br{\"u}ggen}, M., {van Weeren}, R., {et~al.} 2012, \mnras,
  426, 40

\bibitem[{{Boschin} {et~al.}(2006){Boschin}, {Girardi}, {Spolaor}, \&
  {Barrena}}]{2006A&A...449..461B}
{Boschin}, W., {Girardi}, M., {Spolaor}, M., \& {Barrena}, R. 2006, \aap, 449,
  461

\bibitem[{{Botteon} {et~al.}(2016){Botteon}, {Gastaldello}, {Brunetti}, \&
  {Dallacasa}}]{2016MNRAS.460L..84B}
{Botteon}, A., {Gastaldello}, F., {Brunetti}, G., \& {Dallacasa}, D. 2016,
  \mnras, 460, L84

\bibitem[{{Braglia} {et~al.}(2009){Braglia}, {Pierini}, {Biviano}, \&
  {B{\"o}hringer}}]{2009A&A...500..947B}
{Braglia}, F.~G., {Pierini}, D., {Biviano}, A., \& {B{\"o}hringer}, H. 2009,
  \aap, 500, 947

\bibitem[{{Briggs}(1995)}]{briggs_phd}
{Briggs}, D.~S. 1995, PhD thesis, New Mexico Institute of Mining Technology,
  Socorro, New Mexico, USA

\bibitem[{{Brunetti} \& {Jones}(2014)}]{2014IJMPD..2330007B}
{Brunetti}, G., \& {Jones}, T.~W. 2014, International Journal of Modern Physics
  D, 23, 30007

\bibitem[{{Brunetti} \& {Lazarian}(2007)}]{2007MNRAS.378..245B}
{Brunetti}, G., \& {Lazarian}, A. 2007, \mnras, 378, 245

\bibitem[{{Brunetti} {et~al.}(2001){Brunetti}, {Setti}, {Feretti}, \&
  {Giovannini}}]{2001MNRAS.320..365B}
{Brunetti}, G., {Setti}, G., {Feretti}, L., \& {Giovannini}, G. 2001, \mnras,
  320, 365

\bibitem[{{Cassano}(2010)}]{2010A&A...517A..10C}
{Cassano}, R. 2010, \aap, 517, A10+

\bibitem[{{Clarke} \& {En{\ss}lin}(2006)}]{2006AJ....131.2900C}
{Clarke}, T.~E., \& {En{\ss}lin}, T.~A. 2006, \aj, 131, 2900

\bibitem[{{Colless} {et~al.}(2003){Colless}, {Peterson}, {Jackson}, {Peacock},
  {Cole}, {Norberg}, {Baldry}, {Baugh}, {Bland-Hawthorn}, {Bridges}, {Cannon},
  {Collins}, {Couch}, {Cross}, {Dalton}, {De Propris}, {Driver}, {Efstathiou},
  {Ellis}, {Frenk}, {Glazebrook}, {Lahav}, {Lewis}, {Lumsden}, {Maddox},
  {Madgwick}, {Sutherland}, \& {Taylor}}]{2003astro.ph..6581C}
{Colless}, M., {Peterson}, B.~A., {Jackson}, C., {et~al.} 2003, ArXiv
  Astrophysics e-prints

\bibitem[{{Cornwell}(2008)}]{2008ISTSP...2..793C}
{Cornwell}, T.~J. 2008, IEEE Journal of Selected Topics in Signal Processing,
  vol.~2, issue 5, pp.~793-801, 2, 793

\bibitem[{{Cortese} {et~al.}(2007){Cortese}, {Marcillac}, {Richard},
  {Bravo-Alfaro}, {Kneib}, {Rieke}, {Covone}, {Egami}, {Rigby}, {Czoske}, \&
  {Davies}}]{2007IAUS..235..198C}
{Cortese}, L., {Marcillac}, D., {Richard}, J., {et~al.} 2007, in IAU Symposium,
  Vol. 235, Galaxy Evolution across the Hubble Time, ed. F.~{Combes} \&
  J.~{Palou{\v s}}, 198--198

\bibitem[{{de Gasperin} {et~al.}(2015){de Gasperin}, {Ogrean}, {van Weeren},
  {Dawson}, {Br{\"u}ggen}, {Bonafede}, \& {Simionescu}}]{2015MNRAS.448.2197D}
{de Gasperin}, F., {Ogrean}, G.~A., {van Weeren}, R.~J., {et~al.} 2015, \mnras,
  448, 2197

\bibitem[{{Dennison}(1980)}]{1980ApJ...239L..93D}
{Dennison}, B. 1980, \apjl, 239, L93

\bibitem[{{Drury}(1983)}]{1983RPPh...46..973D}
{Drury}, L.~O. 1983, Reports on Progress in Physics, 46, 973

\bibitem[{{Ebeling} {et~al.}(2014){Ebeling}, {Stephenson}, \&
  {Edge}}]{2014ApJ...781L..40E}
{Ebeling}, H., {Stephenson}, L.~N., \& {Edge}, A.~C. 2014, \apjl, 781, L40

\bibitem[{{Eckert} {et~al.}(2016){Eckert}, {Jauzac}, {Vazza}, {Owers}, {Kneib},
  {Tchernin}, {Intema}, \& {Knowles}}]{2016arXiv160302272E}
{Eckert}, D., {Jauzac}, M., {Vazza}, F., {et~al.} 2016, ArXiv e-prints

\bibitem[{{En{\ss}lin} {et~al.}(1998){En{\ss}lin}, {Biermann}, {Klein}, \&
  {Kohle}}]{1998A&A...332..395E}
{En{\ss}lin}, T.~A., {Biermann}, P.~L., {Klein}, U., \& {Kohle}, S. 1998, \aap,
  332, 395

\bibitem[{{En{\ss}lin} \& {Gopal-Krishna}(2001)}]{2001A&A...366...26E}
{En{\ss}lin}, T.~A., \& {Gopal-Krishna}. 2001, \aap, 366, 26

\bibitem[{{Feretti} {et~al.}(2012){Feretti}, {Giovannini}, {Govoni}, \&
  {Murgia}}]{2012A&ARv..20...54F}
{Feretti}, L., {Giovannini}, G., {Govoni}, F., \& {Murgia}, M. 2012, \aapr, 20,
  54

\bibitem[{{Feretti} {et~al.}(2004){Feretti}, {Orr{\`u}}, {Brunetti},
  {Giovannini}, {Kassim}, \& {Setti}}]{2004A&A...423..111F}
{Feretti}, L., {Orr{\`u}}, E., {Brunetti}, G., {et~al.} 2004, \aap, 423, 111

\bibitem[{{Ferrari} {et~al.}(2008){Ferrari}, {Govoni}, {Schindler}, {Bykov}, \&
  {Rephaeli}}]{2008SSRv..134...93F}
{Ferrari}, C., {Govoni}, F., {Schindler}, S., {Bykov}, A.~M., \& {Rephaeli}, Y.
  2008, Space Science Reviews, 134, 93

\bibitem[{Foreman-Mackey(2016)}]{corner}
Foreman-Mackey, D. 2016, The Journal of Open Source Software, 24

\bibitem[{{Foreman-Mackey}(2017)}]{2017ascl.soft02002F}
{Foreman-Mackey}, D. 2017, {corner: Corner plots}, Astrophysics Source Code
  Library

\bibitem[{{Foreman-Mackey} {et~al.}(2013){Foreman-Mackey}, {Hogg}, {Lang}, \&
  {Goodman}}]{2013PASP..125..306F}
{Foreman-Mackey}, D., {Hogg}, D.~W., {Lang}, D., \& {Goodman}, J. 2013, \pasp,
  125, 306

\bibitem[{{Fujita} {et~al.}(2015){Fujita}, {Takizawa}, {Yamazaki}, {Akamatsu},
  \& {Ohno}}]{2015ApJ...815..116F}
{Fujita}, Y., {Takizawa}, M., {Yamazaki}, R., {Akamatsu}, H., \& {Ohno}, H.
  2015, \apj, 815, 116

\bibitem[{{Giacintucci} {et~al.}(2008){Giacintucci}, {Venturi}, {Macario},
  {Dallacasa}, {Brunetti}, {Markevitch}, {Cassano}, {Bardelli}, \&
  {Athreya}}]{2008A&A...486..347G}
{Giacintucci}, S., {Venturi}, T., {Macario}, G., {et~al.} 2008, \aap, 486, 347

\bibitem[{{Giovannini} {et~al.}(1991){Giovannini}, {Feretti}, \&
  {Stanghellini}}]{1991A&A...252..528G}
{Giovannini}, G., {Feretti}, L., \& {Stanghellini}, C. 1991, \aap, 252, 528

\bibitem[{{Giovannini} {et~al.}(1999){Giovannini}, {Tordi}, \&
  {Feretti}}]{1999NewA....4..141G}
{Giovannini}, G., {Tordi}, M., \& {Feretti}, L. 1999, New Astronomy, 4, 141

\bibitem[{{Govoni} {et~al.}(2001){Govoni}, {Feretti}, {Giovannini},
  {B{\"o}hringer}, {Reiprich}, \& {Murgia}}]{2001A&A...376..803G}
{Govoni}, F., {Feretti}, L., {Giovannini}, G., {et~al.} 2001, \aap, 376, 803

\bibitem[{{Hindson} {et~al.}(2014){Hindson}, {Johnston-Hollitt},
  {Hurley-Walker}, {Buckley}, {Morgan}, {Carretti}, {Dwarakanath}, {Bell},
  {Bernardi}, {Bhat}, {Bowman}, {Briggs}, {Cappallo}, {Corey}, {Deshpande},
  {Emrich}, {Ewall-Wice}, {Feng}, {Gaensler}, {Goeke}, {Greenhill}, {Hazelton},
  {Jacobs}, {Kaplan}, {Kasper}, {Kratzenberg}, {Kudryavtseva}, {Lenc},
  {Lonsdale}, {Lynch}, {McWhirter}, {McKinley}, {Mitchell}, {Morales},
  {Morgan}, {Oberoi}, {Ord}, {Pindor}, {Prabu}, {Procopio}, {Offringa},
  {Riding}, {Rogers}, {Roshi}, {Shankar}, {Srivani}, {Subrahmanyan}, {Tingay},
  {Waterson}, {Wayth}, {Webster}, {Whitney}, {Williams}, \&
  {Williams}}]{2014MNRAS.445..330H}
{Hindson}, L., {Johnston-Hollitt}, M., {Hurley-Walker}, N., {et~al.} 2014,
  \mnras, 445, 330

\bibitem[{{Ibaraki} {et~al.}(2014){Ibaraki}, {Ota}, {Akamatsu}, {Zhang}, \&
  {Finoguenov}}]{2014A&A...562A..11I}
{Ibaraki}, Y., {Ota}, N., {Akamatsu}, H., {Zhang}, Y.-Y., \& {Finoguenov}, A.
  2014, \aap, 562, A11

\bibitem[{{Jaffe}(1977)}]{1977ApJ...212....1J}
{Jaffe}, W.~J. 1977, \apj, 212, 1

\bibitem[{{Jauzac} {et~al.}(2016){Jauzac}, {Eckert}, {Schwinn}, {Harvey},
  {Baugh}, {Robertson}, {Bose}, {Massey}, {Owers}, {Ebeling}, {Shan}, {Jullo},
  {Kneib}, {Richard}, {Atek}, {Cl{\'e}ment}, {Egami}, {Israel}, {Knowles},
  {Limousin}, {Natarajan}, {Rexroth}, {Taylor}, \&
  {Tchernin}}]{2016MNRAS.463.3876J}
{Jauzac}, M., {Eckert}, D., {Schwinn}, J., {et~al.} 2016, \mnras, 463, 3876

\bibitem[{{Kalberla} {et~al.}(2005){Kalberla}, {Burton}, {Hartmann}, {Arnal},
  {Bajaja}, {Morras}, \& {P{\"o}ppel}}]{2005A&A...440..775K}
{Kalberla}, P.~M.~W., {Burton}, W.~B., {Hartmann}, D., {et~al.} 2005, \aap,
  440, 775

\bibitem[{{Kang}(2015)}]{2015JKAS...48....9K}
{Kang}, H. 2015, Journal of Korean Astronomical Society, 48, 9

\bibitem[{{Kang} \& {Ryu}(2011)}]{2011ApJ...734...18K}
{Kang}, H., \& {Ryu}, D. 2011, \apj, 734, 18

\bibitem[{{Kang} \& {Ryu}(2015)}]{2015ApJ...809..186K}
---. 2015, \apj, 809, 186

\bibitem[{{Kempner} {et~al.}(2004){Kempner}, {Blanton}, {Clarke}, {En{\ss}lin},
  {Johnston-Hollitt}, \& {Rudnick}}]{2004rcfg.proc..335K}
{Kempner}, J.~C., {Blanton}, E.~L., {Clarke}, T.~E., {et~al.} 2004, in The
  Riddle of Cooling Flows in Galaxies and Clusters of galaxies, ed.
  T.~{Reiprich}, J.~{Kempner}, \& N.~{Soker}, 335--+

\bibitem[{{Kempner} \& {David}(2004)}]{2004MNRAS.349..385K}
{Kempner}, J.~C., \& {David}, L.~P. 2004, \mnras, 349, 385

\bibitem[{{Landau} \& {Lifshitz}(1959)}]{1959flme.book.....L}
{Landau}, L.~D., \& {Lifshitz}, E.~M. 1959, {Fluid mechanics}, ed. {Landau,
  L.~D.~\& Lifshitz, E.~M.}

\bibitem[{{Lotz} {et~al.}(2014){Lotz}, {Mountain}, {Grogin}, {Koekemoer},
  {Capak}, {Mack}, {Coe}, {Barker}, {Adler}, {Avila}, {Anderson}, {Casertano},
  {Christian}, {Gonzaga}, {Ferguson}, {Fruchter}, {Jenkner}, {Jordan},
  {Hammer}, {Hilbert}, {Lawton}, {Lee}, {Lucas}, {MacKenty}, {Mutchler},
  {Ogaz}, {Reid}, {Royle}, {Robberto}, {Sembach}, {Smith}, {Sokol}, {Surace},
  {Taylor}, {Tumlinson}, {Viana}, {Williams}, \&
  {Workman}}]{2014AAS...22325401L}
{Lotz}, J., {Mountain}, M., {Grogin}, N.~A., {et~al.} 2014, in American
  Astronomical Society Meeting Abstracts, Vol. 223, American Astronomical
  Society Meeting Abstracts \#223, 254.01

\bibitem[{{Lotz} {et~al.}(2016){Lotz}, {Koekemoer}, {Coe}, {Grogin}, {Capak},
  {Mack}, {Anderson}, {Avila}, {Barker}, {Borncamp}, {Brammer}, {Durbin},
  {Gunning}, {Hilbert}, {Jenkner}, {Khandrika}, {Levay}, {Lucas}, {MacKenty},
  {Ogaz}, {Porterfield}, {Reid}, {Robberto}, {Royle}, {Smith},
  {Storrie-Lombardi}, {Sunnquist}, {Surace}, {Taylor}, {Williams}, {Bullock},
  {Dickinson}, {Finkelstein}, {Natarajan}, {Richard}, {Robertson}, {Tumlinson},
  {Zitrin}, {Flanagan}, {Sembach}, {Soifer}, \&
  {Mountain}}]{2016arXiv160506567L}
{Lotz}, J.~M., {Koekemoer}, A., {Coe}, D., {et~al.} 2016, ArXiv e-prints

\bibitem[{{Macario} {et~al.}(2011){Macario}, {Markevitch}, {Giacintucci},
  {Brunetti}, {Venturi}, \& {Murray}}]{2011ApJ...728...82M}
{Macario}, G., {Markevitch}, M., {Giacintucci}, S., {et~al.} 2011, \apj, 728,
  82

\bibitem[{{Markevitch}(2010)}]{2010arXiv1010.3660M}
{Markevitch}, M. 2010, ArXiv e-prints

\bibitem[{{Markevitch} {et~al.}(2002){Markevitch}, {Gonzalez}, {David},
  {Vikhlinin}, {Murray}, {Forman}, {Jones}, \& {Tucker}}]{2002ApJ...567L..27M}
{Markevitch}, M., {Gonzalez}, A.~H., {David}, L., {et~al.} 2002, \apjl, 567,
  L27

\bibitem[{{Markevitch} {et~al.}(2005){Markevitch}, {Govoni}, {Brunetti}, \&
  {Jerius}}]{2005ApJ...627..733M}
{Markevitch}, M., {Govoni}, F., {Brunetti}, G., \& {Jerius}, D. 2005, \apj,
  627, 733

\bibitem[{{Markwardt}(2009)}]{2009ASPC..411..251M}
{Markwardt}, C.~B. 2009, in Astronomical Society of the Pacific Conference
  Series, Vol. 411, Astronomical Data Analysis Software and Systems XVIII, ed.
  D.~A. {Bohlender}, D.~{Durand}, \& P.~{Dowler}, 251

\bibitem[{{McMullin} {et~al.}(2007){McMullin}, {Waters}, {Schiebel}, {Young},
  \& {Golap}}]{2007ASPC..376..127M}
{McMullin}, J.~P., {Waters}, B., {Schiebel}, D., {Young}, W., \& {Golap}, K.
  2007, in Astronomical Society of the Pacific Conference Series, Vol. 376,
  Astronomical Data Analysis Software and Systems XVI, ed. R.~A. {Shaw},
  F.~{Hill}, \& D.~J. {Bell}, 127

\bibitem[{{Medezinski} {et~al.}(2016){Medezinski}, {Umetsu}, {Okabe}, {Nonino},
  {Molnar}, {Massey}, {Dupke}, \& {Merten}}]{2016ApJ...817...24M}
{Medezinski}, E., {Umetsu}, K., {Okabe}, N., {et~al.} 2016, \apj, 817, 24

\bibitem[{{Merten} {et~al.}(2011){Merten}, {Coe}, {Dupke}, {Massey}, {Zitrin},
  {Cypriano}, {Okabe}, {Frye}, {Braglia}, {Jim{\'e}nez-Teja}, {Ben{\'{\i}}tez},
  {Broadhurst}, {Rhodes}, {Meneghetti}, {Moustakas}, {Sodr{\'e}}, {Krick}, \&
  {Bregman}}]{2011MNRAS.417..333M}
{Merten}, J., {Coe}, D., {Dupke}, R., {et~al.} 2011, \mnras, 417, 333

\bibitem[{{Mohan} \& {Rafferty}(2015)}]{2015ascl.soft02007M}
{Mohan}, N., \& {Rafferty}, D. 2015, {PyBDSM: Python Blob Detection and Source
  Measurement}, Astrophysics Source Code Library

\bibitem[{{Murgia} {et~al.}(2009){Murgia}, {Govoni}, {Markevitch}, {Feretti},
  {Giovannini}, {Taylor}, \& {Carretti}}]{2009A&A...499..679M}
{Murgia}, M., {Govoni}, F., {Markevitch}, M., {et~al.} 2009, \aap, 499, 679

\bibitem[{{Offringa} {et~al.}(2014){Offringa}, {McKinley}, {Hurley-Walker},
  {Briggs}, {Wayth}, {Kaplan}, {Bell}, {Feng}, {Neben}, {Hughes}, {Rhee},
  {Murphy}, {Bhat}, {Bernardi}, {Bowman}, {Cappallo}, {Corey}, {Deshpande},
  {Emrich}, {Ewall-Wice}, {Gaensler}, {Goeke}, {Greenhill}, {Hazelton},
  {Hindson}, {Johnston-Hollitt}, {Jacobs}, {Kasper}, {Kratzenberg}, {Lenc},
  {Lonsdale}, {Lynch}, {McWhirter}, {Mitchell}, {Morales}, {Morgan},
  {Kudryavtseva}, {Oberoi}, {Ord}, {Pindor}, {Procopio}, {Prabu}, {Riding},
  {Roshi}, {Shankar}, {Srivani}, {Subrahmanyan}, {Tingay}, {Waterson},
  {Webster}, {Whitney}, {Williams}, \& {Williams}}]{2014MNRAS.444..606O}
{Offringa}, A.~R., {McKinley}, B., {Hurley-Walker}, N., {et~al.} 2014, \mnras,
  444, 606

\bibitem[{{Ogrean}(2016)}]{2016AAS...22831709O}
{Ogrean}, G. 2016, in American Astronomical Society Meeting Abstracts, Vol.
  228, American Astronomical Society Meeting Abstracts, 317.09

\bibitem[{{Ogrean}(2017)}]{2017AAS...22943808O}
{Ogrean}, G. 2017, in American Astronomical Society Meeting Abstracts, Vol.
  229, American Astronomical Society Meeting Abstracts, 438.08

\bibitem[{{Orr{\'u}} {et~al.}(2007){Orr{\'u}}, {Murgia}, {Feretti}, {Govoni},
  {Brunetti}, {Giovannini}, {Girardi}, \& {Setti}}]{2007A&A...467..943O}
{Orr{\'u}}, E., {Murgia}, M., {Feretti}, L., {et~al.} 2007, \aap, 467, 943

\bibitem[{{Owers} {et~al.}(2012){Owers}, {Couch}, {Nulsen}, \&
  {Randall}}]{2012ApJ...750L..23O}
{Owers}, M.~S., {Couch}, W.~J., {Nulsen}, P.~E.~J., \& {Randall}, S.~W. 2012,
  \apjl, 750, L23

\bibitem[{{Owers} {et~al.}(2011){Owers}, {Randall}, {Nulsen}, {Couch}, {David},
  \& {Kempner}}]{2011ApJ...728...27O}
{Owers}, M.~S., {Randall}, S.~W., {Nulsen}, P.~E.~J., {et~al.} 2011, \apj, 728,
  27

\bibitem[{{Perley} \& {Butler}(2013)}]{2013ApJS..204...19P}
{Perley}, R.~A., \& {Butler}, B.~J. 2013, \apjs, 204, 19

\bibitem[{{Petrosian}(2001)}]{2001ApJ...557..560P}
{Petrosian}, V. 2001, \apj, 557, 560

\bibitem[{{Pinzke} {et~al.}(2013){Pinzke}, {Oh}, \&
  {Pfrommer}}]{2013MNRAS.435.1061P}
{Pinzke}, A., {Oh}, S.~P., \& {Pfrommer}, C. 2013, \mnras, 435, 1061

\bibitem[{{Planck Collaboration} {et~al.}(2013){Planck Collaboration}, {Ade},
  {Aghanim}, {Arnaud}, {Ashdown}, {Atrio-Barandela}, {Aumont}, {Baccigalupi},
  {Balbi}, {Banday}, \& et~al.}]{2013A&A...554A.140P}
{Planck Collaboration}, {Ade}, P.~A.~R., {Aghanim}, N., {et~al.} 2013, \aap,
  554, A140

\bibitem[{{Rau} \& {Cornwell}(2011)}]{2011A&A...532A..71R}
{Rau}, U., \& {Cornwell}, T.~J. 2011, \aap, 532, A71

\bibitem[{{Robitaille} \& {Bressert}(2012)}]{2012ascl.soft08017R}
{Robitaille}, T., \& {Bressert}, E. 2012, {APLpy: Astronomical Plotting Library
  in Python}, Astrophysics Source Code Library

\bibitem[{{Rybicki} \& {Lightman}(1986)}]{1986rpa..book.....R}
{Rybicki}, G.~B., \& {Lightman}, A.~P. 1986, {Radiative Processes in
  Astrophysics}, 400

\bibitem[{{Sanders}(2006)}]{2006MNRAS.371..829S}
{Sanders}, J.~S. 2006, \mnras, 371, 829

\bibitem[{{Schlickeiser} {et~al.}(1987){Schlickeiser}, {Sievers}, \&
  {Thiemann}}]{1987A&A...182...21S}
{Schlickeiser}, R., {Sievers}, A., \& {Thiemann}, H. 1987, \aap, 182, 21

\bibitem[{{Shimwell} {et~al.}(2014){Shimwell}, {Brown}, {Feain}, {Feretti},
  {Gaensler}, \& {Lage}}]{2014MNRAS.440.2901S}
{Shimwell}, T.~W., {Brown}, S., {Feain}, I.~J., {et~al.} 2014, \mnras, 440,
  2901

\bibitem[{{Shimwell} {et~al.}(2015){Shimwell}, {Markevitch}, {Brown},
  {Feretti}, {Gaensler}, {Johnston-Hollitt}, {Lage}, \&
  {Srinivasan}}]{2015MNRAS.449.1486S}
{Shimwell}, T.~W., {Markevitch}, M., {Brown}, S., {et~al.} 2015, \mnras, 449,
  1486

\bibitem[{{Taylor} {et~al.}(2009){Taylor}, {Stil}, \&
  {Sunstrum}}]{2009ApJ...702.1230T}
{Taylor}, A.~R., {Stil}, J.~M., \& {Sunstrum}, C. 2009, \apj, 702, 1230

\bibitem[{{Trasatti} {et~al.}(2015){Trasatti}, {Akamatsu}, {Lovisari}, {Klein},
  {Bonafede}, {Br{\"u}ggen}, {Dallacasa}, \& {Clarke}}]{2015A&A...575A..45T}
{Trasatti}, M., {Akamatsu}, H., {Lovisari}, L., {et~al.} 2015, \aap, 575, A45

\bibitem[{{Turtle} {et~al.}(1962){Turtle}, {Pugh}, {Kenderdine}, \&
  {Pauliny-Toth}}]{1962MNRAS.124..297T}
{Turtle}, A.~J., {Pugh}, J.~F., {Kenderdine}, S., \& {Pauliny-Toth}, I.~I.~K.
  1962, \mnras, 124, 297

\bibitem[{{Uchida} {et~al.}(2016){Uchida}, {Simionescu}, {Takahashi}, {Werner},
  {Ichinohe}, {Allen}, {Urban}, \& {Matsushita}}]{2016PASJ...68S..20U}
{Uchida}, Y., {Simionescu}, A., {Takahashi}, T., {et~al.} 2016, \pasj, 68, S20

\bibitem[{{Vacca} {et~al.}(2014){Vacca}, {Feretti}, {Giovannini}, {Govoni},
  {Murgia}, {Perley}, \& {Clarke}}]{2014A&A...561A..52V}
{Vacca}, V., {Feretti}, L., {Giovannini}, G., {et~al.} 2014, \aap, 561, A52

\bibitem[{{van Weeren} {et~al.}(2011){van Weeren}, {R{\"o}ttgering}, \&
  {Br{\"u}ggen}}]{2011A&A...527A.114V}
{van Weeren}, R.~J., {R{\"o}ttgering}, H.~J.~A., \& {Br{\"u}ggen}, M. 2011,
  \aap, 527, A114+

\bibitem[{{van Weeren} {et~al.}(2010){van Weeren}, {R{\"o}ttgering},
  {Br{\"u}ggen}, \& {Hoeft}}]{2010Sci...330..347V}
{van Weeren}, R.~J., {R{\"o}ttgering}, H.~J.~A., {Br{\"u}ggen}, M., \& {Hoeft},
  M. 2010, Science, 330, 347

\bibitem[{{van Weeren} {et~al.}(2012{\natexlab{a}}){van Weeren},
  {R{\"o}ttgering}, {Intema}, {Rudnick}, {Br{\"u}ggen}, {Hoeft}, \&
  {Oonk}}]{2012A&A...546A.124V}
{van Weeren}, R.~J., {R{\"o}ttgering}, H.~J.~A., {Intema}, H.~T., {et~al.}
  2012{\natexlab{a}}, \aap, 546, A124

\bibitem[{{van Weeren} {et~al.}(2012{\natexlab{b}}){van Weeren},
  {R{\"o}ttgering}, {Rafferty}, {Pizzo}, {Bonafede}, {Br{\"u}ggen}, {Brunetti},
  {Ferrari}, {Orr{\`u}}, {Heald}, {McKean}, {Tasse}, {de Gasperin},
  {B{\^i}rzan}, {van Zwieten}, {van der Tol}, {Shulevski}, {Jackson},
  {Offringa}, {Conway}, {Intema}, {Clarke}, {van Bemmel}, {Miley}, {White},
  {Hoeft}, {Cassano}, {Macario}, {Morganti}, {Wise}, {Horellou}, {Valentijn},
  {Wucknitz}, {Kuijken}, {En{\ss}lin}, {Anderson}, {Asgekar}, {Avruch}, {Beck},
  {Bell}, {Bell}, {Bentum}, {Bernardi}, {Best}, {Boonstra}, {Brentjens}, {van
  de Brink}, {Broderick}, {Brouw}, {Butcher}, {van Cappellen}, {Ciardi},
  {Eisl{\"o}ffel}, {Falcke}, {Fender}, {Garrett}, {Gerbers}, {Gunst}, {van
  Haarlem}, {Hamaker}, {Hassall}, {Hessels}, {Koopmans}, {Kuper}, {van
  Leeuwen}, {Maat}, {Millenaar}, {Munk}, {Nijboer}, {Noordam}, {Pandey},
  {Pandey-Pommier}, {Polatidis}, {Reich}, {Scaife}, {Schoenmakers}, {Sluman},
  {Stappers}, {Steinmetz}, {Swinbank}, {Tagger}, {Tang}, {Vermeulen}, {de Vos},
  \& {van Haarlem}}]{2012A&A...543A..43V}
{van Weeren}, R.~J., {R{\"o}ttgering}, H.~J.~A., {Rafferty}, D.~A., {et~al.}
  2012{\natexlab{b}}, \aap, 543, A43

\bibitem[{{van Weeren} {et~al.}(2016){van Weeren}, {Brunetti}, {Br{\"u}ggen},
  {Andrade-Santos}, {Ogrean}, {Williams}, {R{\"o}ttgering}, {Dawson}, {Forman},
  {de Gasperin}, {Hardcastle}, {Jones}, {Miley}, {Rafferty}, {Rudnick},
  {Sabater}, {Sarazin}, {Shimwell}, {Bonafede}, {Best}, {B{\^i}rzan},
  {Cassano}, {Chy{\.z}y}, {Croston}, {Dijkema}, {En{\ss}lin}, {Ferrari},
  {Heald}, {Hoeft}, {Horellou}, {Jarvis}, {Kraft}, {Mevius}, {Intema},
  {Murray}, {Orr{\'u}}, {Pizzo}, {Sridhar}, {Simionescu}, {Stroe}, {van der
  Tol}, \& {White}}]{2016ApJ...818..204V}
{van Weeren}, R.~J., {Brunetti}, G., {Br{\"u}ggen}, M., {et~al.} 2016, \apj,
  818, 204

\bibitem[{{van Weeren} {et~al.}(2017{\natexlab{a}}){van Weeren}, {Ogrean},
  {Jones}, {Forman}, {Andrade-Santos}, {Pearce}, {Bonafede}, {Br{\"u}ggen},
  {Bulbul}, {Clarke}, {Churazov}, {David}, {Dawson}, {Donahue}, {Goulding},
  {Kraft}, {Mason}, {Merten}, {Mroczkowski}, {Nulsen}, {Rosati}, {Roediger},
  {Randall}, {Sayers}, {Umetsu}, {Vikhlinin}, \&
  {Zitrin}}]{2017ApJ...835..197V}
{van Weeren}, R.~J., {Ogrean}, G.~A., {Jones}, C., {et~al.} 2017{\natexlab{a}},
  \apj, 835, 197

\bibitem[{{van Weeren} {et~al.}(2017{\natexlab{b}}){van Weeren},
  {Andrade-Santos}, {Dawson}, {Golovich}, {Lal}, {Kang}, {Ryu}, {Br{\`i}ggen},
  {Ogrean}, {Forman}, {Jones}, {Placco}, {Santucci}, {Wittman}, {Jee}, {Kraft},
  {Sobral}, {Stroe}, \& {Fogarty}}]{2017NatAs...1E...5V}
{van Weeren}, R.~J., {Andrade-Santos}, F., {Dawson}, W.~A., {et~al.}
  2017{\natexlab{b}}, Nature Astronomy, 1, 0005

\bibitem[{{Venturi} {et~al.}(2013){Venturi}, {Giacintucci}, {Dallacasa},
  {Cassano}, {Brunetti}, {Macario}, \& {Athreya}}]{2013A&A...551A..24V}
{Venturi}, T., {Giacintucci}, S., {Dallacasa}, D., {et~al.} 2013, \aap, 551,
  A24

\bibitem[{{Vikhlinin} {et~al.}(2005){Vikhlinin}, {Markevitch}, {Murray},
  {Jones}, {Forman}, \& {Van Speybroeck}}]{2005ApJ...628..655V}
{Vikhlinin}, A., {Markevitch}, M., {Murray}, S.~S., {et~al.} 2005, \apj, 628,
  655

\bibitem[{{Williams} {et~al.}(2010){Williams}, {Bureau}, \&
  {Cappellari}}]{2010MNRAS.409.1330W}
{Williams}, M.~J., {Bureau}, M., \& {Cappellari}, M. 2010, \mnras, 409, 1330

\end{thebibliography}
\end{document}